\documentclass[preprint,12pt]{aastex}

\begin{document}

\title{SEGUE: A Spectroscopic Survey of 240,000 stars with $g=$14--20}

\author{
Brian Yanny\altaffilmark{\ref{FNAL}},
Constance Rockosi\altaffilmark{\ref{Lick}},
Heidi Jo Newberg\altaffilmark{\ref{RPI}},
Gillian R. Knapp\altaffilmark{\ref{Princeton}},
Jennifer K. Adelman-McCarthy\altaffilmark{\ref{FNAL}},
Bonnie Alcorn\altaffilmark{\ref{FNAL}},
Sahar Allam\altaffilmark{\ref{FNAL}},
Carlos Allende Prieto\altaffilmark{\ref{Texas},\ref{UCL}},
Deokkeun An\altaffilmark{\ref{OSU}},
Kurt S. J. Anderson\altaffilmark{\ref{APO},\ref{NMSU}},
Scott Anderson\altaffilmark{\ref{UW}},
Coryn A.L. Bailer-Jones\altaffilmark{\ref{MPH}},
Steve Bastian\altaffilmark{\ref{FNAL}},
Timothy C. Beers\altaffilmark{\ref{MSUJINA}},
Eric Bell\altaffilmark{\ref{MPH}},
Vasily Belokurov\altaffilmark{\ref{Cambridge}},
Dmitry Bizyaev\altaffilmark{\ref{APO}},
Norm Blythe\altaffilmark{\ref{APO}},
John J. Bochanski\altaffilmark{\ref{UW}},
William N. Boroski\altaffilmark{\ref{FNAL}},
Jarle Brinchmann\altaffilmark{\ref{Porto}},
J. Brinkmann\altaffilmark{\ref{APO}},
Howard Brewington\altaffilmark{\ref{APO}},
Larry Carey\altaffilmark{\ref{UW}},
Kyle M. Cudworth\altaffilmark{\ref{Chicago}},
Michael Evans\altaffilmark{\ref{UW}},
N. W. Evans\altaffilmark{\ref{Cambridge}},
Evalyn Gates\altaffilmark{\ref{Chicago}},
B. T. G\"ansicke\altaffilmark{\ref{Warwick}},
Bruce Gillespie\altaffilmark{\ref{APO}},
Gerald Gilmore\altaffilmark{\ref{Cambridge}},
Ada Nebot Gomez-Moran\altaffilmark{\ref{Potsdam}},
Eva K. Grebel\altaffilmark{\ref{Heidelberg}},
Jim Greenwell\altaffilmark{\ref{UW}},
James E. Gunn\altaffilmark{\ref{Princeton}},
Cathy Jordan\altaffilmark{\ref{APO}},
Wendell Jordan\altaffilmark{\ref{APO}},
Paul Harding\altaffilmark{\ref{Case}},
Hugh Harris\altaffilmark{\ref{NOFS}},
John S. Hendry\altaffilmark{\ref{FNAL}},
Diana Holder\altaffilmark{\ref{APO}},
Inese I. Ivans\altaffilmark{\ref{Princeton}},
{Z}eljko Ivezi\'{c}\altaffilmark{\ref{UW}},
Sebastian Jester\altaffilmark{\ref{MPH}},
Jennifer A. Johnson\altaffilmark{\ref{OSU}},
Stephen M. Kent\altaffilmark{\ref{FNAL}},
Scot Kleinman\altaffilmark{\ref{APO}},
Alexei Kniazev\altaffilmark{\ref{MPH}},
Jurek Krzesinski\altaffilmark{\ref{APO}},
Richard Kron\altaffilmark{\ref{Chicago}},
Nikolay Kuropatkin\altaffilmark{\ref{FNAL}},
Svetlana Lebedeva\altaffilmark{\ref{FNAL}},
Young Sun Lee\altaffilmark{\ref{MSUJINA}},
R. French Leger\altaffilmark{\ref{FNAL}},
S\'ebastien L\'epine\altaffilmark{\ref{AMNH}},
Steve Levine\altaffilmark{\ref{NOFS}},
Huan Lin\altaffilmark{\ref{FNAL}},
Daniel C. Long\altaffilmark{\ref{APO}},
Craig Loomis\altaffilmark{\ref{Princeton}},
Robert Lupton\altaffilmark{\ref{Princeton}},
Olena Malanushenko\altaffilmark{\ref{APO}},
Viktor Malanushenko\altaffilmark{\ref{APO}},
Bruce Margon\altaffilmark{\ref{UCSC}},
David Martinez-Delgado\altaffilmark{\ref{MPH}},
Peregrine McGehee\altaffilmark{\ref{IPAC}},
Dave Monet\altaffilmark{\ref{NOFS}},
Heather L. Morrison\altaffilmark{\ref{Case}},
Jeffrey A. Munn\altaffilmark{\ref{NOFS}},
Eric H. Neilsen, Jr.\altaffilmark{\ref{FNAL}},
Atsuko Nitta\altaffilmark{\ref{APO}},
John E. Norris\altaffilmark{\ref{ANU}},
Dan Oravetz\altaffilmark{\ref{APO}},
Russell Owen\altaffilmark{\ref{UW}},
Nikhil Padmanabhan\altaffilmark{\ref{LBL}},
Kaike Pan\altaffilmark{\ref{APO}},
R. S. Peterson\altaffilmark{\ref{FNAL}},
Jeffrey R. Pier\altaffilmark{\ref{NOFS}},
Jared Platson\altaffilmark{\ref{FNAL}},
Paola Re Fiorentin\altaffilmark{\ref{Slovenia},\ref{MPH}},
Gordon T. Richards\altaffilmark{\ref{Drexel}},
Hans-Walter Rix\altaffilmark{\ref{MPH}},
David J. Schlegel\altaffilmark{\ref{LBL}},
Donald P. Schneider\altaffilmark{\ref{PSU}},
Matthias R. Schreiber\altaffilmark{\ref{Chile}},
Axel Schwope\altaffilmark{\ref{Potsdam}},
Valena Sibley\altaffilmark{\ref{FNAL}},
Audrey Simmons\altaffilmark{\ref{APO}},
Stephanie A. Snedden\altaffilmark{\ref{APO}},
J. Allyn Smith\altaffilmark{\ref{APeay}},
Larry Stark\altaffilmark{\ref{UW}},
Fritz Stauffer\altaffilmark{\ref{APO}},
M. Steinmetz\altaffilmark{\ref{Potsdam}},
C. Stoughton\altaffilmark{\ref{FNAL}},
Mark SubbaRao\altaffilmark{\ref{Chicago},\ref{Adler}},
Alex Szalay\altaffilmark{\ref{JHU}},
Paula Szkody\altaffilmark{\ref{UW}},
Aniruddha R. Thakar\altaffilmark{\ref{JHU}},
Sivarani Thirupathi\altaffilmark{\ref{MSUJINA}},
Douglas Tucker\altaffilmark{\ref{FNAL}},
Alan Uomoto\altaffilmark{\ref{CarnegieObs}},
Dan Vanden Berk\altaffilmark{\ref{PSU}},
Simon Vidrih\altaffilmark{\ref{Heidelberg}},
Yogesh Wadadekar\altaffilmark{\ref{Princeton},\ref{india}},
Shannon Watters\altaffilmark{\ref{APO}},
Ron Wilhelm\altaffilmark{\ref{TTech}},
Rosemary F. G. Wyse\altaffilmark{\ref{JHU}},
Jean Yarger\altaffilmark{\ref{APO}},
Dan Zucker\altaffilmark{\ref{Cambridge}}
}

\altaffiltext{1}{Fermi National Accelerator Laboratory, P.O. Box 500, Batavia,
IL 60510\label{FNAL}}

\altaffiltext{2}{
UCO/Lick Observatory, University of California, Santa Cruz, CA 95064.
\label{Lick}}

\altaffiltext{3}{Dept. of Physics, Applied Physics and Astronomy, Rensselaer
Polytechnic Institute Troy, NY 12180\label{RPI}}

\altaffiltext{4}{
Department of Astrophysical Sciences, Princeton University, Princeton, NJ
08544.
\label{Princeton}}

\altaffiltext{5}{
McDonald Observatory and Department of Astronomy, The University of
Texas, 1 University Station, C1400,  Austin, TX
78712-0259.\label{Texas}}

\altaffiltext{6}{
Mullard Space Science Laboratory,
University College London, Holmbury St. Mary,
Surrey RH5 6NT, United Kingdom
\label{UCL}
}

\altaffiltext{7}{
Department of Astronomy,
Ohio State University, 140 West 18th Avenue, Columbus, OH 43210.
\label{OSU}}

\altaffiltext{8}{
Apache Point Observatory, P.O. Box 59, Sunspot, NM 88349.
\label{APO}}

\altaffiltext{9}{ Department of Astronomy, MSC 4500, New Mexico State University, P.O. Box 30001, Las Cruces, NM 88003.  \label{NMSU}}

\altaffiltext{10}{
Department of Astronomy, University of Washington, Box 351580, Seattle, WA 98195.
\label{UW}}

\altaffiltext{11}{Max Planck Institute for Astronomy, K\"onigstuhl 17, D-69117 Heidelberg, Germany \label{MPH}}

\altaffiltext{12}{Department of Physics and Astronomy, Center for the Study of Cosmic Evolution, and Joint Institute for Nuclear Astrophysics, Michigan State University, East Lansing, MI 48824\label{MSUJINA}}

\altaffiltext{13}{ Institute of Astronomy, Madingley Road, Cambridge CB3 0HA, UK \label{Cambridge}}

\altaffiltext{14}{ Centro de Astrof{\'\i}sica da Universidade do Porto, Rua das Estrelas - 4150-762 Porto, Portugal.  \label{Porto}}

\altaffiltext{15}{
Department of Astronomy and Astrophysics, University of Chicago, 5640 South
Ellis Avenue, Chicago, IL 60637.
\label{Chicago}}

\altaffiltext{16}{
Department of Physics 
Warwick University
Coventry CV47AL, UK
\label{Warwick}}

\altaffiltext{17}{
Astrophysikalisches Institut Potsdam
An der Sternwarte 16, D-14482 Potsdam, Germany
\label{Potsdam}}

\altaffiltext{18}{
Astronomisches Rechen-Institut, Zentrum f\"ur Astronomie,
University of Heidelberg, M\"onchhofstrasse 12-14,
D-69120 Heidelberg, Germany.\label{Heidelberg}}

\altaffiltext{19}{
Department of Astronomy, Case Western Reserve University,
Cleveland, OH 44106.
\label{Case}}

\altaffiltext{20}{
US Naval Observatory, Flagstaff Station, 10391 W. Naval Observatory Road, Flagstaff, AZ 86001-8521.  \label{NOFS}}

\altaffiltext{21}{ Department of Astrophysics, American Museum of Natural History, Central Park West at 79th Street, New York, NY 10024\label{AMNH}} 

\altaffiltext{22}{
Department of Astronomy and Astrophysics, University of
California, 1156 High Street, Santa Cruz, CA 95064
\label{UCSC}}

\altaffiltext{23}{
IPAC, MS 220-6, California Institute of Technology, Pasadena, CA 91125.\label{IPAC}}

\altaffiltext{24}{
Research School of Astronomy and Astrophysics, Australian National 
University, Weston, ACT, 2611, Australia
\label{ANU}
}

\altaffiltext{25}{
Lawrence Berkeley National Laboratory, One Cyclotron Road,
Berkeley, CA 94720.
\label{LBL}}

\altaffiltext{26}{
Department of Physics, University of Ljubljana, Jadranska 19, 1000
Ljubljana, Slovenia
\label{Slovenia}}

\altaffiltext{27}{
Drexel University, Philadelphia, PA.
\label{Drexel}}

\altaffiltext{28}{Department of Astronomy and Astrophysics, The Pennsylvania State University, University Park, PA 16802\label{PSU}}

\altaffiltext{29} {
Departamento de Fisica y Astronomia, Facultad de Ciencias, 
Universidad de Valparaiso, Valparaiso, Chile
\label{Chile}
}

\altaffiltext{30}{
Department of Physics and Astronomy, Austin Peay State University, P.O. Box 4608, Clarksville, TN 37040.  \label{APeay}}

\altaffiltext{31}{ Adler Planetarium and Astronomy Museum, 1300 Lake Shore Drive, Chicago, IL 60605.  \label{Adler}}

\altaffiltext{32}{
Center for Astrophysical Sciences, Department of Physics and Astronomy, Johns
Hopkins University, 3400 North Charles Street, Baltimore, MD 21218.
\label{JHU}}

\altaffiltext{33}{
Observatories of the Carnegie Institution of Washington,
813 Santa Barbara Street,
Pasadena, CA  91101.
\label{CarnegieObs}}

\altaffiltext{34}{
National Centre for Radio Astrophysics, Post Bag 3, Ganeshkhind, Pune
411007, India
\label{india}}

\altaffiltext{35}{Department of Physics, Texas Tech University, Lubbock, TX \label{TTech}}

\shorttitle{The SEGUE survey}

\shortauthors{Yanny, Rockosi et al.}

\begin{abstract}

%Fix string  of types

The Sloan Extension for Galactic Understanding and Exploration (SEGUE) survey obtained $\approx$ 240,000 moderate 
resolution ($R\sim 1800$)
spectra from 3900\AA\ to 9000\AA\ of fainter Milky Way stars 
($14.0 < g < 20.3$) of a wide variety of spectral types, both
main-sequence and evolved objects, with the 
goal of studying the kinematics and populations of our 
Galaxy and its halo.  The spectra are clustered in 212 regions
spaced over three-quarters of the sky.  Radial velocity accuracies for stars 
are $\sigma(\rm RV) \sim 4 \>\rm km~s^{-1}$ at $g < 18$, degrading 
to $\sigma(\rm RV) \sim 15\rm \>km~s^{-1}$ at $g\sim 20$.  For stars 
with signal-to-noise ratio $> 10$ per resolution element, stellar atmospheric parameters 
are estimated, including metallicity, surface gravity, and effective 
temperature.  SEGUE obtained $3500 \rm deg^2$ of 
additional $ugriz$ imaging (primarily at low Galactic latitudes) 
providing precise multicolor photometry 
($\sigma(g,r,i) \sim 2$\%), ($\sigma(u,z) \sim 3$\%) and 
astrometry ($\approx 0.1''$) for spectroscopic target selection.  
The stellar spectra, imaging data, and derived parameter catalogs
for this survey are publicly available as part of Sloan Digital Sky
Survey Data Release 7.

\end{abstract}


\section{Introduction}

\subsection{Stellar Spectroscopic Surveys}

A large-scale study of the Milky Way is important to our general understanding of 
galaxy structure and formation.  It is only
in our own `backyard' that great numbers of stars, the building block of all galaxies,
may be observed individually, with their collective properties 
serving as constraints on theories of galaxy formation and evolution.
Spectroscopic data of individual stars can provide a much richer variety of 
information on both stellar kinematics and stellar atmospheric parameters than 
is possible with photometric measurements alone.  

%Connie's updated plate items 2336/2337, cluste rplates, table 5 message (3 messages) XX
%Also add stripe numbers and upper axis to figure 1 XX

The first large area spectroscopic
surveys used objective prism plates, including the fundamental surveys
for spectroscopic classification \citep{cannon18,mkk43,h78} and more
specialized surveys for unusual stars
\citep{cameron56,nassau65,sb71,bm73}. Later objective prism surveys
focused on extremely metal-poor stars \citep{bps85,norbert} and on
halo giants \citep{kavan,fm90}.
Objective prism surveys had the advantage of rapidly recording a large number
of stellar spectra over significant solid angles, but had disadvantages 
such as a bright limiting magnitude and an inability to accurately 
calibrate the spectra. 

``Aperture" spectroscopic surveys using modern
spectrographs have been rarer, in part because of the huge investment
of telescope time required to assemble substantial-sized samples. 
They include the ``Spaghetti" survey for
halo giants \citep{metal00}, the SIM Grid Giant Star Survey
\citep{simgrid}, the detailed chemical study of \citet{eetal93}
and its various sequels,  and the monumental survey of \citet{netal04} who
studied nearby, bright F and G stars 
and obtained metallicity, temperature and age information from
Str\"omgren photometry and accurate radial velocities (RVs) from CORAVEL and
other spectrographs.

Multiobject spectroscopic surveys (first implemented with plug
boards, or slit masks, then with automated positioners), which provide a
large gain in efficiency, have also been
done with specific scientific goals in mind.
A number of programs \citep{kg89,ig95,igi94,gwj95,gwn02} have
searched for coherent structures in the halo.
The RAVE Survey \citep{setal06,zetal08}, which focuses on bright stars of
 all colors ($9 < I < 13$), produces accurate velocities and estimates of
stellar parameters from a small spectral region including the Ca II
infrared (IR) triplet. 

Orthogonal to the volume-limited or spectral-type specific surveys
have been the compilations of homogeneous spectroscopic
atlases of a few hundred objects \citep{gs83,jhc84,p85} 
obtained with electronic scanners and imaging tubes.  
The members of these catalogs were selected to sample
objects of all spectral types with at least one example of each
temperature and luminosity class.  These catalogs do not give relative
numbers of stars in the different spectral categories, however, and
they may miss some rare categories, especially the low-metallicity
stars.

Looking to the future, the Gaia spaced-based mission
\citep{petal01} plans to obtain proper motions (and precise positions)
of approximately one billion stars to $g \sim 20$,
with RVs for all of the brighter objects with $g < 17$ 
\citep{katz04,wketal05}. 
Gaia, when underway, will represent a several orders of magnitude leap forward
in our knowledge of the kinematics, structure and evolution 
of our Galaxy.

\subsection {Stars and the SDSS}

The Sloan Digital Sky Survey (SDSS; York et al. 2000)
is primarily an extragalactic survey that has
obtained 2\% multicolor imaging of nearly 8000 $\rm deg^2$ of
filled contiguous sky toward the Northern Galactic Cap, and 700 $\rm deg^2$
in three stripes in the South Galactic Cap near the celestial
equator.  Spectra have been acquired of one million galaxies 
and one hundred thousand quasars.
The major science program consists of 
constructing a large three-dimensional map of the universe and 
constraining cosmological models;  see, e.g. 
\citet{betal01,bletal03,fetal03,thketal04,tetal04,eetal05,retal06}.

A significant product of the SDSS was a large number of Milky Way
stellar spectra combined with deep, accurate multicolor photometry.  
This led to several (initially) serendipitous Galactic structure, Galactic
halo, and M31 halo science results; see  
\citet{ietal00,yetal00,netal02,retal02,wetal05,yetal03,netal03,zetal04a,zetal04b,betal06a,betal06b,apetal06,betal07,betal07a,betal07b,ketal07,xdh07,xetal08,jetal08}.

Near the conclusion of the original SDSS program in 2004, partially as
a result of the productive Galactic science enabled by the SDSS, a set
of three individual surveys (under the umbrella designation of
SDSS-II) were designed: 1) Legacy:  a survey following the same goals of the
original SDSS, to complete the SDSS imaging and spectroscopic
footprint; 2) SN Ia \citep{fetal08}: a well-calibrated, systematic
survey for 200 intermediate redshift ($0.1 < z < 0.4$) type Ia
supernovae, filling an important gap in redshift coverage, and
anchoring the calibrations of higher redshift supernova surveys; and 3) 
Sloan Extension for Galactic Understanding and Exploration (SEGUE), an
imaging and spectroscopic survey of the Milky Way and its surrounding
halo.  SDSS-II operated from 2005 August to 2008 July at Apache Point
Observatory, building on SDSS, which operated from 2000 August until
2005 July.

The SEGUE Survey is the subject of this paper.  The processed,
searchable data archive from SEGUE was made publicly available in the
Fall of 2008 as part of SDSS-II Data Release 7 (DR7).  
With few exceptions, all stellar spectral types are represented in the
SEGUE Survey.  Notable categories which are missing include luminous
Population I early types, such as O, B, and Wolf-Rayet stars, and some 
Population I giants, which are not targeted because they are 
generally too bright for SEGUE observations if they are in 
the solar neighborhood and are too rare or are obscured by 
dust toward the Galactic center to be seen at greater distances.
Samples of spectrophotometrically calibrated stars of a wide variety of
spectral types are presented in Section \S 3.
A defining part of SEGUE is the creation and
release of its public database that can be mined to enable a whole
range of astrophysics projects not conceived of when the
survey was carried out.

\section{Survey Goals and Footprint}

\subsection{SEGUE Goals}

The original, five year SDSS program demonstrated the existence of
significant spatial substructure in the stellar halo of the Milky Way
discovered from photometric data, from which stellar distance estimates
were obtained primarily for bluer (A and F) stars.  These substructures
cast doubt upon previous measurements of a presumed axially symmetric
spheroid stellar component of the Galaxy with a smoothly varying
power law density structure.  Discovery of the substructure also created
a tremendous need for follow-up spectroscopy of each structure, so that
stellar population and orbital information for the debris could be
determined.

SEGUE was designed to sample the stellar spheroid at a variety of
distances, from a few kpc to a hundred kpc, in 200 ``pencil beams,"
spaced around the sky so that they would intersect the largest
structures.  At the time SEGUE was designed, we knew about the
Sagittarius Dwarf spheroidal tidal debris stream and the controversial
Monoceros stream in the Galactic plane.  Both structures were
6--10 kpc across and were believed to extend all of the way
around the Milky Way.  We expected that additional substructure would be
discovered, and that it was most important to identify and
characterize the largest structures; without knowledge of the spatial
variation of the spheroid at 10 kpc scales, it was difficult to
positively identify smaller or lower surface brightness structures.

SEGUE augmented the photometric data from the SDSS/Legacy
Surveys to sample the sky about every 15$^\circ$ across the sky,
in all parts of the sky accessible from the telescope's latitude.  This
included adding photometry at low latitudes ($|b| < 35^\circ$) and 
additional photometry
in the South Galactic Cap.  Two hundred pencil beams were selected for
spectroscopy because these could be arranged to sample the sky at intervals of
10$^\circ$ to 15$^\circ$, and two observations of each pencil beam could be observed in the three year duration of SDSS-II.

Spectroscopic target selection was designed to maximize the science from SEGUE
stellar spectroscopy; in particular we wished to study the Milky Way's
chemical and dynamic formation history and to constrain the Galaxy's gravitational
potential.  

The target selection strategy that was used to achieve these general
goals devoted most of the fibers on each pencil beam line of sight to
sampling the stellar populations of the Galaxy on large (tens of kpc)
scales, including spheroid substructure and global properties of the
thin and thick disk components.  In addition, a small subset of the
fibers was devoted to unusual stars, including those thought to have
low metallicity ([M/H] $< -2$) or to be otherwise unusual based on their
colors and velocities (as determined by photometry and proper motions).
In addition, we specifically targeted
star clusters with a variety of ages and metallicities so that this
important spectral database could be well calibrated.

%The target selection strategy that was used to achieve these
%general goals included:
%(1) sampling the stellar populations of the Galaxy on large (tens
%of kpc) scales, including spheroid substructure and global
%properties of the thin and thick disk components, and (2)
%preferentially selecting unusual stars, including those thought to have
%low metallicity ([M/H] $< -2$), unusual colors, or high velocities 
%(as determined by proper motions).  

To meet these goals, SEGUE has produced 1) an imaging survey of
3500 $\rm deg^2$ of $ugriz$ imaging with the SDSS telescope and
camera \citep{getal98,getal06}, 2) a spectroscopic
catalog that spans the stellar population observable in
magnitude and Galactic latitude range of the data
at a resolution of $R \sim 1800$.
% stars of various types are 
%sampled in numbers representative of their densities on the 
%sky towards various sightlines around the Galaxy.
The spectroscopic catalog includes estimates for the observational 
parameters (position, RV, multicolor photometric 
and spectrophotometric 
magnitudes), as well as the derived, modeled parameters 
(including $[\rm M/H]$, surface 
gravity, and $T_{\rm eff}$) for all observed stars in a systematic 
and well-calibrated fashion.

%Because of the constraints arising from being but one component in a
%three-part, three year survey on a 2.5m telescope, the scope of SEGUE
%is limited.   Both the imaging sampling of the low-latitude sky and the 
%selection of stars by spectral type for spectroscopic followup are 
%restricted.

\subsection{SEGUE Imaging}

The original SDSS imaged most of the North Galactic Cap plus three
stripes of data in the South Galactic Cap; regions of low Galactic latitude
($|b| < 35^\circ$) were largely excluded by design.  
The SEGUE imaging footprint was designed to allow the selection of spectroscopic 
targets in as broad a range of sky directions as possible, to enable 
study of the important transition zones between our Milky Way's disks 
and stellar halo, to include a large and varied sample of Galactic star
clusters that could be used for calibration, and to ensure that
photometric calibration would be feasible (i.e. avoid zones of
extreme and variable extinction).

The low-latitude SEGUE imaging area includes 15 2.5$^\circ$-wide stripes
of data along constant Galactic longitude, spaced by approximately 
$20^\circ$ around the sky.  These
stripe probe the Galaxy at a wide variety of 
longitudes, sampling the changing relative densities 
of global Galactic components (thin disk, thick disk, halo).
The precise longitudes of the SEGUE stripes are not evenly spaced;  
the $l$ of several stripes were shifted by up to $8^\circ$, 
so that several known open clusters near the Galactic plane 
could be optically imaged.  
We added two stripes of data in the South Galactic Cap.
Because spectra of stars toward cardinal Galactic directions 
($l$ near $90^\circ, 180^\circ, 270^\circ, 360^\circ$) are important for
generating a simplified kinematic analysis of a very complex dynamic Galaxy,  
the two SEGUE stripes (at $l=94^\circ$ and $l=178^\circ$) that nearly 
coincide with cardinal pointings were extended to give
nearly complete pole-to-pole imaging and more complete spectroscopic 
plate coverage than at other longitudes.
Where possible, the SEGUE stripes were designed to cross other SDSS
imaging data at multiple locations to facilitate photometric calibration.

Figure 1 shows the constant longitude and two Southern 
imaging stripes chosen to augment the original SDSS footprint in 
Equatorial (top) and Galactic coordinates (bottom).  
We sample the sky in all directions that are accessible
to the Apache Point Observatory; since the observatory is at
a Northern latitude of $32^\circ$, essentially no SEGUE data is obtained
with the Equatorial coordinates $\delta < -20^\circ$. 
The Galactic anticenter ($\delta = 29^\circ$) is 
well sampled, but the Galactic center ($\delta = -29^\circ$) is not.
The stellar population of the bulge is largely inaccessible and
obscured by dust in this optical survey.  

%These clusters will be are cataloged in a separate forthcoming 
%paper (McGehee et al., in preparation).

The SEGUE imaging scans (i.e., data not associated with the Legacy SDSS Survey)
are tabulated in Table 1.  All SEGUE imaging was 
obtained between 2004 August and 2008 January.  Note that each SEGUE stripe is
$2.5^\circ$ wide.  Stripes 72 and 79 follow standard numbering conventions
of the original SDSS imaging survey.  Stripes with four digit numbers run
along constant $l$, running for variable extents in $b$.
The formula for converting the SDSS Survey coordinates $(\mu,\nu)$ for a 
particular node and inclination to the
Equatorial coordinates J2000 $(\alpha, \delta)$ may be 
found in \citet{setal02}.

Except in regions of high stellar density, the processing and calibration 
of the SEGUE imaging data are the same as that of the SDSS imaging data 
\citep{betal03b,fetal96,hetal01,petal03,smetal02,setal02,tetal06}.
A modified version of the SDSS PHOTO processing pipeline 
software (R. H. Lupton et al., 2010 in preparation) was combined 
with the Pan-STARRS \citep{m06} object detection code, and 
was run on all lower latitude SEGUE imaging scans.  The key 
modifications were:  
1) the PHOTO code was optimized for primarily stellar objects by truncating fits to the wings of extended sources at essentially the point-spread function (PSF) radius,  
2) the PHOTO code's object deblender was allowed to deblend groups of closely
spaced objects into more `child' objects than in standard high latitude
(less crowded) SDSS fields, and 3) Pan-STARRS threshold object detection 
code was used.  This code generates more complete object lists in regions of
high stellar density, and was used to supplement PHOTO's object detector.  
Imaging scans which participated in 
this low-latitude processing are tagged in the data archive with a 
rerun (reprocessing version) number of 648.
This is in contrast to standard SDSS and SEGUE PHOTO reprocessing numbers
$40 \le \rm rerun \le 44$.  
Except in regions of high stellar density, the magnitudes from this
version of PHOTO are interchangeable (within the errors) with the 
version used to process the rest of the SDSS and SEGUE imaging data.  
At high stellar density the choice depends on the needs of the particular
investigation.  Both the PHOTO PSF magnitudes 
and the Pan-STARRS aperture magnitudes are available for comparison in 
the DR7 data archive for low latitude regions of sky.  We refer
the reader to the DR7 paper \citep{strauss09} for more discussion of this
reprocessing of the imaging data, and to the documentation on the
DR7 Web site.  At the density extreme, several globular clusters
present in the SDSS and SEGUE footprints are analyzed by \citet{aetal08} 
using independent software (see below).

As in SDSS DR6 \citep{dr6}, the zeropoint photometric
calibration SEGUE imaging has been enhanced 
by the calibration procedure described in \citet{petal08}.
This procedure finds a simultaneous global fit for the 
individual imaging scans' photometric zeropoints, extinction 
coefficients and flat-field characteristics (of all 30 SDSS camera CCDs),
relying on the overlap between SEGUE and SDSS Legacy scans to improve
the absolute zeropoints accuracies in the $gri$ filters to $<1\%$ in most areas
around the sky.  
%Of course, at very low latitudes, extinction corrections 
%due to dust dominate the accuracy to which one can correct apparent optical 
%stellar magnitudes and colors for foreground reddening \citep{sfd98}.

\subsection{SEGUE Spectroscopy}

SEGUE leveraged the unique features of
the SDSS telescope and spectrographs (namely the ability to
go deep and wide, with broad spectral coverage and high spectrophotometric
accuracy) to acquire spectra of $\sim$ 240,000 stars of 
a wide variety of spectral types 
(from white dwarfs (WDs) on the blue end to M and L subdwarfs
 on the red end), probing a wide range of 
distances ($<10 \rm \> pc\>to \> > 100 \>kpc$).
The SDSS spectrographs used for SEGUE are a pair of highly efficient dual
CCD camera, fiber-fed spectrographs, with
wavelength coverage from 3900\AA\ to 9000\AA\  at resolving power 
$R \sim 1800$.  The twin spectrographs can simultaneously record
data from 640 fibers in a 7 $\rm deg^2$ field of view; 7\% -- 12\% of the fibers
are reserved for sky signal and
other calibration fibers (such as spectrophotometric standard stars,
generally chosen from color-selected F subdwarfs with $16 < g < 18.5$).

SEGUE took spectra of stars in the magnitude range $14 < g < 20.3$.
At $g\sim 14$ the SDSS spectrographs saturate in a 300 second exposure.
Objects
down to $r = 18.5$ can be routinely obtained with signal-to-noise ratio (S/N) $>$ 30, sufficient
for RVs good to 4 $\rm km\>s^{-1}$ and metallicity $[\rm
M/H]$ measurements accurate to 0.2 dex for a wide variety of spectral
types (A-K).  
At $g \sim 20.3$ we were able to obtain $ \rm S/N \sim 3$ in 
2 hours integration time under photometric conditions 
with seeing of $2''$ or better (all S/N quotations are per 
$\rm 150 \>km~s^{-1}$ resolution element). 

%The RAVE project \citep{setal06,zetal08} focuses on bright
%objects ($9 < I < 13$) and thus on objects in the extended solar neighborhood.
%By contrast, SEGUE obtains spectra of BHBs down to $g \sim
%20$, to distances $> 50\>$ kpc from the Sun.  
%Spectra of objects brighter than $g \sim 14$ will saturate in SEGUE 
%observations, and as RAVE is a Southern hemisphere survey, RAVE and SEGUE 
%are largely non-overlapping.

Spectroscopic plate pointings sparsely sample all areas of the sky with available
imaging (Figure 1), probing all the major known Galactic structures (thin
and thick disk, halo, and streams) with the exception of the bulge, which is
below our Southern declination limit.
To study the
detailed structure of these Galactic components, the density of targets
is made high enough that the velocity distribution of one homogeneous 
subset of stars (the G dwarfs) in one distance bin (an interval 
of one apparent magnitude) may be determined to be either 
consistent or distinct from a Gaussian.  This requires at least 
40 targets of one spectral type (G dwarfs) per magnitude interval
(14th through 20th) per plate pointing.  It is this scientific
goal which drives the assignment of over 300 fibers per plate pair
to G star candidates toward each SEGUE line of sight.

% The above placement of G dwarf cuts is out of place, can we move it? XXX

The required RV accuracy is driven by the scientific goal
of separating stellar streams with dispersion 
$\sigma \sim 10\rm \>km~s^{-1}$ from field disk and halo
stars with dispersions of $\sigma\sim$ 30 and 100 $\rm \>km~s^{-1}$
respectively.  
Figure 2 shows the actual relative RV accuracy obtained 
from SEGUE spectra using quality assurance (QA) stars.  
These are simply stars at $r\sim 18$ that are observed twice, once on each 
of the two plates that make a complete observation of 
a SEGUE pencil beam.  There are $\sim 20$ such pairs for
every pencil beam.  We restrict our QA sample to those with $S/N > 10$.  
RVs are measured by cross-correlating each spectrum against a
set of $\sim 900$ selected templates taken from the ELODIE high resolution
spectroscopic survey \citep{moultaka04}. The ELODIE templates span a wide range of
spectral types and metallicities. Very early and very late types, however,
are nearly absent.  We note that the correlation is done by shifting 
each spectrum repeatedly, 
stepping through wavelength space (rather 
then via fast Fourier transform techniques), and, while time consuming, 
this appears to result in somewhat higher accuracy.
The top panel shows the measured RV difference histogram between 
all QA stars and a second observation of the same
stars on a different plate. 
Individual errors are $\rm \sigma \sim 4.4 \> km~s^{-1}$.
The lower panel shows how the RV accuracy degrades with color and S/N: 
in general bluer objects have worse RV errors than red, with blue horizontal branch stars (BHBs) and their
broad Balmer lines being hardest to measure accurately.  We expand on the
content of Figure 2 in Table 2, where we order the set of spectra 
with multiple independent observations of the same object by 
S/N and divide the data for each of the six color 
ranges into four quartiles.
For each color range from blue to red, we tabulate N and list by quartile the
average (dereddened) magnitude $\bar g_0$, average S/N and the 
$1\sigma$ velocity error (divided by $\sqrt{2}$ to compensate for the 
fact that we have two independent measurements).
The errors are well behaved as S/N decreases from $>50$ to $<10$.  We
do not characterize velocity errors for extreme spectral types (WDs on
the blue end and late M and L type stars on the red end), as these 
samples have very large systematic errors, due to very broad spectral
features and a lack of standard templates with which to cross-correlate.  
Follow-up reductions of the SEGUE spectra will include synthetic templates for 
deriving velocities for stars with unusual Carbon enhancement (S. Thirupathi, 2008, private communication).

% S/N range on K stars...i.e. is S/N the same?

%One component of the radial velocity error are the plate-to-plate 
%systematic errors in these fiber spectra.  To set the zero-point
%of the RV calibration, a set of 100 bright field stars were observed both
%with SEGUE and at higher resolution on the Hobby-Eberly telescope
%\citep{apetal08}.  SEGUE spectra of stars in the outskirts of globular clusters
%with known radial velocity also provided checks on the RV zeropoint.
%This analysis resulted in an offset of $\rm 7.3\>km~s^{-1}$, applied
%to SEGUE RVs in the final DR7 catalog. 
%After this correction, systematic errors of 
%3 $\rm \>km~s^{-1}$ plate-to-plate remained, quantified by measuring
%the mean of the QA stars on each plate in a pair.  

One component of the total RV error is the systematic
error in zeropoint of each plate. The scatter in that offset between
plates, measured from the mean offset of the QA stars for each pencil
beam, has zero mean and standard deviation 1.8 $\rm km~s^{-1}$. This scatter of
1.8 $\rm km~s^{-1}$ contributes to the RV uncertainty of all SEGUE stars, and
sets the lower bound of our RV errors. To check the
overall zeropoint of the RV calibration we used a set of 100 bright
field stars observed both with SEGUE and at higher resolution on the
Hobby-Eberly telescope \citep{apetal08}. SEGUE spectra of
stars in the outskirts of globular clusters with known RV
also provided checks on the RV zeropoint. This analysis resulted in an
offset of 7.3 km/s with very small scatter.  This 7.3 km/s offset was
applied to SEGUE RVs (but not to the $z$ redshift measurements directly from the
spectroscopic pipelines) in the final DR7 catalog.  The origin of
these systematics are not completely understood, but they are thought to 
be associated with telescope flexure or night sky line fitting errors when 
computing wavelength solutions for 
each merged SEGUE plate, which consists of a numbers of exposures taken 
over several hours, often spanning multiple nights (see below).

Our scientific goals require that for a large fraction of 
the stars with $g < 18.5$, sufficient S/N be obtained so it is possible 
to reliably estimate the 
metallicities $\rm [M/H]$ and luminosity classes (dwarf vs. subgiant
vs. giant) for stars of spectral types A-M well enough to separate
stream stars from field disk populations from halo populations.
An S/N $> 10$ is required to measure these stellar atmospheric 
parameters, with more accurate measurements at higher S/N.  This drove
the integration time for the SDSS spectrographs
to about 2 hours for a $g=18.5$ object and 1 hour for a $g=17$ object.

The actual SEGUE spectroscopic survey selected pointings are shown as 
blue circles
in Figure 1.   
There are 212 SEGUE pointings on the sky, all listed in Table 3.
These pointings are divided into the following categories: 1) `Survey',
172 approximately evenly spaced pointings around the 
SDSS and SEGUE imaging area,
separated by no more than $20^\circ$ from the next nearest pointing,
sampling all directions without regard for known structures or streams.
 2) `LLSurvey', 12 pointings at low latitude,
where a separate target selection algorithm is used that functions even
in highly reddened lines of sight; 3) `Strm', 16 pointings toward
five previously discovered stream-like structures around the halo 
of the Milky Way, such as the Sagittarius or Orphan streams;
4) `Cluster', 12 pointings towards known globular or open clusters
of known $\rm [M/H]$ for purposes of calibrating the metallicity and luminosity
pipelines and 5) `Test', five early SEGUE pointings to test the target 
selection algorithms at a variety of latitudes and to test the RV 
accuracy of the survey.  Six pointings are duplicated.
Note that a significant fraction of
SEGUE spectroscopy relies upon the SDSS Legacy Survey imaging in the North
Galactic Cap.  

%The capabilities of the SDSS spectrographs are described in detail in (ref XX).

A SEGUE (and SDSS) spectroscopic plate is a circular disk
of machined aluminum, with a diameter of $0.75$ m, corresponding to an 
angular on-the-sky radius of $1.49^\circ$.  A small hole, holding one 
fiber, is drilled at the position of each object of interest.   
Each plate may have up to 640 object holes, which are fed to 
twin 320-fiber spectrographs.  Target holes are 
restricted to being no closer together 
than $55''$ on the sky \citep{setal02}.  The total area of 
each plate is approximately 7 $\rm deg^2$ on the sky.  

%To adequately sample the number of stars available per 
%square degree on the sky at moderate latitudes $|b| \sim 45^\circ$ at 
%each of the 212 SEGUE pointing directions, two SDSS-style plates 
%with a maximum of 640 fibers each are designed.  
Because there are many more than 640 stars per 7 $\rm deg^2$ to
g=20.3 and because the SEGUE magnitude limits $14< g < 20$ span more than
a factor of 100 in apparent brightness, we observe each SEGUE pointing
with two SDSS-style plates, each with a maximum of 640 fibers. 
Recall that for SDSS extragalactic observations, only one plate at 
each position was designed, to match the sampling of approximately
100 galaxies per square degree for objects with $r_{\rm extended} < 17.77$.

One plate of the pair is called the SEGUE bright (or SEGUE regular) plate, 
and consists of holes targeting stars with $14.0 < r < 17.8$, 
exposed for typically
 1 hr.  
The bright magnitude limit is set by the saturation of the spectrographs
for a 300~s exposure, as well as cross-talk considerations between the 
brightest and faintest objects in adjacent fibers on a given plate.
The second plate of 640 fibers, designated the SEGUE 
faint plate, primarily consists of stars with $17.8 < r < 20.1$, exposed
for a total of typically 2 hr.

About 20 stars per pointing with $r \sim 18$ are targeted twice, on both
the bright and faint plates.  These objects are called QA 
objects, and, as mentioned above,
are used to determine the systematic reproducibility of 
RVs and other derived parameters from plate to plate.

For the special case of `Cluster' plates, where it is desired to
obtain spectra of bright nearby globular cluster giant branch stars 
for calibration, the SDSS spectrograph saturation limit 
is extended by taking short (1--2 minute) exposures, allowing 
one to sample stars as bright as $g\sim 11$.
These short exposures, however, have little sky signal, and it is
thus difficult to do an accurate wavelength calibration. This is because 
the final step in the calibration process depends on 
using the positions of fixed, known night
sky lines such as Hg, Na and [O I]).  
%(list the lines AA XX) 

The bright plates have 32 fibers reserved for blank sky and 16 reserved for
spectrophotometric standards; the faint plates have 64 fibers reserved
for blank sky and 16 fibers reserved for spectrophotometric standards. This approach leaves
approximately 1152 fibers available for science targets in each 7 $\rm deg^2$ pointing.  
The number of sky fibers was determined by the need to maximize target
S/N in fiber-fed multiobject spectrographs, as discussed in \citet{wg92}.
The spectrophotometric standards (primarily halo F subdwarfs) from SEGUE and 
the SDSS Survey constitute a valuable sample in themselves for numerous
Galactic structure studies \citep{apetal06,cetal07}.

%Table 3 has SDSS jargon, spell it out. XX

Altogether, there are 416 plates in the SEGUE database; 
all but 17 of the 212 SEGUE pointings have a bright and faint 
plate of $\sim 576$ targets each.
Individual 10--30 minute exposures were obtained, 
sometimes on successive nights, until the desired S/N for each SEGUE plate
was reached.  All common exposures for a given plate and plugging 
(fibers are plugged into the metal plates by hand and the plugged
plates can be moved into and out of the focal plane as many times as
necessary to reach a desired S/N value) were combined
and uniquely identified by four digit plate number and by the 
Modified Julian Date (MJD) of the last night on which 
a given plate-plugging was observed.  The plate names and 
identifying observation dates (MJDs) are all in Table 3.  
The total number of unique SEGUE spectra is approximately 240,000.

SEGUE spectra are processed with the same basic pipelines used to process the
SDSS data \citep{setal02}.  The pipelines have been modified slightly to
enhance the radial velocity accuracy of stellar spectra.  It should be noted
that a handful of SEGUE plates were obtained under moon illumination 
fractions of greater than 0.85, and a careful analysis indicates 
that the wavelength solutions of these 10 plates
are systematically off by as much as $10 \rm \>km~s^{-1}$.   
These plates are marked in Table 3 with asterisks.

%Spectroscopic observations 
%of globular clusters with known cataloged radial velocities allowed SEGUE
%to determine it's absolute radial velocity accuracy 
%oto be $\sim 3 \rm km~s^{-1}$.  

%% We should mark all those plates in Table 3. XX

In addition to the standard extraction and RV reduction
pipelines of SDSS, the SEGUE plates have been processed through 
an additional `SEGUE Stellar Parameters Pipeline' (SSPP) that estimates the 
metallicity $\rm [M/H]$, surface gravity ($\rm log~g$) and effective 
temperature $\rm T_{\rm eff}$, along with associated errors, for each star 
with sufficient S/N.  
Details of the design, operation and error analysis
of the SSPP as run on SEGUE spectra are described in 
\citet{letal08a,letal08b} and \citet{apetal08}.
The uncertainties in the SSPP parameters were determined by
analysis of star clusters with known metallicities, reddenings, and
distances and by comparison of SSPP-derived parameters with
parameters derived from higher resolution spectra of the same sample
of field stars. For spectra with $\rm S/N ~ > 30$, which are usually obtained
for stars with $g < 18.5$, the errors are 
$\sigma(\rm [M/H]) \sim 0.2 $ dex, 
$\sigma(\rm log~g) \sim 0.3 $ dex, and $\sigma(\rm T_{\rm eff}) 
\sim 200 $ K. These uncertainties are
valid for stars with $T_{\rm eff}$ between 4500 K and 8500 K.
For stars outside this temperature range, and for stars of lower $\rm S/N$, 
atmospheric parameters are still computed, but with appropriately higher
error estimates.   Table 4 contains a list of SSPP quality flags which are set
for every spectrum.  These flags, when set,
indicate something unusual about a given spectrum, i.e. one with 
unusually strong Balmer, Mg or Na lines, or one where there is 
a mismatch between the photometric color [derived from the 
$(g-r)_0$ color], and the spectroscopic type
of the star.   Within the context of DR7, a practical quality
cut used to select only stars (and avoid galaxies, quasars, and low-S/N spectra 
of the sky) is to insist that 
a given spectrum has the error on the ELODIE template cross-correlation RV
$elodierverr$ strictly  $ > 0$.

\section {TARGET SELECTION BY CATEGORY} 

\subsection {General SEGUE Target Selection Principles}

SEGUE targets are selected primarily from photometry in the $ugriz$
SDSS filter system \citep{fetal96}.  For a few categories 
(cool WD, K giant, and M subdwarf candidates), 
the presence (or absence) and amplitude of a proper 
motion measurement from a match to an astrometric 
catalog \citep{metal04,l08} is also used.

The broad science goal of characterizing large-scale stellar structures 
in the Galaxy, 
combined with the specific goal of studying halo streams, 
informed the target selection algorithms' design.  
The additional goals of finding rare but scientifically interesting samples of
 objects of unusually low metallicity, odd spectral type, or extreme kinematics 
were also factors.

The variety of science goals dictated a target selection algorithm 
that sampled stars at a variety of distances, favoring those 
at large distances.
% as the realm that SEGUE could uniquely excel in probing.  
The SEGUE Survey targets objects 
at a variety of colors and apparent magnitudes to probe distances 
from 10 pc (with WDs and M and K dwarfs) to 
the outer halo at $d \sim 100 $ kpc (with BHB and red K giant stars).
At intermediate distances, a target selection category 
denoted `G star' (which contains some G IV subgiants) 
is sampled over the entire SEGUE magnitude range $14.5 < r_0 < 20.2$, and 
effectively probes the thick disk to inner halo transition region 
around the Galaxy with a large, unbiased sample.

SEGUE targets were divided into 15 different target categories,
which spanned the range from the 
bluest (WD) to reddest (spectral type L 
brown dwarfs).  
Table 5 lists the SDSS/SEGUE Primary target bit in hexadecimal in Column 3.
A search of the database may be done for objects with Primtarget matching
these bits in the SpecObjAll table of the CAS (see below).
The fourth column of Table 5 lists the magnitude, color, and proper 
motion cuts for each
target type.  When more stars than fibers in a given category are available for
targeting, weighting mechanisms are used to subselect
from candidates within a given category.
%, along with weighting mechanisms used to subselect objects when
%more stars than desired were available in a given category 
%for a specific pointing.  
These weighting mechanisms generally were 
designed to randomly subselect from all possible stars in 
a given category in a given 7 $\rm deg^2$ 
field, with the probability of selection weighted by either magnitude (favoring brighter objects over 
fainter) or by color (favoring blue objects over red, except in the case
of K giants).  
The colors used for target selection
were in some cases generalized linear combinations of the SDSS $ugriz$ 
generated colors.  These generalized colors were designed to run parallel
to or perpendicular to the color-color space stellar locus of (dereddened)
Galactic stars \citep{lenz,helmi03}.  
There is a maximum number of targets accepted 
in any given pointing in each category (see Column 5 of Table 5 which
lists the maximum number of targets allowed per pointing, the approximate total number targeted by category during SEGUE, and a rough estimate of 
the fraction of the spectra in each category that turn out to be 
of the type that was targeted).

The SEGUE target selection algorithms were not perfected immediately, 
and for some of the categories, several versions of the algorithm 
exist, as indicated in Column 2 of Table 5.  
The final version of SEGUE target selection
is designated v4.6, with earlier revisions having lower version numbers.  The
significant changes from earlier versions of the target selection cuts 
for each target type are tabulated, also in Table 5.
Table 6 indicates the range of SEGUE plate numbers which correspond to each
version of SEGUE target selection.

%Boris comment on cool white dwarfs.XX
In general, colors of all objects are dereddened by 
their full \citet{sfd98} extinction value before applying the various
target selection cuts described in Table 5 and below.
Exceptions to this are the white dwarf/main-sequence (WD/MS) binary,
the esdM, and the legacy brown dwarf and legacy WD categories,
where applying the full dust correction, assuming the star lies behind 
the dust screen, is not correct, and thus uncorrected colors are preferred.

For each pointing, candidate lists of all objects 
which match the color and reduced proper motion cuts for each of the 15
categories are generated.  
Each of the 15 candidate lists is then sorted, usually randomly, 
but by magnitude in some cases (red K giants, low-metallicity categories).
After guide stars, blank sky patches and spectrophotometric 
and reddening standards are assigned hole positions on each plate, science 
target assignment begins in a round-robin fashion:  
the lists of possible objects in each of the 15 categories are examined 
in turn and the first object in each list is assigned a fiber 
(assuming no $55''$ collision with prior targets).  After selecting a target
from the fifteenth category, the algorithm returns to the first category and
the process repeats until all fibers are assigned.  Categories are eliminated
from consideration for target assignment when they reach their maximum
fiber allocation or when there are no more candidates on a given list.
This allows the categories with only a few targets per plate pair
(cool WD, sdM, brown dwarf) to always have their 
candidates targeted, while
several large categories (BHB, F, low-metal candidates, G, K giant) take up the bulk of the fibers on each plate pair.

The overall picture of where target categories are located in color and 
proper motion space is shown in Figure 3.  
The $(g-r)_0$ optical color can be used as proxy 
for effective temperature, and it provides a reasonable estimate 
of spectral type.
For a star with proper motion $\mu$ in $\rm arcsec ~yr^{-1}$, we define 
a reduced proper motion in the $g$ filter: $H_g = g + 5{\rm log} ~\mu + 5.$

Targets are selected as PRIMARY detections (duplicates and objects from
overlapping scans removed) of objects in the DR7 database with 
stellar (not extended) PSFs (the ``STAR" table in the database).  
Candidates are required to pass the following SDSS flag cuts on 
the quality of their photometric imaging data:
objects must not be saturated (flag bit: SATURATED), not close to 
the EDGE, not have an interpolated PSF, 
(INTERP\_PSF), and not have an inconsistent flux 
count (BADCOUNTS).  In addition, if
the center is interpolated (INTERP\_CENTER), there should not be a 
cosmic ray (CR) hit indicated.  These flags are set by the PHOTO pipeline for
every object, and details on all possible flags, and their meaning
can be found on the SDSS Web site: http://www.sdss.org.
A few categories below, such as the L brown dwarf category, 
require stricter flag cuts.

%Do we need to list photo flags??? XXX

We now give details of each target selection algorithm 
by category, as summarized in Table 5, and present representative spectra of
the various target types.  The sample spectra are plotted
with flux in units of $\rm 10^{-17}~ergs~cm^{-2}~s^{-1}~\AA ^{-1}$ on the y-axis
versus wavelength in \AA\ on the x-axis.
If more than one spectrum appears in a plot, then additional spectra are offset
by an arbitrary amount for readability.  Spectra are smoothed from
1.5 to three $150 ~\rm km~s^{-1}$ resolution elements, depending on the S/N. 
Common spectral features are indicated by 
line name (and the occasional night-sky feature by `NS'). 
Spectra are labeled with their
unique SDSS/SEGUE three part id (plate-mjd-fiberId), as well as with
relevant magnitudes, colors or atmospheric quantities from the SSPP analysis.
All example spectra, along with their measured parameters, can be found 
in the DR7 data release by using their three-part id to look them up.

\subsection {White Dwarfs, sdB, sdO}

	%PrimTarget Bit pattern: 0x8008000
	%Color cuts:
%$g_0 < 20.3, -1 < g-r_0 < -0.2, -1 < u-g_0 < 0.7, u-g_0 +2(g-r)_0 < -0.1$

%Weighting: magslope 1.2
%Target counts Goal: 25 per plate pair

WDs are important for absolute calibration of the
astrophysical temperature and flux scales, including calibration
of filter systems that span ultraviolet (UV), optical and IR wavelengths \citep{hb06,b07}.
The SDSS WD catalogs of \citet{ketal04} and \citet{eetal06} contain 
an extensive list
of WDs discovered in the SDSS imaging and spectroscopy.

A goal of SEGUE is to continue this survey of hot WDs
by obtaining spectra of most WDs with $\rm T_{\rm eff} > 14000~K $
and most hot subdwarf stars, while excluding most QSOs.
SEGUE obtained spectra of 4069 hot WD candidates, of which
about 62\% appear to be DA type WDs while roughly 15\% are
other types of WDs.
Other hot stars, designated sdB and sdO types 
\citep{gsl86}, most of which are extreme horizontal branch stars,
are also selected by this SEGUE WD color box.  
Roughly 10\% of the objects selected as hot WD targets
are QSO or emission-line galaxy contaminants.
Figure 4 shows sample SEGUE DA WD, sdB and sdO spectra.

\subsection {Cool White Dwarfs}

Cool WD stars, the fossil remains of an ancient
stellar population,  offer a window into the early stages of the
Galaxy and its formation.  They can be used to place lower limits
on the ages of various Galactic components, extend our knowledge of stellar
evolution, and provide hints of star formation processes
during the Galaxy's earliest epochs. 
Very cool (ultracool) WDs with hydrogen in their 
atmospheres exhibit a unique spectral signature due to
collision-induced absorption (CIA) by molecular hydrogen.  
In pure H-atmosphere WDs, 
CIA is mediated by H2--H2 collisions 
that produce a flux suppression in the IR at temperatures
below about 5000 K, resulting in objects of very unusual color.

SDSS has proven to be an excellent database in which to search for 
ultracool WDs.  
To date, 14 new ultracool WDs have been discovered 
in SDSS spectral data (Harris et al. 2001; Gates et al. 2004; 
Harris et al. 2008; Hall et al. 2008), constituting the majority of 
known ultracool WDs. Additional cool WD candidates 
have been identified in SDSS photometric data
(Kilic et al. 2006; Vidrih et al. 2007).   Several extremely
faint high proper motion cool WDs have recently 
confirmed in the SDSS repeat-scan (stripe 82) data \citep{scholz08}.

SEGUE presents a unique opportunity to find more of these rare objects.
Recent SDSS studies 
of WDs (Kilic et al. 2006 and Harris et al. 2006) 
have demonstrated the usefulness  of a reduced proper motion cut
for selecting candidates from the photometric data, and
a similar procedure is used to identify targets in the SEGUE
imaging data.  
Objects which satisfy the Table 5 selection cuts in color and 
reduced proper motion are targeted as cool WDs and assigned spectral fibers.
All selected targets are required to have a good proper motion match
(as defined in Kilic et al. 2006) in order to obtain a reliable reduced proper 
motion $H_g$.

For target selection versions prior to v3.3, cool WDs were allotted
a maximum of 10 fibers per plate pair; however, analysis of preliminary
SEGUE data observed in 2004 revealed that selection cuts frequently 
yielded fewer
than this, with occasional fields containing 11 to 15 target objects while
the overall average remained less than 10.  As a result
the final target selection algorithm targets all objects which satisfy
the selection criteria (increasing the number of ultracool WD fibers to 
more than 10 if necessary for a specific plate pair), ensuring that 
we obtain spectra for all candidates.  
This reduced proper motion selection algorithm also allows us
to target all low luminosity WDs and most high-velocity WDs  
and identify cool WD candidates that exhibit 
milder CIA suppression. 

SEGUE targeted about 1187 cool WD candidates.
While an analysis of the full SEGUE cool WD set is currently
underway, preliminary results show that the selection by
reduced proper motion is yielding a high return of cool and
high-velocity WDs.  Out of 16 plate-pairs studied,
60\% of the targets are cool DA type WDs, 18\% are DQ or DZ type,
and 12\% are DC type, while 10\% are contaminating objects of
nondegenerate stars or QSOs.  Of the rare ultracool WDs
that are intended as a goal of this selection category,
two of the 15 published SDSS ultracoool WDs (SDSS J0310-01 and
SDSS J2239+00) were targeted in this category and found in
early SEGUE spectra.  Figure 5 shows a sample cool WD
spectrum from SEGUE.

%While an analysis of the full SEGUE cool white dwarf set is currently underway, 
%the target algorithm has already proven successful -- two of 
%the 14 published SDSS ultracool white dwarfs, SDSS J0310-01 
%and SDSS J2239+00, were targeted by SEGUE as a cool white 
%dwarfs and found in a search of the early SEGUE data.  
%SEGUE targeted about 1187 CWD candidates, with about a 
%0.5\% success rate.  Figure 5 shows a sample SEGUE CWD star.  

%Cool White Dwarfs \citep{getal04,hetal08} represent a rare and interesting
%population of objects which may help us to date the Milky Way's disks
%and halo independently of other methods.

%What about a better example, or several examples...

\subsection {BHB, A main sequence, Blue Stragglers}

%Weight objects with low v=0 four times as likely
%to get fiber as objects of v= 0.4.
%This 'v' parameter weighting as of v3.3 and later.

%This section needs work XX

BHB stars in the halo are important distant 
standard candles for
mapping structure in the outer halo \citep{yetal00}.  They may be 
seen to distances of 80 kpc or more and, along with red K giants, 
are our most distant stellar halo probe of kinematics \citep{netal03}.
The recent work of \citet{xetal08} shows how this sample probes large
enough distances in the halo to constrain the mass of the Milky Way
at $d \sim 50 $ kpc.  
A linear combination of the SDSS colors, 
$v = -0.283(u-g) -0.354(g-r) +0.455(r-i) + 0.766(i-z)$  \citep{lenz} is
somewhat helpful in separating BHBs from higher surface gravity blue
stragglers (BS), and it is used to weight selection of targets toward BHBs.
BS are surprisingly common in the halo, 
and are thought to be the result of binary evolution.  They are an interesting
population in their own right, though halo members can be difficult to 
separate from disk populations at brighter magnitudes \citep{mom07}.
SEGUE placed fibers on 24,688 BHB/BS candidates, and about 66\% of the 
resulting spectra are auto-classified as BHB/BS types.
Figure 6 shows examples of BHB and BS
spectra.  

%v color weighting... of lenz???

%RR Lyraes here or F

\subsection {F turnoff, Plate Spectrophotometic Standards}

%Note: there was an s-color in versions prior to v3.3, $s >= -0.065$,
%no longer present for v3.3 and later.

%Weighting: colorslope and mag weighting P1 min/max slope = 5

%Target count goal: 200 per plate pair

F turnoff stars are an extremely numerous, relatively luminous category,
and their spectra are very clean and amenable to accurate spectroscopic
analysis \citep{rzetal06}.  
The number of F turnoff stars available toward a given SEGUE pointing 
far exceeds the number of fibers available, thus, in targeting this category,
SEGUE used linear combinations of the ($(u-g)_0,(g-r)_0$) colors 
to favor the selection of lower metallicity, halo objects.  This is done by
using the `s' (perpendicular to the stellar locus) and P1(s) (parallel
to stellar locus) colors as described in \citet{helmi03}.  These
colors essentially ``straighten-out" the stellar locus in the vicinity of
the turnoff and allow a simple cut on P1(s) to favor halo subdwarfs,
even as the relative density of thick disk versus halo turnoff stars 
is changing rapidly as a function of magnitude.
For stars with $g < 19$, the $S/N$ is generally high enough for the
SSPP to derive atmospheric parameters with relatively small uncertainties.
In the faintest magnitude bin ($19 < g < 20.5$), only the RVs
are still accurate, but it is possible to use this large number
of F dwarfs to probe halo substructure at distances of 
$10 < d < 18 $ kpc from the Sun \citep{netal02,netal07}.
Due to their large numbers and relative absolute brightnesses, SEGUE targeted
37,900 F subdwarfs, plus about 6500 spectrophotometric and
reddening standards.  About 70\% of these candidates
yielded spectra classified as type F, with some indication of lower
metallicity ([M/H] $< -0.7$).

Kinematics of SEGUE F turnoff stars has recently allowed \citet{wetal09}
to tightly constrain the orbit of a halo stream \citep{gd06}.

%quantify indication of lower metallicity. sspp Fe/H < -1?? XX

\subsection {Low-Metallicity Candidates}

%Note: prior to v3.4, l-color cut was $l-color > 0.15$, but this was too
%restrictive, and so it was loosend to $l-color > 0.135$ in v3.4 and later.
%Prior to v3.3, this l-color cut was at $> 0.12$, but that was a bit too loose.

%Weighting: magslope 1.8 and faint plate ($r_0 > 17.8$ targets sorted by 
%magnitude to 'pack' them so that brightest ones targeted before fainter ones.

%Quote howard bond 1970, 1980 XX

The $u-g$ color of F, G, and K stars with $0.4<(g-r)_0<0.8$ can be used as a
metallicity proxy, in the sense that bluer stars tend to have lower
metallicity \citep{lenz}.   We can employ this color cut to restrict
the number of spectroscopic targets.   For $(g-r)_0>0.8$, $(u-g)_0$ fails as an
effective discriminant.

Another item to consider when searching for very low metallicity stars is
the volume sampled.  While low metallicity K and M dwarf stars 
live for much longer on the main sequence than F and G stars, their 
intrinsic faintness relative to the other spectral types means that our
magnitude-limited sample of these stars is dominated by disk stars 
that are near the Sun.  Relatively few spheroid stars are expected to
be observed in the small volume of the Galaxy that we probe with
late type dwarfs.
Since F ``turnoff" stars are significantly brighter (they can be 1--2 
mag brighter
than their zero-age main-sequence luminosity), and therefore can be 
seen to further
distances,  many of the lowest metallicity [Fe/H] $< -3.0$ 
candidate objects identified to date have been found to have colors of
F turnoff stars.   See \citet{apetal08} and \citet{cetal07} for 
extensive studies of halo F turnoff stars.

SEGUE targeted 30,998 candidates in the low metallicity category.  
About 12\% (4600) of them show an SSPP 
metallicity of $\rm [M/H] < -2$. About 0.1\% (32)
indicate metallicity $\rm [M/H] < -3$.  The very lowest metallicity candidates
will need to be follow-up up on larger telescope at higher resolution.
We show in Figure 7 a set of turnoff stars, all with similar effective
temperatures, that have metallicities ranging from $\rm [M/H] < -3$ 
to super-solar (the higher metallicity stars in this
sequence were not selected from the low-metallicity targeting category).

%Plus Many in K giant and Fsd, Stds categories.

It should be noted that the present version of the SSPP, 
described in detail by \citet{letal08a,letal08b},
Allende Prieto et al. (2008), and Re Fiorentin et al. (2007) 
produces conservative estimates of
[Fe/H] for stars with metallicities below roughly [Fe/H] = -2.7.  A
recent preliminary analysis of over 80 SDSS/SEGUE stars with SSPP
metallicity determinations [Fe/H] $<$ -2.7, based on high-resolution
spectroscopy obtained with the Subaru/HDS (W. Aoki, 2008, private
communication), indicates that the actual metallicity can be 0.3
dex lower than the level determined by the SSPP.  This analysis shows
that the
lowest metallicity stars from this category have [Fe/H] = -3.7.
With this new recalibration,
T. C. Beers et al. (2009, in preparation) assemble
some 15,000 stars with [Fe/H] $< -2.0$, and several hundred with
[Fe/H] $< -3.0$.  These totals
include low-metallicity stars from all of the various target
categories used in SDSS/SEGUE.

\subsection {F/G stars}

The F/G target category represents an unbiased random subsampling 
of the range of stars with colors $0.2 < (g-r)_0 < 0.48$.  This
distinguishes it from the F subdwarf category (above), which
is biased toward objects of lower metallicity.
This category was only used in target selection 
versions v3.3 and later.
SEGUE targets 6939 of these, of which 90\% are classified as 
type F or G.  Figure 8 shows an example spectrum.

%Can we get more spectra examples here XX with Teff, log g, etc?
%Also, table of spectral type vs. Teff, g-r conversions

\subsection {G dwarfs and Sgr subgiants}

The G dwarf sample represents SEGUE's largest single homogeneous stellar
spectral category.  The target selection is very simple, just a color cut in 
$(g-r)_0$, and thus is very close to unbiased.
%A simple description of the G-dwarf problem says that more 
%low-metallicity G stars should be found in the solar neighborhood than are
%actually observed. A more in-depth analysis shows that kinematics plays
%an important role \citep{wg95,w86}.  
With the SEGUE unbiased G dwarf sample, researchers will
be able to address the metallicity distribution function (MDF) as well
as the kinematic distribution of G dwarfs in a much larger 
volume of the Galaxy than has been previously attempted \citep{wg95,w86}.

This sample will also be extremely useful for
probing the structure of the Galaxy's major components, especially using
the brighter stars with S/N  $> 20$, for which surface gravities can be
determined by the SSPP.  Subgiant stars (spectral type G IV) stars in the
Sagittarius dwarf tidal debris stream can also be isolated from stars in this
selection category.

SEGUE targeted 62,784 G star candidates based on the simple color selection.
At least 96\% of these yield G star spectra. A significant
population (roughly 7\%) of evolved (subgiant or giant) spectra are 
indicated by the SSPP $\rm log~g< 3.75$ indicator.
Figure 9 shows a sample G dwarf star spectrum (lower) and a G giant spectrum (upper).

\subsection {K Giants}

K giants are the most luminous tracer available for old stellar
populations. They can (albeit rarely) be found with absolute
magnitudes $M_g$ as high as $-2.0$.  Such stars at $g=18$
are then located at a distance of 100 kpc.

It is not possible to use a simple selection criterion such as that
used by SEGUE for F/G stars or G dwarfs to select K giants, because
in SEGUE's apparent magnitude range,
giants are swamped in number by foreground dwarfs with the same
$(g-r)$ color.  We can use the G dwarf category, which overlaps the 
blue edge of the K giant color range, to demonstrate this.  Only
3\% of the targeted G dwarf stars with $g=15-17$ in the SEGUE data are
giants. By contrast, 50\% of the stars with similar magnitudes that were 
targeted with the selection criteria described below are giants.

In past surveys, K giants have been identified using the
pressure-sensitive feature near 5200\AA\ , produced by a
blend of the Mg I$b$ triplet and the
MgH band, which is strong in dwarfs and weak in giants, 
but unfortunately also has some sensitivity to stellar
metallicity and temperature \citep{bonner}. This feature is a good
luminosity criterion for late G and K stars redward of $(g-r)_0 = 0.5$
($B-V \sim 0.75$). \citet{kavan} and \citet{fm90} used objective prism
spectra to isolate stars with weak Mg$b$/H features in the G/K color
range. As it is quite a broad feature, it is also possible to
measure their strength using intermediate-band photometry
\citep{metal00}.  

%This photometric survey technique is quite
%efficient, but requires a large amount of imaging time because of the
%160\AA wide filter.

Since the SDSS $ugriz$ filters do not have a narrow filter in the
Mg$b$/H region, we use a more indirect method of finding giants.  In
SEGUE's magnitude range, many of the giants that we observe are
halo stars, so we use a photometric metallicity indicator to
remove foreground disk dwarfs.  Briefly, we select stars with a
UV-excess in the $ugriz$ system, and use the power of our fiber
spectroscopy to eliminate the remaining foreground dwarfs. 

In more detail, in the $(g-r)_0,(u-g)_0$ diagram, metal-poor stars appear bluer
in $(u-g)_0$ for a given $(g-r)_0$ because they have less line-blanketing in the $u$
filter. Two different colors have been defined to measure the
deviations from the mean stellar locus in this diagram: the $s$-color of
\citet{helmi03} and the $l$-color of \citet{lenz}. Large values of the
$l$-color correspond to metal-poor stars, and we have chosen to target
stars with $l>0.07$.
Our selection strategy is complicated by the fact that the Mg$b$/MgH
feature is within the SDSS $g$ passband (it is in the blue wing of the $V$
filter).  Giants (with weaker Mg$b$/MgH) will be bluer in $g-r$ than dwarfs.

Figure 10 shows how the competing effects of metallicity and gravity
play out. Our K giant selection region extends from $(g-r)_0 = $ 0.5 to
1.2. The main locus, from the accurate averaged photometry from Stripe
82 \citep{zeljko} is shown in gray: stars with $l$-color greater than
0.07 are shown in light gray. Spectroscopically confirmed dwarfs are
shown with large crosses, and giants with large filled circles. Giants
range in metallicity from near solar to [Fe/H] less than -2.0. It can be
seen that for $(g-r)_0$ between 0.5 and 0.8, the metallicity sensitivity
of the $(u-g)_0$ color is the dominant effect, and metal-weak giants are
clustered toward the blue side of the locus. For redder stars, the
luminosity sensitivity of $(g-r)_0$ dominates, and giants appear above the
main locus.

The details of our target selection for this category have changed as
we have learned more about the behavior of giants, particularly the
rare red giants and dwarfs in
$ugriz$ colors, particularly the rare red giants.
The efficiency of our discovery of K giants depends on magnitude and
color. Using an $l$-color cut of 0.10 rather than 0.07, success rates
are as follows: For $g=17-18$, $0.5 < (g-r)_0 < 0.6$, our success
rate is 45\%. For the same magnitude range and $0.6 < (g-r)_0 < 0.8$, 
the success rate is 28\%.

A modification to the K giant algorithm occurred for SEGUE target-selection
versions v4.2 and later to extend the selection to very red ($0.8 < (g-r)_0 < 1.2$)
K giants, which are referred to as red K giants in the
SEGUE target selection tables 
(there are few bona fide M giant spectra, with TiO bands and 
$(g-r)_0 > 1.3$, in SEGUE).  Although only the last 
quarter or so of SEGUE plates targeted very red K giants, 5948 fibers
were placed on these very red stars, of which 466 (8\%) yielded a 
low gravity spectral indicator from SSPP.  The selection algorithm included
$(g-r)_0$ color, lack of proper motion, and a weighting toward brighter magnitudes.
For effective random sampling
with only a proper motion cut for $(g-r)_0$ between 0.8 and 1.2, the
success rate is 2.5\%.

%Weighting: magslope 2.0, Target count Goal:
%95 per plate pair (expected density 4/sq degree = 28/plate pair)

%Note weight towards what are actually GIV subgiants,
%Actual red K/M giants didn't come until late versions of
%the code v4.X and later.

%Extra section, or drop >1.3 discussion XX

This important category received 22,814 fibers, of which about 30\% yielded 
a low surface gravity K giant, red K giant or subgiant spectrum.  
Figure 11 shows a classic K giant (top panel) and a very red K giant 
(lower panel).

%Note: plates 2336,2337 have special K giant selection,
%$0.35 < (g-r)_0 < 0.9$, proper motion consistent with 0.

%Likely to contain Monoceros Giants

%The K giant category allows us to probe the most distant ($d \sim 100 \rm kpc$)  regions of the Milky Way halo kinematically.  The Spaghetti survey \citep{metal00} has also focused on this category to probe structure.  The 2MASS survey
%has revealed a large number of distant M-giants (the very reddest K-giants,
%with $g-r_0 > 1.1$) associated with the Sagittarius tidal stream 
%\citep{metal03}. These stars have well defined SDSS colors, $1.1 < g-r_0 < 1.3, 3.03 < u-g_0 < 3.47, 0.366 < r-i_0 < 0.61,  0.23 < i-z_0 < 0.36$, and
%for $g < 18$, there is little contamination from K dwarfs. The Sgr stream stars
%with these colors have $M_g \sim -1.0$.
%The very brightest K(M) giants have the very reddest colors,
%$1.1 < g-r_0 < 1.3$, objects with somewhat bluer $g-r_0$ colors perhaps
%more properly called subgiants, as they are generally below the
%point on the color magnitude diagram where the horizontal branch
%intersects the red giant branch.   Initial targeting for SEGUE
%(and SDSS) obtained very few very red giants, and many more
%sub giants.  The algorithm for plates obtained with v4.2 and beyond
%has a special selection to get many of the reddest K(M) giants, and
%has found numerous objects, many of them apparently associated with
%the Sgr dwarf tidal stream.
%
%Failed category.

%AGB category 499 targeted, not sure how many were useful actual AGBs.
%XX

\subsection {AGB}

An early SEGUE target selection category, designated ``asymptotic giant branch'' (AGB), that was intended to select the 
very reddest giants in Figure 10, targeted 1343 objects.
Only about 8\%, however, yielded actual red giant (and possible AGB
) spectra.  In hindsight, the choice 
of $s$-color cuts, and $(g-r)_0$ limits were not optimal, 
and the low latitude AGB category, 
(LL AGB) below, and the red K giant category above 
eventually superseded this AGB category for SEGUE target selection of
this type of object.  

\subsection {K dwarfs, early M dwarfs}

%Science goal: full phase info of ppm+RV + dist+ xy of stars within
%1 kpc.
The K dwarf and early M dwarf category generated
a significant  unbiased sample of relatively nearby 
Galactic objects with complete six-dimensional phase space information:
proper motions, RVs, positions, and photometric parallax
distance estimates are all available for this data set.
SEGUE targeted 18,358 stars in this category, of which 80\% yield useful spectra.
Figure 12 shows sample K and M dwarfs.

\subsection {M subdwarfs, high proper motion M stars}

The aim of this category is to obtain a sample of low metallicity M 
(sub)dwarfs, which are strongly correlated with halo M dwarfs.
Originally, 1012 sdMs were targeted, based on a version of 
the \citet{wetal08} color cuts.
However, for a variety of reasons, 
only a handful of these candidates were actual sdMs.
A second window onto this category was opened by using
the American Museum of Natural History (AMNH) high proper motion catalog 
and careful color selection criteria \citep{l08,ls08}.
Approximately 40 fibers per plate pair were allocated to this
search for sdM, esdM, and usdM objects.
This category obtained a very high ($\sim 20$\%) success rate
on an allocation of 9420 candidate fibers.
Figure 13 shows a low metal M dwarf and a high proper motion selected 
extreme subdwarf M star (esdM2.5).  

\subsection {WD/MS binaries}

%fix the refs as bibitems XX
This category, designated WD/MS binary,  
has been used successfully to improve our understanding of close compact binary 
evolution.  If the SEGUE WD/MS
Survey is combined with follow-up observations identifying the short
orbital period systems among the WD/MS binary stars, important processes
such as the common envelope phase (see Webbink 2007, for a recent 
review) or angular momentum loss by magnetic wind braking can be 
constrained (see Politano \& Weiler 2006, Schreiber et al. 2007). 
Already, SDSS-I efficiently identified new WD/MS binaries. \citet{setal04}
identified a new stellar locus, the WD/MS binary bridge.
\citet{siletal06,siletal07} published lists of
more than 1400 spectroscopically identified WD/MS binaries.
These samples, however, mainly consisted of young systems
containing hot WDs.  

The selection criteria used here have been designed to identify a large 
sample of old WD/MS binaries containing cold WDs that, according 
to Schreiber \& G\"ansicke (2003), should represent the dominant population 
of WD/MS binaries.  
On 240 spectroscopic plates in DR7, the WD/MS color selection a
lgorithm chose 9531 candidates of which 431 have been observed 
spectroscopically.  Among these we confirm 244 WD/MS objects (with 25 other
 possible candidates) resulting in a success rate of $\sim 56\%$.
A first analysis shows that indeed for the first time a large sample of old 
systems with WD temperatures below $\sim\,12000$\,K could be 
identified (for more details see M. R. Schreiber et al. 2008, in preparation). 
Follow-up observations to further constrain compact binary evolution 
are well underway (Rebassa-Mansergas et al. 2007, 2008, 
Schreiber et al. 2008, A. Schwope et al., 2009 in preparation).
The total SEGUE allocation to this WD/MS category is about 500 fibers, with
a 56\% success rate expected.
Figure 14 shows one example WD/MS spectrum.

\subsection {Brown Dwarfs, L and T dwarfs}

Very red objects are photometrically detected in the $z$ image, but not
in the bluer $ugri$ SDSS images.  These are rare and interesting objects,
and the only difficulty in targeting them was to assure that the
one detection was a solid detection and not a spurious CR hit
to the $z$-band CCD.  Additional flag checking was performed for
candidates in this category, similar to that done by \citet{fetal99} for 
selecting very high redshift quasars that are detected in $i, ~z$ or $z$ only.  
SEGUE devoted 1277 fibers to 
this category, with about a 7\% initial estimated yield of objects with
spectral type later than M8.  Figure 15 shows a SEGUE L dwarf spectrum.

\subsection {The Low-Latitude (LL) Algorithms}

For 12 pointings at $|b| < 20$ (all toward regions of high
reddening, $E(B-V) > 0.3$), the previously mentioned SEGUE target 
selection algorithms were not effective.  
The long lines of sight through the Galactic disk, at low latitudes,
often rendered invalid the implicit assumption that all of the dust lies
{\it in front} of the target stars. 
For these pointings, marked as `LLSurvey' in Table 3, we use
an alternative target selection scheme that targeted three categories:
1) bluest object candidates (mostly BHB and F stars): selecting the 
bluest ($g-r$) objects in any given pointing without regard for absolute colors;
2) K giant candidates: using the absence of proper motion as the primary 
selection criterion; and 3) AGB type objects: singling out objects 
in a fashion similar to that used for selecting red K giants, but with 
a brighter $g$ magnitude limit.  This latter category was assigned
only a small fraction of fibers.

Because of dust and crowding, recovering the selection function 
at low latitudes is problematic, and optical
spectrophotometry of these low-latitude stars with SEGUE should be regarded as
experimental, with an eye toward future surveys.

About 12,241 fibers were devoted to low-latitude algorithm plates.
Each of the three primary LL categories, K giant (3220 candidates), 
AGB (499 candidates), and Blue-tip (8522 candidates) 
yielded about 30\% success rate based on SSPP analysis using methods
which fit normalized spectra (flattened without continuum slope, and 
are thus much less dependent on reddening).  
Figure 16 shows three spectra from one LL algorithm plate,
one from each category.

%indicate which plates have this algorithm used in table (LLSurvey)

%Additional changes for early versions documented on the
%http://segue.uchicago.edu/targetsel.html web page.

\subsection{The SDSS Legacy Stellar Target Selection algorithm}

In addition to SEGUE, the SDSS and SDSS-II Legacy Surveys obtained spectra for
nearly 200,000 (additional) stars, which were allocated fibers on 
the main SDSS Legacy Survey plates that observed primarily galaxies.  
The target selection categories were briefly
described in \citet{setal02}.  We list in Table 7 the
color and magnitude cuts used to select stars by category in the
SDSS Legacy Survey.   It should be noted that, unlike SEGUE, Legacy's stars
targeting algorithms may assign multiple target type bits to the same
target, that is, an object may be targeted based on its colors, as
both a SERENDIPITY\_RED object and a ROSAT\_E object.  These bits are
tabulated in the PrimTarget field of the CAS database. The final
column of Table 7 lists the approximate number of candidates which received
a fiber in each target category.

\subsection {Cluster Plates}

In order to calibrate the SSPP's metallicity, luminosity and effective temperature scales for
all stars, a significant number of known globular clusters and a few open clusters were
targeted with one or more SEGUE plates.  These 12 pointings are indicated 
with the cluster name followed by the word `Cluster' in Table 3.  Because many of these
clusters are relatively nearby, they have giant branches with stars brighter than
the SEGUE spectroscopic saturation limit of $r\sim 14$.  Additionally, due to the
extreme density of the globular cluster fields, even the Pan-STARRS assisted PHOTO processing
of SDSS and SEGUE imaging scans failed to resolve individual stars in the centers of
the globulars.  For these reasons, the following procedures were followed for most
cluster plate observations: 1) the target list for each cluster was generated individually.  Proper motion membership criteria were used for
many clusters, allowing the targeting of stars in the dense cores of clusters which were saturated in the regular SDSS imaging.
K. Cudworth (2006, private communication) provided membership lists for a significant fraction
of the clusters targeted.  Some of these targets do not, therefore, have standard SDSS
run-rerun-camcol-field-id numbers, and are identified only by their R.A., decl. positions.  
2) Shorter exposures were obtained for many of the clusters.  One or two minute exposures allowed us to obtain nonsaturated spectra of stars
as bright as $r \sim 11$.  

These cluster spectra were used to calibrate
the SSPP for all types of stars \citep{letal08a,letal08b}.  The
spectra of cluster stars are available on an as-is basis in the DR7 database.  
Users should be aware that the spectrophotometry and in some 
cases the RV
solutions of these plates may be subject to large systematic errors due 
to the extremely short exposure times,
and that occasional bright targets may be saturated.

A major effort to process SDSS and SEGUE imaging data for clusters has 
been undertaken by \citet{aetal08}, using the DAOPHOT/ALLFRAME suite 
of programs.  They  reduced imaging data for crowded cluster fields
where PHOTO did not run, and presented accurate photometry for 20
globular and open clusters.\footnote{Available at
{\tt http://www.sdss.org/dr7/products/value\_added/anjohnson08\_clusterphotometry.htm}}
This effort has led to fiducial sequences for clusters over a wide range
of metallicities in the SDSS $ugriz$ filter system.

%Mention Cudworth pm selections

%A few plates of known globular were populated with targets which didn't
%have SDSS photometry (cluster center saturated or not scanned by
%the 2.5m imaging camera).

%Marked in table with "X Cluster"

\subsection{Special plates centered on halo substructure}

Studying halo streams is an important goal of SEGUE, and SEGUE is well placed to
obtain new kinematic information on the stars in these streams.
Five previously known streams were specially targeted at various 
positions along their extent with SEGUE plates, for a total of 16 pointings. 
The five streams are Sagittarius \citep{yetal00}, Monoceros \citep{netal02}, 
Orphan \citep{betal07b}, Virgo \citep{vetal01,jetal08} and 
GD-1 \citep{gd06}.  Plates with these pointings are 
marked with the stream name and `Strm' in Table 3.  The regular target selection
algorithms described above were used on these streams, i.e. no special targets
within the stream were given specific fibers.  Some BHB, BS, 
F turnoff, K giant, and G subgiant spectra are clearly confirmed 
stream members on these `Strm' plates.  Identification of a particular star
as a stream member versus a field star is of course, often difficult, and
left to the interested researcher.

%Marked in table, used same algorithm

%Should this be in the appendix?

\section{ The Data Archive and an Example Query}

All of the spectra and associated imaging from SEGUE were made 
public as a part of DR7
of the SDSS-II Survey on 2008 October 31.  The
calibrated magnitudes and positions of objects as determined by
the SDSS PHOTO pipeline software are available
in the `photoobjall' and `star' tables of the CAS database, 
which is available through interfaces 
at http://cas.sdss.org and described in detail at http://www.sdss.org.
We note the SEGUE and SDSS Legacy imaging and spectra are all in
a single large database, so it is possible to obtain SDSS Legacy 
photometry and SEGUE spectroscopic information for stars in 
the SDSS and SDSS-II footprints as part of a single query, or closely
related set of queries.
The SSPP and SPECTRO outputs (RV, $\rm [M/H]$, $\rm log~g$, $\rm T_{\rm eff}$), are available 
in tabular form in the `sppParams' and `specObjAll' tables of 
the CAS database.  
FITS format files containing all of the $\sim $240,000 extracted, co-added, 
sky-subtracted, flux-calibrated spectra are available at http://das.sdss.org for
interested researchers.  Similar files for all 200,000 stellar objects detected
by SDSS and SDSS-II Legacy Surveys are also available at the same location.

To highlight the usefulness of the data archive, we present an example 
SQL query
from the CAS database to help construct plots showing the extent and
scientific usefulness of SEGUE data.

We design a query to select SEGUE-targeted G stars from the database,
and return each object's photometric and tabulated spectroscopic 
information.

The following SQL query is presented to the CAS DR7 database 
(http://cas.sdss.org/CasJobs):

\begin{verbatim}
select 
       dbo.frun(targetid) as run,
       dbo.frerun(targetid) as rerun,
       dbo.fcamcol(targetid) as camcol,
       dbo.ffield(targetid) as field,
       dbo.fobj(targetid) as obj,
       g0,umg0,gmr0,spp.ra,spp.dec,spp.l,spp.b,
       spp.plate,spp.mjd,spp.fiberid,elodierv,elodierverr,
       feha,fehaerr,fehan, logga,loggaerr,
	elodierv+10.1*cos(b*3.14159/180)*cos(l*3.14159/180)+
		224.0*cos(b*3.14159/180)*sin(l*3.14159/180)+
			6.7*sin(b*3.14159/180) as vgsr
 	from sppParams spp,specobjall sp, platex p 
	where spp.specobjid = sp.specobjid and sp.plateid = p.plateid and 
	p.programname like 'SEGUE%' and 
	gmr0 between 0.48 and 0.55 and elodierverr > 0
\end{verbatim}

This query is written in `SQL', a standard database query language. 
Some details of the query are as follows.

The first six lines select photometric quantities for the desired objects,
include the unique five-part SDSS imaging ID (run,rerun,camcol,field,obj), 
dereddened photometry ($g_0,(u-g)_0,(g-r)_0$), (R.A.,decl.) J2000,
and Galactic $(l,b)$.

The next lines select spectroscopic outputs for each object:
unique three-part SDSS/SEGUE spectroscopic ID (plate,mjd,fiberid),
RV and error (obtained by cross-correlation with ELODIE RV templates),
elodierv, elodierverr in $\rm km\>s^{-1}$, and SSPP estimates of [M/H] (feha) with 
error (fehaerr) and an indication of how many estimators went in to the [M/H] estimate
(fehan). Similarly, the stellar surface gravity and error is retrieved (logga, loggaerr).

A significant fraction of the spectra will have feha or logga set to -9.999, which
indicates that the SSPP did not have sufficient S/N to estimate this value confidently.
These values are mostly for fainter $g > 19$ spectra.

The next lines take each object's heliocentric 
RV and Galactic $(l,b)$ and uses
the database to compute the Galactocentric $v_{\rm gsr}$ velocity using
standard values for the solar motion.

The selection is done with an SQL `join' between the SSPP table `sppParams', 
and the spectroscopic id table `specobjall' and the plate list table `platex', 
requiring that the type of spectroscopic program matches the key phrase `SEGUE' (to exclude LEGACY SDSS galaxy plates) and that the dereddened color of 
the objects fall in the G dwarf color range ($0.48 < (g-r)_0 < 0.55$).  
Additionally the RV error must be greater than zero, 
indicating a minimal level of quality of the spectrum.  This also
excludes galaxies and quasars, leaving only objects with stellar kinematics.
 
This query yields a table (in comma separated value format, which may be 
downloaded to a user's local computer for further manipulation) of 
61,343 objects.

Individual images of spectra in this data set may be examined by fetching
from the DAS, with a link like.

{\tt wget http://das.sdss.org/spectro/1d\_26/1880/gif/spPlot-53262-1880-014.gif}

where the object may be verified to be a G star.

A FITS data file of the calibrated 1D spectrum is available from

{\tt wget http://das.sdss.org/spectro/1d\_26/1880/1d/spSpec-53262-1880-014.fit}

for detailed further manipulation.

Figure 17 shows steps in isolating an interesting subpopulation of these
G stars.  The topmost panel plots Galactic $v_{gsr}$ versus $l$ for all G stars
selected with the CAS database query above.
A sinusoidal curve with amplitude 120 $\rm km~s^{-1}$ shows the average 
path of stars rotating with the Sun about the Galactic center.
Several knots of G stars stick out from the general disk population(s). 
We focus on one set of stars with $-175 < v_{gsr} < -111 \rm \> km\>s^{-1}$ and
$168^\circ < l < 182^\circ$.  Galactic ($l,b$) for this subset is plotted 
in the second panel.  A histogram of the apparent magnitudes of the further subset of stars 
with $-52^\circ < b < -37^\circ$ is plotted in the third panel.
If these stars are dwarfs at $M_g \sim 5.5$, their implied distance from the 
Sun is 5 kpc.  However, they may be giants or subgiants at much further distances. We plot the histogram of SSPP surface gravities in panel 4.  There is a clear peak at subgiant and giant $\rm log~g < 3.75$ (all G dwarfs have $\rm log~g > 4.25$).
Therefore this population consists of subgiants with $M_g \sim 1.5$, placing
the stars at $d = 30 $ kpc from the Sun.  A histogram of the quantity feha (adopted average [Fe/H] from the SSPP pipeline) for the subgiants is shown in the lowest panel.  The estimated metallicity
of these objects is $\rm [M/H] = -1.4\pm 0.5$.  The location of these objects
on the sky and their implied distances and velocities are consistent with 
that of Sagittarius Southern tidal stream stars, such as RR Lyraes and BHBs seen in \citet{ietal00,yetal00}.

This example just scratches the surface of the interesting science
that can be obtained with the SEGUE G star sample.

%Explain what strip/strip, etc are.

%Remove 'possible' strips and plates from table, as they it won't get done.

\section{Summary}

The SEGUE Survey provides a large sample of more than 240,000 spectra
of stars at ($14 < g < 20.3$) covering 212
sightlines out to distances of 100 kpc. It supplements the SDSS Legacy imaging
survey with an additional 3500 $\rm deg^2$ of $ugriz$ imaging
at $|b| < 35^\circ$.
Each 7 $\rm deg^2$ sightline obtains 1150 well-calibrated 
resolution $R = 1800$ spectra, analyzed by a uniform set of software pipelines
to generate tables of RVs for all stars 
and estimates of [Fe/H], $\rm log~g$ and $\rm T_{\rm eff}$
for stars with S/N $> 10$.
The selected targets in each pointing cover all major spectroscopic types
from WDs to L brown dwarfs in numbers sufficient to sample the
kinematic structure of all major stellar components of the Galaxy 
(except the bulge) at distances from the solar neighborhood (probed with
M, L, and T dwarfs) to 100 kpc from the Sun (probed with BHB and K/M
giant stars).  The SEGUE sample is useful for isolating stellar substructures,
particularly in the stellar halo.

The unbiased sample of 
over 60,000 G dwarf spectra presents a unique way to study the Galaxy's 
structure in detail.
Selected populations of rare targets ranging from cool WDs to
high proper motion subdwarf M stars to stars with metallicity $\rm [M/H] < -3$
allow theories of formation and evolution of the Milky Way to be 
newly constrained.

A follow-up to the SEGUE Survey, entitled SEGUE-2, is now underway with
the same instrument and telescope configuration and aims to obtain 
targeted spectra of a sample of similar size and quality to SEGUE.

All SEGUE data are calibrated and publicly accessible,
now enabling a SEGUE to many productive science explorations beyond the 
dreams of the designers.

\acknowledgments

Funding for the SDSS and SDSS-II has been provided by the Alfred
P. Sloan Foundation, the Participating Institutions, the National
Science Foundation, the U.S. Department of Energy, the National
Aeronautics and Space Administration, the Japanese Monbukagakusho, the
Max Planck Society, and the Higher Education Funding Council for
England. The SDSS Web site is http://www.sdss.org/.

The SDSS is managed by the Astrophysical Research Consortium for the
Participating Institutions. The Participating Institutions are the
American Museum of Natural History, Astrophysical Institute Potsdam,
University of Basel, Cambridge University, Case Western Reserve University, 
University of Chicago, Drexel University, Fermilab, the
Institute for Advanced Study, the Japan Participation Group, Johns
Hopkins University, the Joint Institute for Nuclear Astrophysics, the
Kavli Institute for Particle Astrophysics and Cosmology, the Korean
Scientist Group, the Chinese Academy of Sciences (LAMOST), Los Alamos
National Laboratory, the Max-Planck-Institute for Astronomy (MPIA),
the Max-Planck-Institute for Astrophysics (MPA), New Mexico State
University, Ohio State University, University of Pittsburgh,
University of Portsmouth, Princeton University, the United States
Naval Observatory, and the University of Washington.

C. Allende Prieto acknowledges support from NASA grants NAG 5-13057 and
NAG 5-13147.  T.C. Beers, Y.S. Lee, and S. Thirupathi acknowledge  partial 
funding of this work from grant PHY 02-16783: Physics Frontiers 
Center / Joint Institute for Nuclear Astrophysics (JINA), awarded by the 
U.S. National Science Foundation.  P. Re Fiorentin acknowledges support through the Marie Curie Research Training Network ELSA MRTN-CT-2006-033481.
We acknowledge useful discussions with Steve Majewski on the 
G dwarf target selection design.  We acknowledge several useful suggestions from the referee.

%update refs on the SEGUE biblio page -- and give more SEGUE news

%expand low metallicity F star sequence to include Mg I at 5130 or more.

\clearpage

%Add run numbers which make up these stripes in practice to this table

\begin{deluxetable}{rrrrrrrrr}
\tabletypesize{\scriptsize}
\tablecolumns{9}
\footnotesize
\tablecaption{SEGUE Imaging Area Stripes\tablenotemark{a}}
\tablewidth{0pt}
\tablehead{
\colhead{Stripe\tablenotemark{b}} & \colhead{$l$} &  \colhead{node} & \colhead{incl.\tablenotemark{c}} & \colhead{$\mu_{\rm start\tablenotemark{d}}$} & \colhead{$\mu_{\rm end}$} & \colhead{$b_{\rm start}$}& \colhead{$b_{\rm end}$} & \colhead{area}  \\
\colhead{} & \colhead{$^\circ$} & \colhead{$^\circ$} & \colhead{$^\circ$} & \colhead{$^\circ$} & \colhead{$^\circ$} & \colhead{$^\circ$}& \colhead{$^\circ$} & \colhead{$\rm deg^2$} 
}
\startdata
72 & ---  & 95.000 & $-25.000$ & 311.0 & 419.0 & $-14.9$ & $-27.1$ & 270.0 \\
79 & ---  & 95.000 & $-7.500$ & 311.0 & 419.0 & $-22.4$ & $-35.3$ & 270.0  \\
1020 & 10  & 60.004 & 34.950 & 242.3 & 277.3 & 35.0 & 0.0 & 87.5  \\
1062 & 31  & 98.629 & 27.192 & 247.4 & 312.4 & 35.0 & $-30.0$ & 162.5  \\
1100 & 50  & 136.813 & 31.702 & 252.0 & 332.0 & 35.0 & $-45.0$ & 200.0  \\
1140 & 70  & 161.744 & 44.754 & 257.1 & 337.1 & 35.0 & $-45.0$ & 200.0  \\
1188 & 94  & 178.713 & 64.498 & 249.1 & 384.1 & 50.0 & $-85.0$ & 337.5 \\
1220 & 110  & 186.882 & 78.511 & 269.6 & 349.6 & 35.0 & $-45.0$ & 200.0  \\
1260 & 130  & 196.095 & 276.287 & 33.8 & 128.8 & $-45.0$ & 50.0 & 237.5  \\
1300 & 150  & 205.976 & 293.890 & 41.1 & 121.1 & $-45.0$ & 35.0 & 200.0  \\
1356 & 178  & 225.998 & 316.856 & 14.2 & 129.2 & $-80.0$ & 35.0 & 287.5  \\
1374 & 187  & 236.019 & 323.166 & 66.5 & 131.5 & $-30.0$ & 35.0 & 162.5  \\
1406 & 203  & 261.852 & 331.241 & 70.5 & 135.5 & $-30.0$ & 35.0 & 162.5  \\
1458 & 229  & 315.139 & 328.784 & 94.8 & 146.8 & $-12.0$ & 40.0 & 130  \\
1540 & 270  & 176.405 & 61.064 & 152.8 & 187.8 & 35.0 & 70.0 & 87.5  \\
1600 & 300  & 191.522 & 87.391 & 171.7 & 198.7 & 43.0 & 70.0 & 67.5  \\
1660 & 330  & 25.976 & 66.110 & 195.0 & 230.0 & 38.0 & 70.0 & 80.0  \\
\enddata
\tablenotetext{(a)}{Each great-circle scan of the SDSS camera is
designated as a strip.  Given the gaps between the CCDs in the camera,
the 2.5 deg wide region in the focal plane must be imaged with two scans
to completely cover the area.  The filled 2.5 deg region is called a stripe
(see Stoughton et al 2002).}
\tablenotetext{(b)}{Stripes numbered below 1000 are part of the Legacy survey, those
numbered above 1000 are special SEGUE stripes (as can be seen by the fixed
Galactic longitudes of the SEGUE stripes).
}
\tablenotetext{(c)}{Inclination of stripe relative to the celestial equator, node is at $\alpha = 95^\circ$ for SDSS (non-SEGUE) stripes.}
\tablenotetext{(d)}{$\mu$ is the analog of Right Ascension in the great circle
coordinate system, $\nu$ is the analog of Declination.}
\end{deluxetable}

\clearpage
{
\begin{deluxetable}{rrrrrrrrrrrrrr}
\tabletypesize{\scriptsize}
\tablecolumns{14}
\footnotesize
\tablecaption{Velocity Errors by color and S/N Quartile}
\tablewidth{0pt}
\tablehead{
\colhead{$(g-r)_0$} & \colhead{N} & \colhead{$\bar g_0$}& \colhead{Q1} & \colhead {$\sigma$}&\colhead{$\bar g_0$} &\colhead{Q2}   &\colhead {$\sigma$}&\colhead{$\bar g_0$}& \colhead{Q3} &\colhead {$\sigma$}&\colhead{$\bar g_0$}& \colhead{Q4} &\colhead {$\sigma$}  \\
\colhead{mag} & \colhead{} & \colhead{} & \colhead{S/N} & \colhead{$\rm km~s^{-1}$}  & \colhead{} & \colhead{S/N} & \colhead{$\rm km~s^{-1}$}  & \colhead{} & \colhead{S/N} & \colhead{$\rm km~s^{-1}$}  & \colhead{} & \colhead{S/N} & \colhead{$\rm km~s^{-1}$} 
}
\startdata
$(g-r)_0 < 0.10$ & 488 & 19.3 & 6.8 & 23.9 & 17.4 & 15.7 & 9.5 & 16.6 & 29.1 & 6.3 & 15.3 & 54.4 & 4.2 \\
$0.1<(g-r)_0<0.30$ & 2504 & 19.5&7.6&18.0&17.7&16.1&7.6&16.9&30.0&4.7&16.0&54.2&3.1\\
$0.3<(g-r)_0<0.48$ & 2600 & 19.0& 9.9& 11.4& 17.5&21.6&5.1& 17.0&35.7&3.4&16.2&61.7&2.3\\
$0.48<(g-r)_0<0.55$&4348&19.1&8.4&9.0&17.8&21.1&3.9&17.1&36.4&2.7&16.0&62.2&2.0\\
$0.55<(g-r)_0<0.75$&2260&19.0&8.7&7.6&18.1&17.7&4.1&16.9&29.9&2.8&16.1&58.5&2.3\\
$0.75<(g-r)_0<1.00$&700&19.7&6.7&6.6&18.7&12.6&3.6&16.7&24.8&2.8&16.2&64.6&1.9\\
\enddata
\end{deluxetable}
}

\begin{deluxetable}{rrrrrrrrrrrrr}
\tabletypesize{\scriptsize}
\tablecolumns{12}
\footnotesize
\tablecaption{List of SEGUE Plate Pairs}
\tablewidth{0pt}
\tablehead{
\colhead{pid} & \colhead{bplate} & \colhead{fplate} & \colhead{bmjd} & \colhead{fmjd} & \colhead{stripe} & \colhead{$\alpha_{\rm J2000}$} & \colhead{$\delta_{\rm J2000}$}& \colhead{$l$} & \colhead{$b$} & \colhead{E(B-V)} & \colhead{Vers} & \colhead {category} 
}
\startdata
202&2803&2824&54368&54452&1220&0.64&28.14&110.00&-33.50&0.05&v4.3&Survey\\
149&2624&2630&54380&54327&1188&1.02&-4.82&94.00&-65.00&0.03&v4.2&Survey\\
200&2801&2822&54331&54389&72&1.25&24.95&109.77&-36.73&0.07&v4.3&Survey\\
017&1912&1913&53293&53321&86&6.02&-10.00&100.99&-71.69&0.05&v2.0&Survey\\
216&2848&----&54453&-----&1188&7.78&-18.29&94.00&-80.00&0.02&v4.4&Survey\\
075&2312&2327&53709&53710&79&9.03&7.48&116.28&-55.19&0.04&v3.3&Survey\\
031&2038&2058&53327&53349&72&10.51&24.90&120.23&-37.92&0.03&v3.0&Survey\\
013&1904&1905&53682&53706&82&11.00&0.00&118.86&-62.81&0.02&v2.0&Survey\\
009&1896&1897&53242&53242&76&11.21&14.92&120.55&-47.93&0.08&v2.0&Survey\\
076&2313&2328&53726&53728&82&17.00&0.00&131.95&-62.58&0.03&v3.3&Survey\\
203&2804&2825&54368&54439&1260&17.86&15.60&130.00&-47.00&0.07&v4.3&Survey\\
217&2849&2864&54454&54467&86&18.70&-9.72&141.60&-71.74&0.04&v4.4&Survey\\
028&2040&2060&53384&53706&1260&19.14&25.74&130.00&-36.79&0.12&v3.0&Survey\\
029&2041&2061&53387&53711&1260&20.00&31.69&130.00&-30.79&0.07&v3.0&Survey\\
077&2314&2329&53713&53725&79&21.13&7.21&137.25&-54.74&0.04&v3.3&Survey\\
023&2042&2062&53378&53381&1260&21.15&38.63&130.00&-23.79&0.05&v3.0&Survey\\
030&2043&2063&53351&53359&1260&21.33&39.62&130.00&-22.79&0.07&v3.0&Survey\\
085&2336&----&53712&-----&1260&21.33&39.62&130.00&-22.79&0.07&vt.t&Survey\\
018&1914&1915&53729&53612&86&24.27&-9.45&156.44&-69.30&0.04&v2.0&Survey\\
032&2044&2064&53327&53341&72&24.72&23.70&136.73&-37.90&0.12&v3.0&Survey\\
215&2816&----&54397&-----&86&25.28&-9.39&158.75&-68.73&0.03&v4.3&Survey\\
218&2850&2865&54461&54497&86&25.28&-9.39&158.75&-68.73&0.03&v4.4&Sgr Strm\\
014&1906&1907&53293&53315&82&26.00&0.00&150.04&-60.08&0.03&v2.0&Survey\\
010&1898&1899&53260&53262&76&26.67&13.98&142.70&-46.76&0.06&v2.0&Survey\\
024&2045&2065&53350&53678&82&30.00&0.00&157.01&-58.26&0.03&v3.0&Survey\\
219&2851&2866&54485&54478&82&30.00&0.00&157.01&-58.26&0.03&v4.4&Survey\\
033&2046&2066&53327&53349&72&32.23&22.52&145.47&-36.94&0.11&v3.0&Survey\\
069&2306&2321&53726&53711&79&33.20&6.62&156.16&-50.93&0.07&v3.3&Survey\\
026&2047&2067&53732&53738&86&37.40&-8.47&178.72&-60.22&0.03&v3.0&Survey\\
089&2379&2399&53762&53764&1300&38.17&25.50&150.00&-32.00&0.11&v3.4&Survey\\
109&2442&2444&54065&54082&1300&39.70&28.17&150.00&-29.00&0.14&v4.0&Mon Strm\\
999&1664&1664&52965&52973&82&42.00&0.00&173.65&-51.02&0.04&vt.t&RV Test\\
088&2378&2398&53759&53768&1300&43.58&34.33&150.00&-22.00&0.12&v3.4&Survey\\
070&2307&2322&53710&53727&79&45.24&5.74&171.39&-44.61&0.10&v3.3&Survey\\
108&2441&2443&54065&54082&1300&46.07&37.79&150.00&-18.00&0.12&v4.0&Survey\\
025&2048&2068&53378&53386&82&47.00&0.00&179.01&-47.44&0.07&v3.0&Survey\\
084&2335&2340&53730&53733&79&48.25&5.48&174.65&-42.75&0.20&v3.3&Survey\\
107&2397&2417&53763&53766&1300&48.79&41.20&150.00&-14.00&0.13&v3.4&Survey\\
999&1665&1666&52976&52991&-999&49.96&41.53&150.57&-13.24&0.16&vt.t&Test\\
083&2334&2339&53730&53729&79&51.24&5.20&177.71&-40.80&0.13&v3.3&Survey\\
035&2049&2069&53350&53376&82&53.00&0.00&184.53&-42.87&0.09&v3.0&Survey\\
027&2050&2070&53401&53729&86&55.43&-6.41&193.71&-44.60&0.05&v3.0&Survey\\
178&2679&2697&54368&54389&1356&57.20&10.31&178.00&-33.00&0.21&v4.2&Survey\\
034&2051&2071&53738&53741&86&59.42&-5.86&195.91&-40.94&0.11&v3.0&Survey\\
162&2680&2698&54141&54140&1356&63.35&15.61&178.00&-25.00&0.44&v4.2&Survey\\
204&2805&2826&54380&54389&1374&64.85&6.56&187.00&-29.50&0.23&v4.3&Survey\\
154&2669&2673&54086&54096&1374&71.00&10.96&187.00&-22.00&0.38&v4.2&Survey\\
256&----&2942&-----&54521&1406&71.40&-5.68&203.00&-30.48&0.07&v4.6&Survey\\
163&2681&2699&54397&54414&1356&71.50&21.98&178.00&-15.00&0.50&v4.2&LLSurvey\\
153&2668&2672&54084&54085&1374&79.49&16.61&187.00&-12.00&0.46&v4.2&LLSurvey\\
065&2300&2302&53682&53709&1300&82.64&62.07&150.00&15.00&0.25&v3.2&Survey\\
040&2052&2072&53401&53430&82&83.20&0.00&203.68&-17.42&0.26&v3.0&Survey\\
231&2887&2912&54521&54499&1374&91.30&23.40&187.00&1.00&0.77&v4.6&NGC2158/M35 Cls\\
133&2540&2548&54110&54152&1260&91.83&83.51&130.00&25.71&0.06&v4.0&Survey\\
064&2299&2301&53711&53712&1300&92.63&64.25&150.00&20.00&0.11&v3.2&Survey\\
179&2678&2696&54173&54167&1374&98.14&26.67&187.00&8.00&0.23&v4.2&LLSurvey\\
166&2682&2700&54401&54417&1356&101.28&37.61&178.00&15.00&0.14&v4.2&Survey\\
157&2676&2694&54179&54199&1374&102.21&28.39&187.00&12.00&0.09&v4.2&Survey\\
180&2712&2727&54409&54414&1406&105.56&12.44&203.00&8.00&0.09&v4.2&LLSurvey\\
086&2337&----&53740&-----&1300&106.24&65.80&150.00&25.94&0.04&vt.t&Survey\\
257&2938&2943&54526&54502&1356&107.24&39.41&178.00&20.00&0.07&v4.6&Survey\\
158&2677&2695&54180&54409&1374&110.74&31.44&187.00&20.00&0.08&v4.2&Survey\\
260&2941&2946&54507&54506&38&111.29&37.62&180.89&22.44&0.05&v4.6&Mon Strm\\
036&2053&2073&53446&53728&37&112.51&35.99&182.90&22.87&0.06&v3.0&Survey\\
181&2713&2728&54397&54416&1406&113.02&15.86&203.00&16.00&0.05&v4.2&Survey\\
167&2683&2701&54153&54154&1356&113.49&40.89&178.00&25.00&0.05&v4.2&Survey\\
043&2078&2079&53378&53379&30&114.60&21.57&198.11&19.64&0.04&v3.2&NGC2420 Cluster\\
038&2054&2074&53431&53437&27&116.00&18.18&201.97&19.54&0.03&v3.0&Survey\\
234&2890&2915&54495&54497&27&116.00&18.30&201.85&19.59&0.03&v4.6&Mon Strm\\
258&2939&2944&54515&54523&1300&116.19&66.11&150.00&30.00&0.04&v4.6&Survey\\
037&2055&2075&53729&53737&32&116.88&28.02&192.41&23.86&0.04&v3.0&Survey\\
235&2891&2916&54507&54507&29&118.00&23.20&197.73&23.16&0.07&v4.6&Mon Strm\\
182&2714&2729&54208&54419&1406&118.29&18.06&203.00&21.50&0.04&v4.2&Survey\\
259&2940&2945&54508&54505&19&119.00&9.50&211.61&18.62&0.02&v4.6&Mon Strm\\
041&2056&2076&53463&53442&16&121.20&6.75&215.24&19.38&0.03&v3.0&Survey\\
205&2806&2827&54425&54422&1458&122.77&-7.39&229.00&14.00&0.10&v4.3&Survey\\
039&2057&2077&53816&53846&10&124.00&0.00&222.93&18.72&0.05&v3.0&Survey\\
155&2670&2674&54115&54097&33&124.50&38.00&183.37&32.64&0.04&v4.2&Survey\\
134&2541&2549&54481&54523&1260&127.07&83.27&130.00&29.71&0.03&v4.0&Survey\\
078&2315&2330&53741&53738&26&127.73&24.40&199.78&31.96&0.03&v3.3&Survey\\
206&2807&2828&54433&54438&1458&127.96&-4.33&229.00&20.00&0.04&v4.3&Survey\\
079&2316&2331&53757&53742&37&129.62&53.91&164.26&37.20&0.03&v3.3&Survey\\
080&2317&2332&54152&54149&14&132.57&6.14&221.47&29.17&0.06&v3.3&Survey\\
152&2667&2671&54142&54141&17&132.82&10.94&216.61&31.53&0.04&v4.2&M67 Cluster\\
232&2888&2913&54529&54526&12&134.00&3.20&225.20&29.01&0.04&v4.6&Mon Strm\\
090&2380&2400&53759&53765&30&134.44&37.13&185.88&40.31&0.04&v3.4&Survey\\
091&2381&2401&53762&53768&26&139.44&30.43&195.57&43.49&0.02&v3.4&Survey\\
067&2304&2319&53762&53763&22&139.89&22.17&206.64&41.95&0.04&v3.3&Survey\\
092&2382&2402&54169&54176&14&141.56&7.30&225.30&37.58&0.05&v3.4&Survey\\
233&2889&2914&54530&54533&25&144.00&30.05&197.01&47.32&0.02&v4.6&Survey\\
094&2384&2404&53763&53764&34&144.67&52.86&163.48&46.20&0.01&v3.4&Survey\\
093&2383&2403&53800&53795&37&146.38&62.07&150.92&43.62&0.03&v3.4&Survey\\
220&2852&2867&54468&54479&10&150.00&0.00&239.10&40.72&0.04&v4.4&Survey\\
096&2386&2406&54064&54084&22&152.38&25.93&205.39&53.92&0.03&v3.4&Orph Strm\\
097&2387&2407&53770&53771&26&152.52&35.29&189.36&54.80&0.01&v3.4&Survey\\
221&2853&2868&54440&54451&18&156.53&17.74&220.87&55.27&0.03&v4.4&Orph Strm\\
222&2854&2869&54480&54454&14&156.63&8.82&234.18&51.20&0.03&v4.4&Survey\\
142&2557&2567&54178&54179&29&158.57&44.34&171.74&57.63&0.02&v4.2&GD Strm\\
099&2389&2409&54213&54210&10&162.00&0.00&250.28&49.82&0.04&v3.4&Orph Strm\\
144&2559&2569&54208&54234&10&162.00&0.00&250.28&49.82&0.04&v4.2&Orph Strm\\
100&2390&2410&54094&54087&30&163.80&48.01&162.38&59.24&0.02&v3.4&GD Strm\\
223&2855&2870&54466&54534&22&165.56&28.56&203.12&65.87&0.03&v4.4&Survey\\
224&2856&2871&54463&54536&26&166.97&38.59&178.45&65.54&0.02&v4.4&Survey\\
174&2690&2708&54211&54561&1540&167.16&-16.21&270.00&40.00&0.07&v4.2&Survey\\
103&2393&2413&54156&54169&14&168.77&9.61&245.98&61.30&0.02&v3.4&Survey\\
225&2857&2872&54453&54533&18&169.09&19.29&227.63&66.83&0.02&v4.4&Survey\\
104&2394*&2414&54551&54526&34&169.30&59.05&143.49&54.16&0.01&v3.4&Survey\\
227&2859&2874&54570&54561&1540&169.73&-11.87&270.00&45.00&0.07&v4.4&Survey\\
226&2858&2873&54498&54505&37&172.12&66.98&134.92&48.17&0.01&v4.4&Survey\\
229&2861&2876&54583&54581&1540&172.22&-7.52&270.00&50.00&0.04&v4.4&Survey\\
230&2862&2877&54471&54523&10&174.00&0.00&266.09&57.37&0.02&v4.4&Survey\\
261&2963&2965&54589&54594&14&180.94&9.98&267.43&69.50&0.02&v4.6&Survey\\
236&2892&2917&54552&54556&10&181.00&0.00&278.20&60.57&0.02&v4.6&Survey\\
237&2893&2918&54552&54554&18&181.81&19.97&245.85&77.61&0.03&v4.6&Survey\\
238&2894&2919&54539&54537&30&181.89&49.96&140.22&65.67&0.03&v4.6&Survey\\
117&2452&2467&54178&54176&26&182.39&39.97&154.34&74.50&0.03&v4.0&Survey\\
143&2558&2568&54140&54153&10&186.00&0.00&288.15&62.08&0.03&v4.2&Vir Strm\\
239&2895&2920&54567&54562&10&189.00&0.00&294.52&62.62&0.02&v4.6&Survey\\
241&2897&2922&54585&54612&9&191.00&-2.50&299.18&60.32&0.03&v4.6&Survey\\
173&2689&2707&54149&54144&1600&191.16&-7.83&300.00&55.00&0.03&v4.2&Survey\\
122&2457&2472&54180&54175&22&191.46&29.84&147.00&87.02&0.02&v4.0&Survey\\
242&2898&2923&54567&54563&30&192.75&49.74&123.11&67.39&0.01&v4.6&Survey\\
111&2446&2461&54571&54570&34&192.96&59.76&122.84&57.37&0.01&v4.0&Survey\\
243&2899&2924&54568&54582&18&194.57&19.74&315.26&82.45&0.02&v4.6&Survey\\
244&2900&2925&54569&54584&26&197.96&39.27&104.92&77.13&0.02&v4.6&Survey\\
245&2901&2926&54652&54625&10&198.00&0.00&314.09&62.43&0.04&v4.6&Survey\\
126&2476&----&53826&-----&17&199.03&17.01&333.60&78.38&0.03&v4.0&NGC5053 Cluster\\
110&2445*&2460&54573&54616&37&202.79&66.49&116.77&50.16&0.02&v4.0&Survey\\
247&2903&2928&54581&54614&14&205.28&9.39&338.75&68.73&0.03&v4.6&Survey\\
125&2475&----&53845&-----&22&205.55&28.40&42.31&78.70&0.01&v4.0&M3 Cluster\\
248&2904*&2929&54574&54616&22&206.68&28.21&41.20&77.72&0.01&v4.6&Survey\\
249&2905*&2930&54580&54589&18&207.22&18.62&3.16&74.29&0.02&v4.6&Survey\\
184&2716&----&54628&-----&1660&210.10&-9.21&330.00&50.00&0.04&v4.2&Survey\\
250&2906&2931&54577&54590&26&212.81&36.58&67.14&70.65&0.01&v4.6&Survey\\
112&2447&2462&54498&54561&34&214.83&56.35&100.68&56.81&0.01&v4.0&Survey\\
251&2907*&2932&54580&54595&30&217.15&45.26&82.47&63.49&0.01&v4.6&Survey\\
252&2908&2933&54611&54617&14&217.40&8.47&358.72&60.22&0.02&v4.6&Survey\\
132&2539&2547&53918&53917&35&217.73&58.24&100.60&54.36&0.01&v4.0&GD Strm\\
253&2909&2934&54653&54626&10&222.00&0.00&353.65&51.02&0.04&v4.6&Survey\\
254&2910&2935&54630&54652&26&226.36&32.20&51.02&60.57&0.02&v4.6&Survey\\
192&2724&2739&54616&54618&14&229.44&7.18&9.84&50.00&0.04&v4.2&Survey\\
246&2902&2927&54629&54621&14&229.44&7.18&9.84&50.00&0.04&v4.6&Survey\\
255&2911&2936&54631&54626&30&231.38&39.43&63.98&55.84&0.02&v4.6&Survey\\
114&2449&2464&54271&54272&34&231.76&49.88&81.08&52.66&0.02&v4.0&Survey\\
262&2964*&----&54632&-----&18&231.78&14.00&20.99&51.46&0.04&v4.6&Survey\\
124&2459*&2474&54544&54564&26&238.51&26.52&42.88&49.49&0.04&v4.0&Survey\\
045&2175&2186&54612&54327&14&241.41&5.57&16.92&39.10&0.06&v3.2&Survey\\
048&2178&2189&54629&54624&13&242.33&4.07&15.88&37.53&0.07&v3.2&Survey\\
046&2176&2187&54243&54270&37&242.51&52.37&81.35&45.48&0.02&v3.2&Survey\\
047&2177&2188&54557&54595&22&243.77&16.67&31.37&41.94&0.05&v3.2&Survey\\
135&2550&2560&54206&54205&1188&247.15&62.85&94.00&40.00&0.03&v4.2&Survey\\
044&2174&2185&53521&53532&34&250.01&36.20&58.61&41.22&0.02&v3.2&M13 Cluster\\
063&2255&----&53565&-----&34&250.01&36.20&58.61&41.22&0.02&v3.2&M13 Cluster\\
195&2796&2817&54629&54627&1062&253.10&12.49&31.00&32.00&0.05&v4.3&Survey\\
050&2180&2191&54613&54621&30&253.12&23.95&43.95&36.06&0.06&v3.2&Survey\\
051&2181&2192&53524&54232&37&254.92&39.65&63.60&37.73&0.02&v3.2&Survey\\
207&2808&2829&54524&54623&1100&255.75&28.39&50.00&35.00&0.07&v4.3&Survey\\
055&2247&2256&54169&53859&39&258.78&42.03&66.95&35.10&0.02&v3.2&M92 Cluster\\
128&2535*&----&54632&-----&1020&259.41&-13.07&10.00&14.00&0.45&v4.0&LLSurvey\\
052&2182&2193&53905&53888&1100&261.18&27.01&50.00&30.00&0.04&v3.2&Survey\\
196&2797&2818&54616&54616&1062&262.52&8.11&31.00&21.75&0.11&v4.3&Survey\\
136&2551&2561&54552&54597&1188&262.57&64.37&94.00&33.00&0.03&v4.2&Survey\\
061&2253*&2262&54551&54623&39&263.11&33.16&57.36&30.08&0.04&v3.2&Survey\\
198&2799&2820&54368&54599&1140&263.44&44.15&70.00&32.00&0.02&v4.3&Survey\\
053&2183&2194&53536&53904&1100&266.47&25.43&50.00&25.00&0.09&v3.2&Survey\\
054&2184&2195&53534&54234&1100&271.61&23.67&50.00&20.00&0.12&v3.2&Survey\\
127&2534&2542&53917&53919&1100&277.60&21.33&50.00&14.00&0.16&v4.0&LLSurvey\\
201&2802&----&54326&-----&1220&278.04&78.51&110.00&27.50&0.07&v4.3&Survey\\
137&2552&----&54632&-----&1188&278.72&64.15&94.00&26.00&0.04&v4.2&Survey\\
197&2798&2819&54397&54617&1140&279.34&41.30&70.00&20.00&0.07&v4.3&Survey\\
211&2812*&2833&54633&54650&1100&283.38&18.79&50.00&8.00&0.39&v4.3&Survey\\
129&2536&2544&53883&53884&1140&286.66&39.11&70.00&14.00&0.18&v4.0&LLSurvey\\
999&1857&1858&53182&53271&-999&290.34&78.20&110.00&25.00&0.05&vt.t&Test\\
199&2800&2821&54326&54393&1140&290.57&37.68&70.00&10.62&0.15&v4.3&NGC6791 Cluster\\
138&2553&2563&54631&54653&1188&291.69&62.61&94.00&20.00&0.05&v4.2&Survey\\
212&2813&----&54650&-----&1100&298.01&11.26&50.00&-8.00&0.31&v4.3&Survey\\
081&----&2338&-----&53679&-999&298.44&19.08&57.00&-4.41&0.45&v3.3&M71 Cluster\\
066&2303&2318&54629&54628&1062&301.86&-11.46&31.00&-22.00&0.11&v3.3&Survey\\
139&2554&2564&54263&54275&1188&302.97&60.01&94.00&14.00&0.22&v4.2&LLSurvey\\
999&1660&1661&53230&53240&-999&303.62&77.18&110.0&22.00&0.19&vt.t&Test\\
056&2248&2257&53558&53612&72&306.44&13.67&56.45&-13.78&0.09&v3.2&Survey\\
057&2249&2258&53566&54328&72&309.33&14.73&58.97&-15.53&0.09&v3.2&Survey\\
015&1908&1909&53239&53261&82&311.00&0.00&46.64&-24.82&0.07&v2.0&Survey\\
049&2179&2190&53555&54386&1220&311.16&76.18&110.00&20.00&0.28&v3.2&Survey\\
019&1916&1917&53269&53557&86&311.58&-6.00&41.08&-28.23&0.05&v2.0&Survey\\
058&2250&2259&53566&53565&72&312.25&15.76&61.53&-17.25&0.08&v3.2&Survey\\
140&2555&2565&54265&54329&1188&312.39&56.59&94.00&8.00&0.93&v4.2&LLSurvey\\
214&2815&----&54414&-----&1100&312.72&2.52&50.00&-25.00&0.11&v4.3&Survey\\
020&1918&1919&53240&53240&82&317.00&0.00&50.11&-29.97&0.11&v2.0&Survey\\
006&1890&1891&53237&53238&76&319.01&10.54&61.22&-25.64&0.06&v2.0&Survey\\
068&2305&2320&54414&54653&86&320.56&-7.18&44.84&-36.65&0.23&v3.3&Survey\\
021&1960&1962&53289&53321&-999&322.67&11.33&64.41&-27.98&0.10&v2.1&M15 Cluster\\
131&2538&2546&54271&54625&1220&323.07&73.64&110.00&16.00&0.53&v4.0&LLSurvey\\
022&1961&1963&53299&54331&-999&323.36&-0.20&54.00&-35.43&0.04&v2.1&M2 Cluster\\
141&2556&2566&54000&54333&1188&330.15&45.06&94.00&-8.00&0.30&v4.2&LLSurvey\\
059&2251&2260&53557&53638&72&332.50&21.47&80.06&-27.66&0.07&v3.2&Survey\\
072&2309&----&54441&-----&86&332.60&-8.47&51.37&-47.64&0.06&v3.3&Survey\\
071&2308&2323&54379&54380&79&332.78&6.36&67.76&-38.78&0.10&v3.3&Survey\\
130&2537&2545&53917&53915&1220&334.17&69.39&110.00&10.50&0.48&v4.0&LLSurvey\\
145&2620&----&54397&-----&1188&335.67&39.37&94.00&-15.00&0.13&v4.2&Survey\\
007&1892&1893&53238&53239&76&340.25&13.68&81.04&-38.36&0.05&v2.0&Survey\\
060&2252&2261&53613&53612&72&340.97&23.07&88.35&-31.10&0.06&v3.2&Survey\\
011&1900&1901&53262&53261&82&341.00&0.00&69.20&-49.10&0.06&v2.0&Survey\\
146&2621&2627&54380&54379&1188&342.14&30.88&94.00&-25.00&0.05&v4.2&Survey\\
016&1910&1911&53321&53295&86&344.72&-9.39&61.32&-58.07&0.04&v2.0&Survey\\
073&2310&2325&53710&54082&79&344.84&7.05&80.43&-46.37&0.07&v3.3&Survey\\
148&2623&2629&54328&54087&1188&347.53&22.12&94.00&-35.00&0.17&v4.2&Survey\\
999&1662&1663&52970&52973&-999&351.41&52.73&110.00&-7.98&0.31&vt.t&Test\\
005&1888&1889&53239&53240&1220&353.41&48.91&110.00&-12.00&0.21&v2.0&Survey\\
147&2622&2628&54095&54326&1188&354.52&8.71&94.00&-50.00&0.13&v4.2&Survey\\
004&1886&1887&53237&53239&1220&354.71&46.04&110.00&-15.00&0.13&v2.0&Survey\\
008&1894&1895&53240&53242&76&355.69&14.81&99.18&-44.87&0.03&v2.0&Survey\\
003&1884&1885&53228&53230&1220&355.89&43.16&110.00&-18.00&0.12&v2.0&Survey\\
012&1902&1903&53271&53357&82&356.00&0.00&89.32&-58.40&0.03&v2.0&Survey\\
074&2311&----&54331&-----&86&356.88&-9.90&78.72&-67.11&0.03&v3.3&Survey\\
002&1882&1883&53262&53271&1220&357.30&39.30&110.00&-22.00&0.13&v2.0&Mon Strm\\
001&1880&1881&53262&53261&1220&358.26&36.40&110.00&-25.00&0.11&v2.0&Survey\\
087&2377&----&53991&-----&100&359.35&56.71&115.53&-5.39&0.41&v3.4&NGC7789 Cluster\\
\enddata
\tablenotetext{(*)}{Indicates plate obtained in bright moon, velocities off by up to 10 $\rm km~s^{-1}$}
\end{deluxetable}

\begin{deluxetable}{rr}
\tabletypesize{\scriptsize}
\tablecolumns{2}
\footnotesize
\tablecaption{SSPP Flags}
\tablewidth{0pt}
\tablehead{
\colhead{flag} & \colhead{meaning} 
}
\startdata
n &  All is normal for this flag position  \\
d &  Possible WD \\
D & Apparent WD \\
E & Emission, possible QSO \\
h & Helium line detected\\
H & Apparent Teff too Hot for parameter est. \\
l & Sodium line, possibly late type \\
N & Noise spectrum \\
S & Sky fiber -- no signal \\
V & RV mismatch  \\
C & Color mismatch (spectroscopic color Teff far from implied $(g-r)_0$ color Teff) \\
B & Balmer flag: unusual Balmer eq. widths for star \\
g & g band : unusual G-band eq. width for star -- Carbon star?  \\
G & Carbon star?  \\
\enddata
\end{deluxetable}

%plate ranges
%v2_0 1880-1919
%v2_1 1960-1963
%v3_0 2038-2077
%v3_1 2147-2163
%v3_2 2174-2302
%v3_3 2303-2338
%v3_4 2377-2476
%v4_0 2534-2549
%v4_2 2550-2741
%v4_3 2796-2837
%v4_4 2848-2877
%v4_6 2887-2965

\clearpage

%\voffset 1.7in
%\rotate

{\hoffset -0.8in
\begin{deluxetable}{rrrrr}
\tabletypesize{\scriptsize}
\tablecolumns{5}
\footnotesize
\tablecaption{SEGUE Selection by Target Type}
\tablewidth{0pt}
\tablehead{
\colhead{Target Type} & \colhead{Vers} & \colhead{PrimTargBits} & \colhead{Mag,Color,PM Cuts} &  \colhead{Num/PP,Cand.,f} 
}
\startdata
WD/sdO/sdB & v4.6 & 0x80080000 & $g < 20.3, -1 < g-r < -0.2, -1 < u-g < 0.7, u-g + 2(g-r) < -0.1 $  & 25,4069,0.62  \\
 & v3.0 & &  $-1 < u-g < 0.7, u-g + 2(g-r) < -0.1$  &  \\
 & v2.0 & &  $g < 20.3, -1 < g-r < 0.2, -1 < u-g < 0.5$ &  \\
CWD & v4.6 & 0x80020000 & $ r < 20.5, -2 < g-i < 1.7  \tablenotemark{b}, H_g > 16.05+2.9(g-i)$ & 10,1187,0.005  \\
 & v3.1 & &  allowed number of targets/pointing to exceed 10 on occasion &  \\
 & v3.0 & &   $ r < 20.5, -2 < g-i < 1.7, H_g > 16.05+2.9(g-i)$ &  \\
 & v2.0 & &  $15 < r  20, -0.1 < g-r < 1.1, g-r > 2.4(r-i) + 0.5, i-z < 0$ &  \\
BHB/BS/A & v4.6 & 0x80002000 &  $g < 20.5, 0.8 < u-g < 1.5, -0.5 < g-r < 0.2, v\tablenotemark{c} \rm weight$   & 150,24688,0.66 \\
 & v3.3 & &  Added v 'Luminosity color' weighting &  \\
 & v3.2 & &  Sort by color, favor bluest BHBs &  \\
 & v3.0 & &  $ 0.6 < u-g < 1.4, -0.5 < g-r < 0.2$ &  \\
 & v2.0 & &  $g < 20.5, 0.5 < u-g < 1.4, -0.8 < g-r < 0.2, s < -0.065$ &  \\
F & v4.6 & 0x80100000 &  $g < 20.3, -0.7 < P1(s)\tablenotemark{d} < -0.25, 0.4 < u-g < 1.4, 0.2 < g-r < 0.7$   & 200,37900,0.68  \\
 & v3.0 & &  $g < 20.3, -0.7 < P1(s) < -0.25, 0.4 < u-g < 1.7, 0.2 < g-r < 3 $ & \\
 & v2.0 & &  $g < 20.3, -0.7 < P1(s) < -0.3, 0.4 < u-g < 1.7, -0.3 < g-r < 3 $ & \\
Low Metal & v4.6 & 0x80010000 &  $r < 19, -0.5 < g-r < 0.75, 0.6 < u-g < 3.0, l\tablenotemark{e} > 0.135$  & 150,29788,0.12  \\
 & v3.4 & &  changed l-color cut to $l > 0.135$ &  \\
 & v3.3 & &  changed l-color cut to $l > 0.15$ &  \\
 & v3.1 & &  weighted by l-color and magnitude (bright targets favored) &  \\
 & v3.0 & &  $r<19.5, -0.5 < g-r < 0.75, l > 0.12$ &  \\
 & v2.0 & &  $r<20.2, -0.5 < g-r < 0.9, 0.3 < u-g < 3, l > 0.15$ &  \\
F/G & v4.6 & 0x80000200 &  $g < 20.2, 0.2 < g-r < 0.48 $ & 50,6939,0.9  \\
 & v3.3 & &  First appearance of F/G category &  \\
G & v4.6 & 0x80040000 &  $r < 20.2, 0.48 < g-r < 0.55$   & 375,62784,0.96 \\
  & v3.0 &            &   $0.48 < g-r < 0.55$    &  \\
  & v2.0 &            &  $ r < 20.2$ , $0.50 < g-r < 0.55$    &  \\
K giant & v4.6 & 0x80004000 &  $ r < 18.5, 0.5 < g-r < 0.8 $ & 70,16866,0.3  \\
         &      &            &                $l\tablenotemark{e} > 0.07 $  $\rm pm < 11 mas~yr^{-1}$  &   \\
  & v4.2 &            &  $g < 19.5, 0.5 < g-r < 0.9$   &  \\
  & v3.0 &            &  $  0.35 < g-r < 0.8, l > 0.07$, pm $<$ 11 mas/yr    &  \\
  & v2.0 &            &  $ r < 20.2 , 0.7 < u-g < 4, 0.4 < g-r < 1.2, 0.15 < r-i < 0.6, l > 0.1$    &  \\
  & vt.t &            &  $ r < 20.2 , 0.5 < g-r < 1.3$    &  \\
Red K giant & v4.6 & 0x80004000 &  $ r < 18.5, 0.8 < g-r < 1.2, \rm pm < 5 mas~yr^{-1}$   & 30,5948,0.08  \\
  & v4.4 &            &  $g < 18.5$,  weight brighter, redder  objects higher   &  \\
  & v4.3 &            &  $g < 19.5, 0.9 < g-r < 1.2$   &  \\
AGB & v4.6 & 0x80800000 &  $ r<19.0, 2.5 < u-g < 3.5, 0.9 < g-r <1.3, \rm s\tablenotemark{f} < -0.06 $  & 10,1343,0.08  \\
dK , dM& v4.6 & 0x80008000 &  $  r < 19, 0.55 < g-r < 0.75$    &50,18358,0.80   \\
  & v3.0 &            &  $ r < 19.0 ,  0.55 < g-r <  0.75, 0.3 < r-i < 0.8 $    &  \\
  & v2.0 &            &  $ r < 19.5 ,  g-r > 0.7, 0.3 < r-i < 0.8 $    &  \\
sdM& v4.6 & 0x80400000 &  $  r < 20$ , $g-r > 1.6, 0.9 < r-i < 1.3 $   & 25,1012,0.003  \\
esdM\tablenotemark{a,g}& v4.6 & 0x81000000 &  $ (g-r)0.787-0.356>(r-i) , r-i <0.9, H_r >17, 2.4 >g-i>1.8 $  & 40,9420,0.20  \\
WD/MS\tablenotemark{a}& v4.6 & 0x80001000 &  $ g < 20, (u-g) < 2.25, -0.2 < g-r < 1.2, 0.5 < r-i < 2.0$   &5,431,0.56 \\
&  &  & $g-r > -19.78(r-i)+11.13, g-r < 0.95(r-i)+0.5$      &   \\
& &      &   $i-z > 0.5 \rm ~for~r-i > 1$        &  \\
& &      &   $i-z > 0.68(r-i)-0.18~ \rm for~ r-i <= 1$      &   \\
&v3.4 &      &   algorithm uses non-dereddened colors   &   \\
&v3.3 &      &   First appearance MSWD category   &   \\
L\tablenotemark{h}& v4.6 & 0x80200000 &  $ z < 19.5$ ,  $i-z > 1.7$     & 5,1277,0.07  \\
 &      &            &     $u > 21$    &  \\
  & v3.0 &            &  hi-z QSO style flag checks added   &  \\
LL Blue\tablenotemark{i}& v4.6 & 0x80000800 &$g-r<0.25$       &800,8522,0.3  \\
  & v4.0 &            &  First appearance all low-latitude algs   &  \\
LL AGB& v4.6 &  0x88000000 & $s\tablenotemark{f}>-0.13,3.5>u-g>2.6,0.8 <g-r <1.3$      &50,499,0.3 \\
LL KIII& v4.6 & 0x80000400 &   $0.55<g-r<0.9,g<19,\rm pm<11 mas~yr^{-1}$    &300,3220,0.3 \\
\enddata
\tablenotetext{(a)}{For this category, colors and magnitudes are NOT dereddened. }
\tablenotetext{(b)}{If neighbor with $g < 22$ within $7''$, else $g-i < 0.12$.}
\tablenotetext{(c)}{$v = -0.283(u-g) -0.354(g-r) +0.455(r-i) + 0.766(i-z)$, $-0.3 < g-r < 0.1$, gravity sensitive color for BHB/BS separation.}
\tablenotetext{(d)}{$P1(s) = 0.91(u-g) + 0.415(g-r) -1.28$}
\tablenotetext{(e)}{$l = -0.436u+1.129g-0.119r-0.574i+0.1984$, $0.5 < g-r < 0.8$ }
\tablenotetext{(f)}{$s= -0.249u+0.794g-0.555r+0.234$ }
\tablenotetext{(g)}{The quoted color cuts are approximate.}
\tablenotetext{(h)}{Additional flag checks are performed.}
\tablenotetext{(i)}{This cut varies with reddening.}
\end{deluxetable}
}

\begin{deluxetable}{rr}
\tabletypesize{\scriptsize}
\tablecolumns{2}
\footnotesize
\tablecaption{Versions of SEGUE Target Selection}
\tablewidth{0pt}
\tablehead{
\colhead{version} & \colhead{plate range} 
}
\startdata
vt.t&1660-1858\\
v2.0&1880-1919\\
v2.1&1960-1963\\
v3.0&2038-2077\\
v3.1&2147-2163\\
v3.2&2174-2302\\
v3.3&2303-2338\\
v3.4&2377-2476\\
v4.0&2534-2549\\
v4.2&2550-2741\\
v4.3&2796-2837\\
v4.4&2848-2877\\
v4.6&2887-2965\\
\enddata
\end{deluxetable}

{\hoffset -1.4in

\begin{deluxetable}{rrrr}
\tabletypesize{\scriptsize}
\tablecolumns{4}
\footnotesize
\tablecaption{Legacy Stellar Target Selection by Type}
\tablewidth{0pt}
\tablehead{
\colhead{Target Type} & \colhead{PrimTargBits} & \colhead{Mag,Color,PM Cuts} &  \colhead{Cand.\tablenotemark{a}} 
}
\startdata
SPECTROPHOTO STD &0x20  & $0.1<g-r<0.4,16<g<17.1$      &20320 \\
REDDEN STD& 0x2 &    $0.1<g-r<0.4,17.1<g<18.5$   &22337 \\
HOT STD &0x200  &$g<19,u-g<0,g-r<0$       &3370 \\
ROSAT C  & 0x800 & $r<16.5$ ROSAT X-ray source within $60''$& 8000\\
ROSAT C  & 0x800 &$u-g <1.1$ ROSAT within $60''$\\
ROSAT E  & 0x8000000 & ROSAT within $60''$ & 8000\\
STAR BHB & 0x2000 & $-0.4 < g-r < 0, 0.8 < u-g < 1.8$ & 19887\\
STAR CARBON  & 0x4000 & $g-r > 0.85, r-i > 0.05, i-z > 0, r-i < -0.4+0.65(g-r)  $ & 4453\\
&   & $ (g-r) > 1.75 $ & \\
STAR BROWN DWARF  & 0x8000 & $z<19, \sigma(i) < 0.3, r-i > 1.8, i-z > 1.8$  &  667\\
STAR SUB DWARF  & 0x10000 & $g-r > 1.6, 0.95 < r-i < 1.6, \sigma(g) < 0.1$ & 1482\\
STAR CATY VAR\tablenotemark{b}  & 0x20000 & $g<19.5,u-g < 0.45,g-r < 0.7,r-i>0.3,i-z>0.4 $  & 8959  \\
&   0x20000 & $(u-g)-1.314(g-r) < 0.61 , r-i > 0.91, i-z > 0.49$\\
STAR RED DWARF  & 0x40000 & $i < 19.5, \sigma(i) < 0.2, r-i > 1.0, r-i > 1.8$ & 14649\\
STAR WHITE DWARF  & 0x80000 & $g-r < -0.15, u-g+2(g-r) < 0,r-i < 0$ & 6059\\
&   & $H_g\tablenotemark{c} > 19 , H_g > 16.136 + 2.727(g-i)$ \\
&   & $ g-i > 0,  H_g > 16.136 + 2.727(g-i)$ \\
STAR PN & 0x10000000 & $g-r > 0.4, r-i < -0.6, i-z > -0.2, 16 < r_0 < 20.5$ & 20 \\
SERENDIP BLUE  & 0x100000 &$u-g <0.8, g-r < -0.05 $& 81937 \\
SERENDIP FIRST  & 0x200000 & FIRST radio source within $1.5''$ & 14689 \\
SERENDIP RED\tablenotemark{b} & 0x400000 & $r-i > 2.7, i-z >1.6$ & 4179\\
SERENDIP DISTANT\tablenotemark{b} & 0x800000 & $g-r>2, r-i <0$ & 11820 \\
\enddata
\tablenotetext{(a)}{An object may be a Legacy target in multiple categories}
\tablenotetext{(b)}{For this category, colors and magnitudes are NOT dereddened }
\tablenotetext{(c)}{$\rm H_g = g + 5log(\mu) + 5$, $\mu$ in $\rm arcsec~yr^{-1}$ }
\end{deluxetable}
}

\clearpage

\setcounter{page}{1}

%footprintadlb
\begin{figure}
\centering
\includegraphics[scale=0.76]{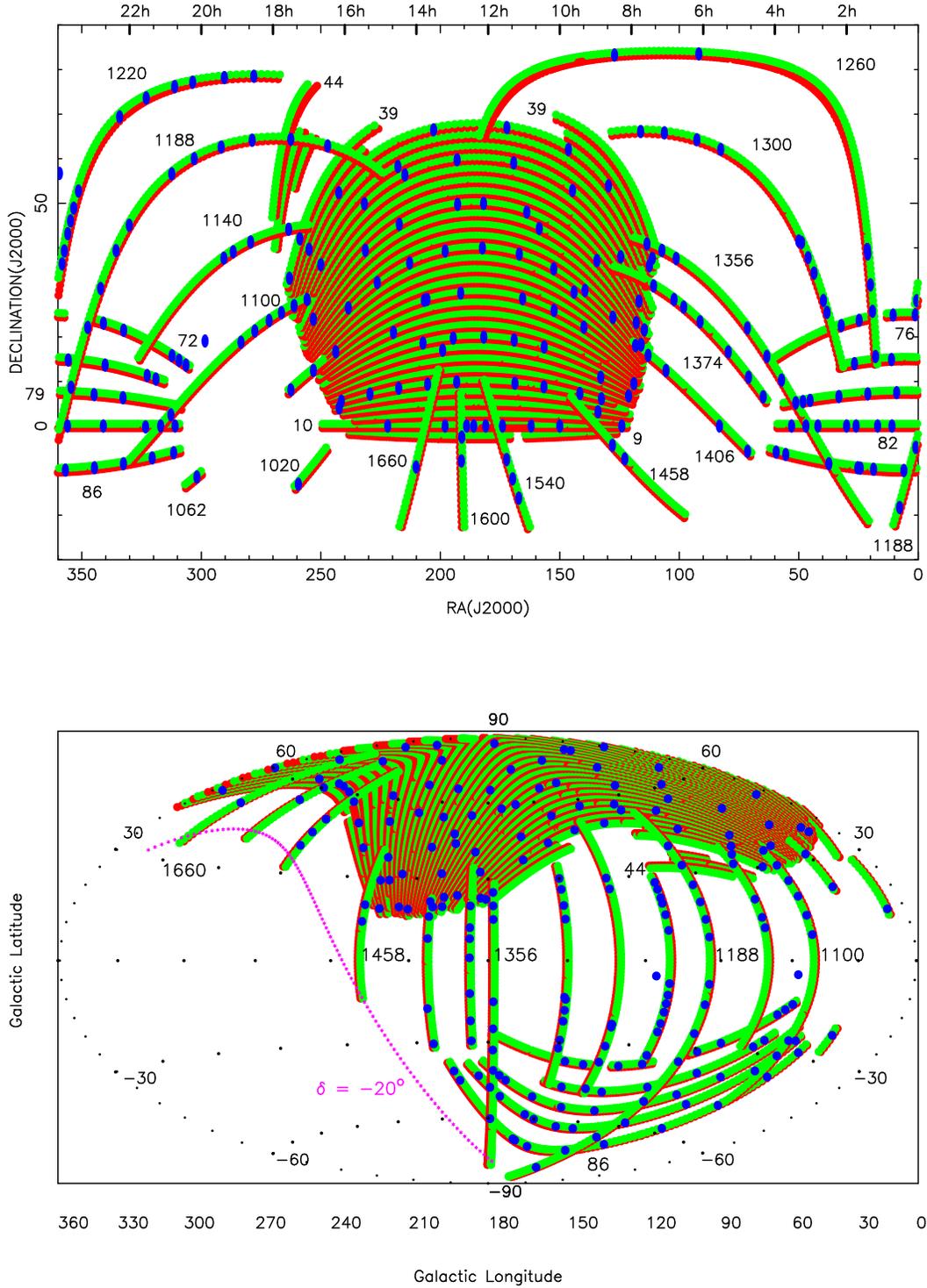}
\caption [SEGUE survey footprint] {
\footnotesize
Upper panel: the SEGUE Survey footprint in the Equatorial coordinates
from $360^\circ$ to $0^\circ$ (left to right) and $-26^\circ$ to $90^\circ$
(bottom to top).  Selected stripes are labeled with their stripe number.
The SDSS North Galactic Cap stripes are numbered from 9 to 44.  Southern
SDSS stripes are numbered 76, 82 and 86. SEGUE fills in Southern stripes 72
and 79.   SEGUE's constant Galactic longitude stripes are numbered
with $stripe = 1000 + 2l$ where $l$ is the Galactic longitude.
Each SEGUE plate pointing (usually representing a pair of 640 hole plates),
is indicated with a blue circle.
Lower panel: the SEGUE Survey footprint in $(l,b)$ in Aitoff projection, centered
on the Galactic anticenter.   The line marking the Southern limit
of the telescope observing site $\delta = -20^\circ$ is indicated in 
magenta.  Red and green filled areas represent South and North SDSS and
SEGUE strips respectively.
\label{footprintadlb}}
\end{figure}

%RV QA
\begin{figure}
\centering
\includegraphics[angle=-90,scale=0.65]{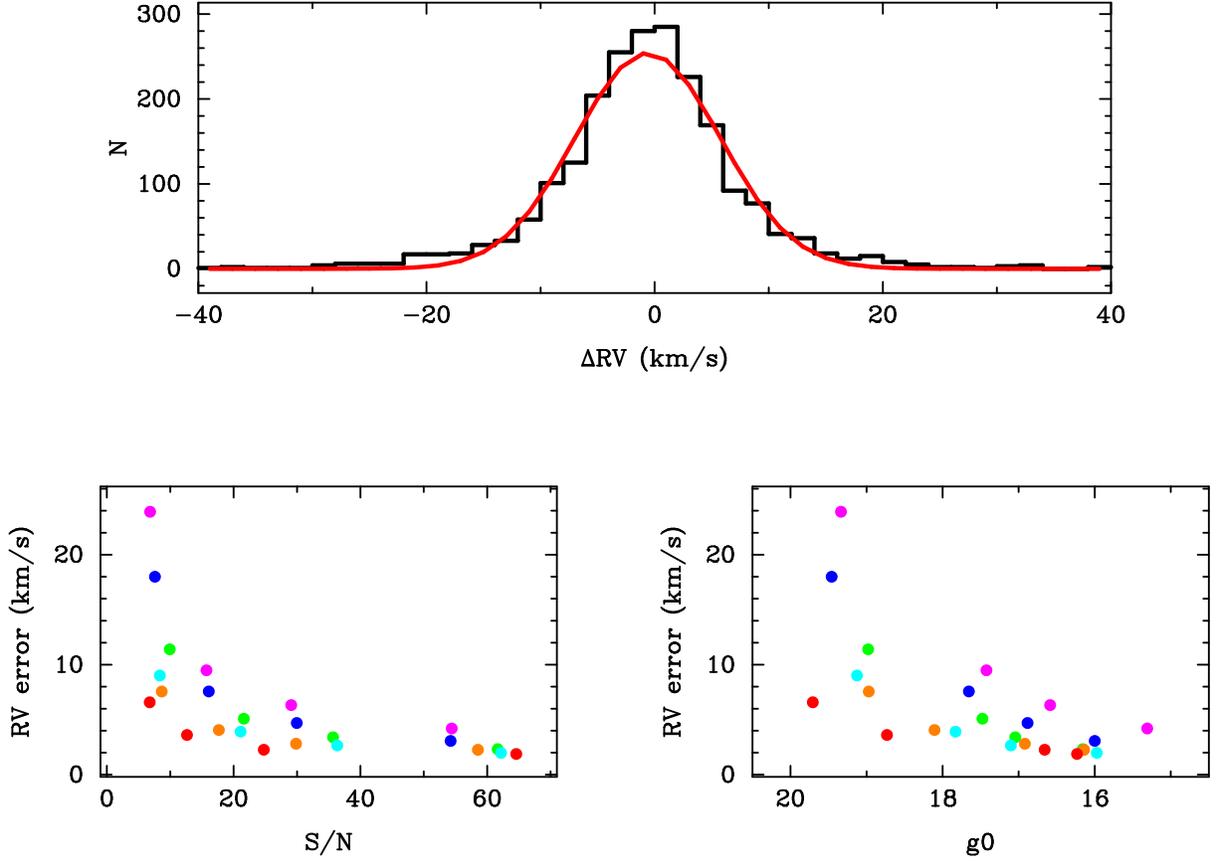}
\caption[SEGUE RV accuracy] {
\footnotesize
top panel shows the histogram of differences in spectroscopic pipeline 
measurements of RV for two separate observations of the 
same star on different survey SEGUE plates for 2200 stars with $r \sim 18$ and S/N 
$> 10$.  The quoted sigma is divided by $\sqrt{2}$ to estimate the
measurement error for a single observation.  The red curve is the best fit
Gaussian, it has a mean offset of $< 0.3 \rm ~km~s^{-1}$ and a 
$\sigma/\sqrt{2} = 4.4 ~\rm km~s^{-1}$.
Lower panel: RV accuracy by S/N and $(g-r)_0$ color.  The color code is
magenta: $(g-r)_0 < 0.1$ (BHB/BS/A), blue: $0.1 < (g-r)_0 < 0.3$ (F), 
green: $0.3 < (g-r)_0 < 0.48 $ (F/G), cyan: $0.48 < (g-r)_0 < 0.55$ (G, orange: $(0.55 < (g-r)_0 < 0.75$ (G/K) and red: $0.75 < (g-r)_0 < 1.0 $ (K).
\label{rv}}
\end{figure}

%figts
\begin{figure}
\centering
\includegraphics[scale=0.7]{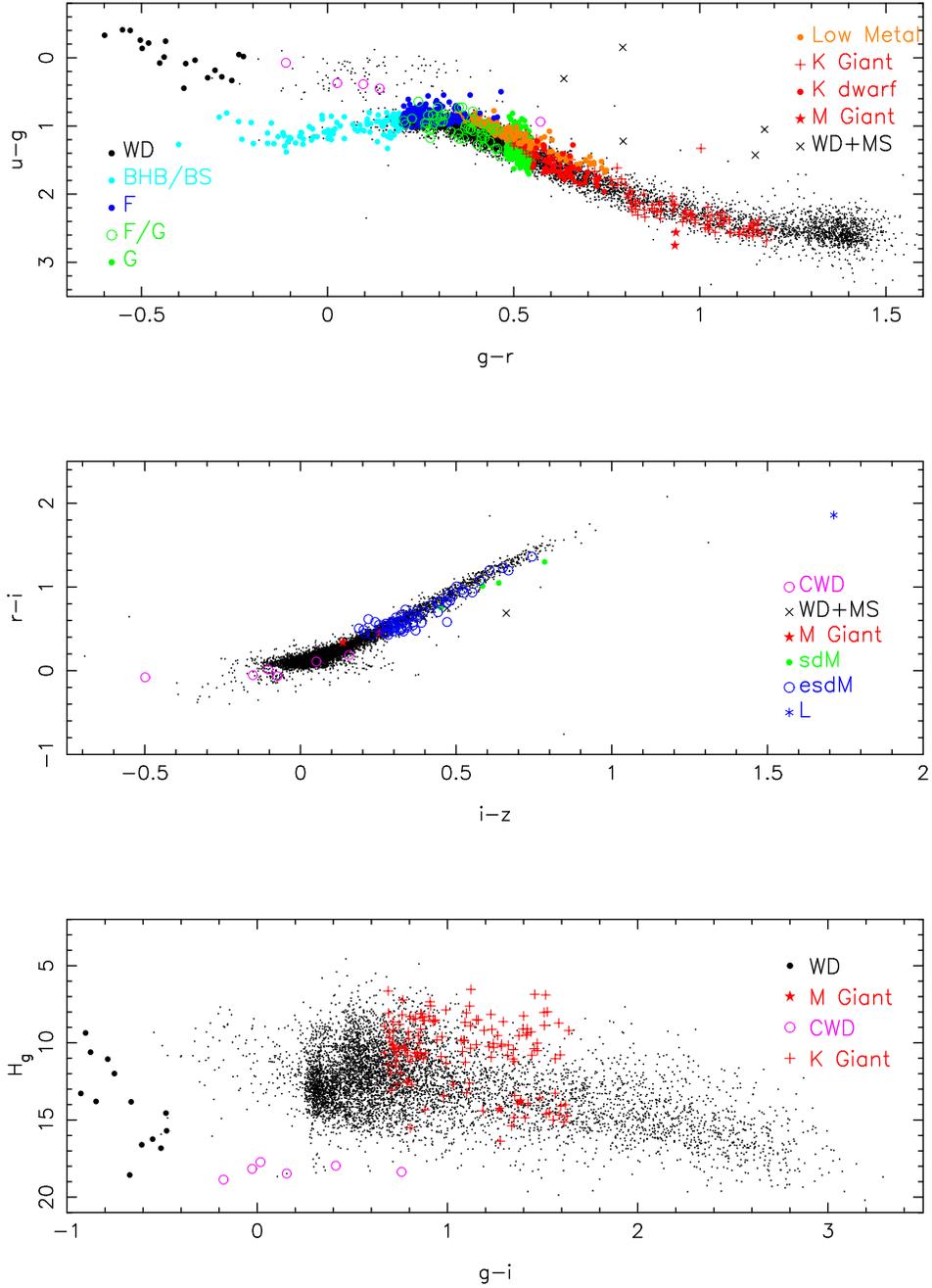}
\caption[SEGUE target selection at a glance] {
\footnotesize
target selection categories in SEGUE.
Top panel: $(g-r)_0,(u-g)_0$ color-color diagram showing different SEGUE
target categories in different colors/symbols.  Note the `Low Metal' category
hugs the blue side (in $(u-g)_0$) of the stellar locus, and a substantial
fraction of F stars with redder $(u-g)_0$ and $0.2 < (g-r)_0 < 0.48$ are
not targeted, except by the F/G category.
Middle panel: the same as above, except categories which use redder
$(i-z)_0,(r-i)_0$ colors are highlighted.  Note the L dwarf candidate at
$(i-z,r-i)_0 = (1.72,1.9)$.  The proper motion selected extreme M subdwarf
candidates are shown as open blue circles.
Lower panel:  a $(g-i)_0, H_g$ selection diagram for categories which
use USNO-B proper motion in their selection.  $H_g = g + 5\rm log_{10}(\mu/1000)+5$
where $\mu$ is the proper motion in $mas\>yr^{-1}$.  Note the cool WD
candidates (high proper motion) as magenta circles and the K giant 
candidates (consistent with 0 proper motion).
\label{figts}
}
\end{figure}

%figwdsdbsdo
\begin{figure}
\centering
\includegraphics[scale=0.7,angle=-90]{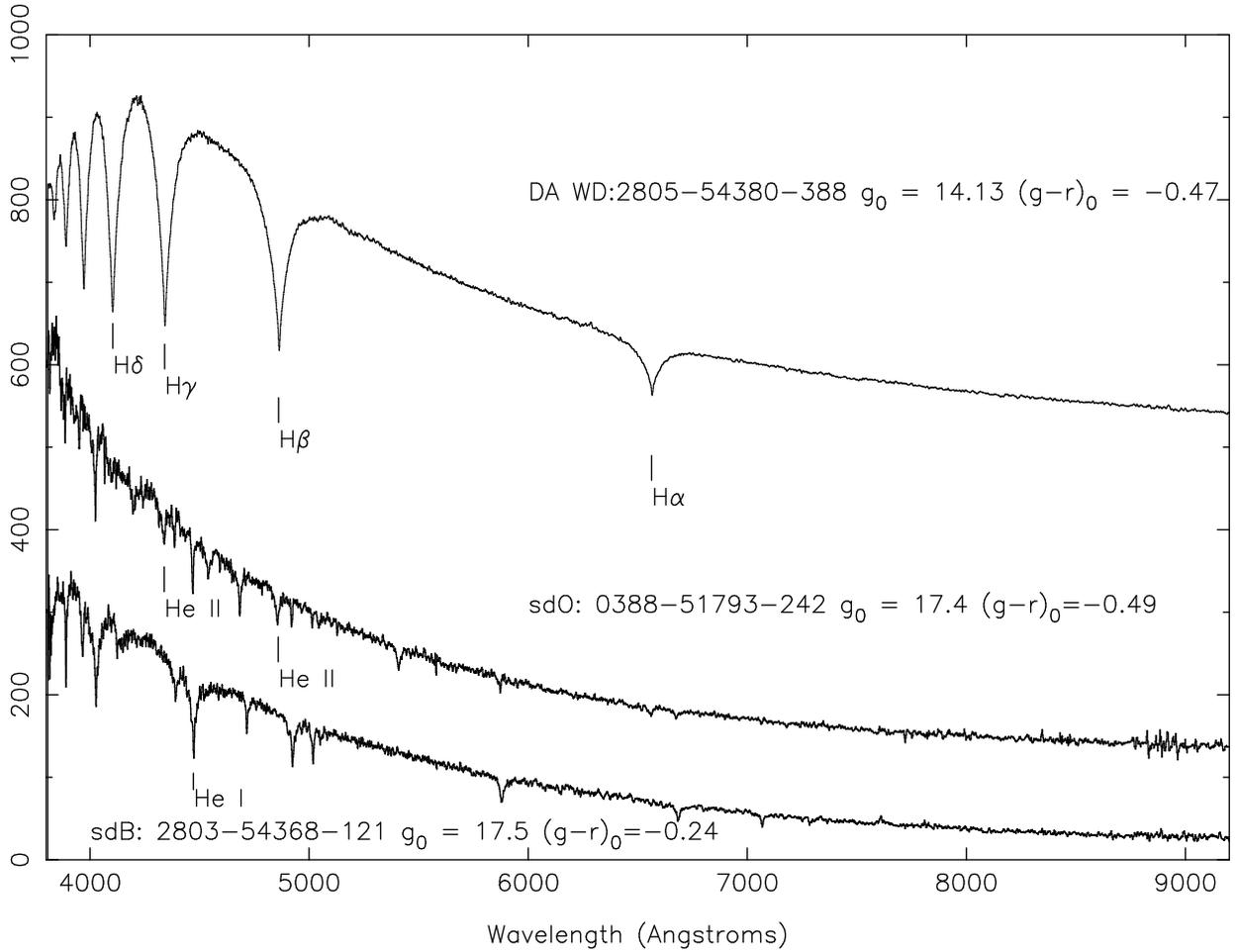}
\caption[WD] {
\footnotesize
top: a sample SEGUE DA (hydrogen) WD.
middle: a candidate sdO star showing a characteristic hot spectrum.
lower:  a candidate sdB star, which has colors similar to the WD.
\label{figwd}}
\end{figure}

%fig CWD
\begin{figure}
\centering
\includegraphics[scale=0.7,angle=-90]{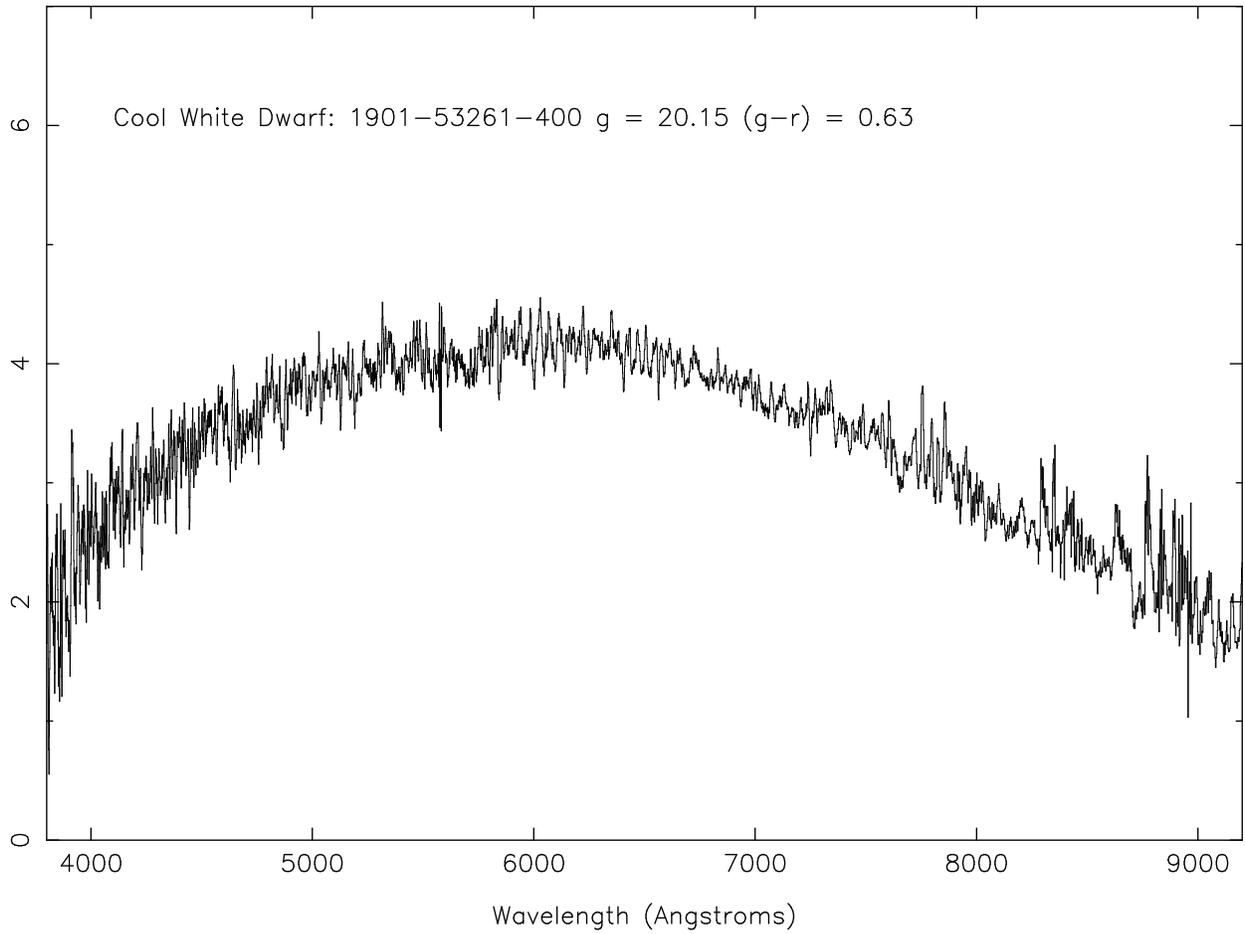}
\caption[CWD] {
\footnotesize
sample SEGUE cool WD where CIA due to $H_2$ has removed 
flux redward of 6000\AA, leaving a spectrum with very unusual colors.
\label{figcwd}}
\end{figure}

%fig BHBA
\begin{figure}
\centering
\includegraphics[scale=0.7,angle=-90]{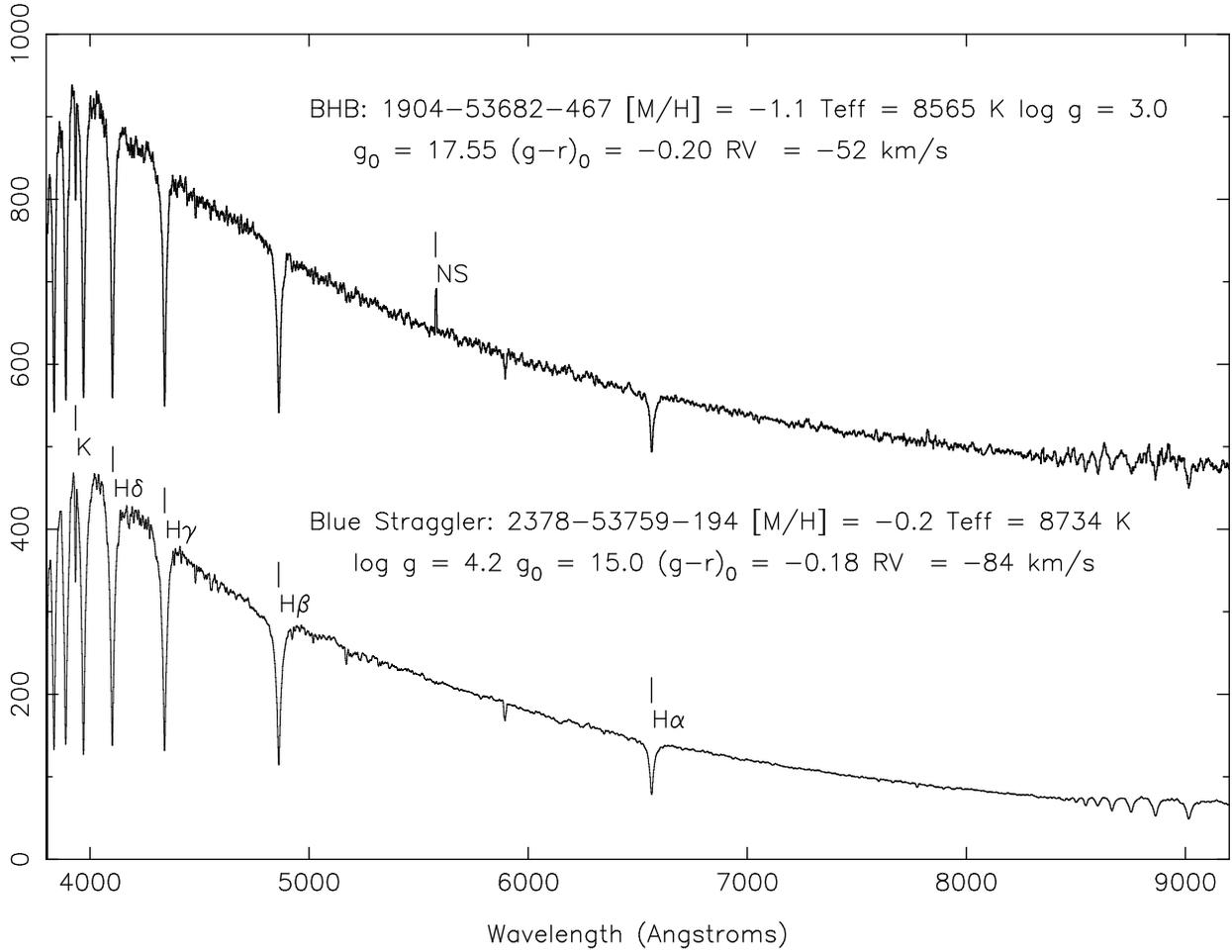}
\caption[BHB/BS] {
\footnotesize
Upper: a sample SEGUE BHB star.  Note the low surface gravity, $\rm log~g = 3.0$, deep and narrow Balmer lines characteristic of this
type.
Lower: a sample SEGUE BS star.  Note the somewhat broader Balmer
lines, the strong Paschen series lines in the IR and the higher measured
surface gravity $\rm log~g = 4.2$.
\label{figbhb}}
\end{figure}

%metalseq  F stars
\begin{figure}
\centering
\includegraphics[scale=0.8]{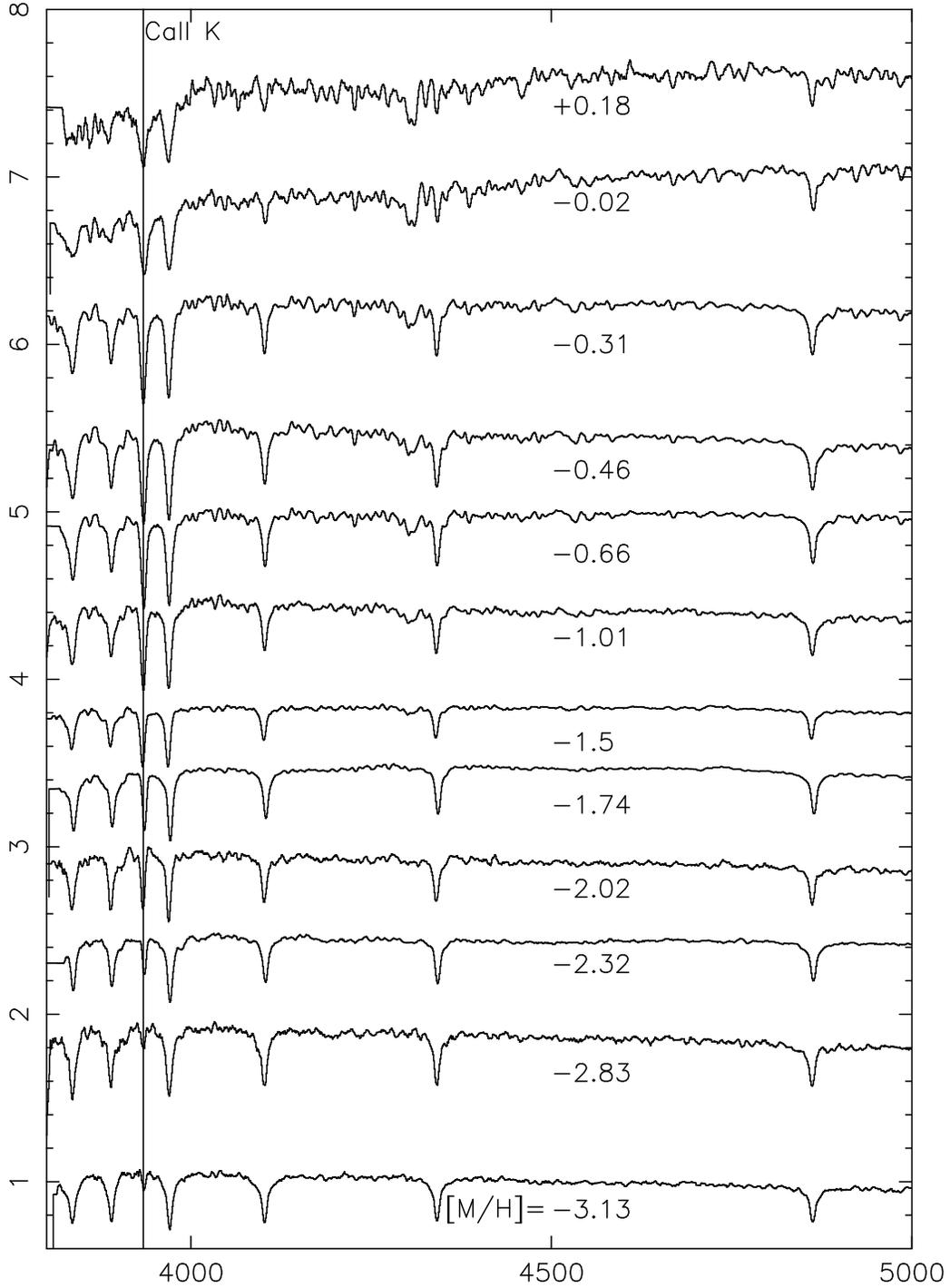}
\caption[F star metal sequence] {
\footnotesize
a set of SEGUE F stars, selected to show the range of metallicities sampled by
the F subdwarf, F/G, spectrophotometric standard 
and reddening standard categories.  All 13 stars have similar effective
temperatures, near 6500 K, but the 
strength of the Ca K line at $\lambda 3933$ indicates metallicities ranging
from less than 0.001 to 1.5 times Solar.
\label{metalseq}}
\end{figure}

%fig F/G
\begin{figure}
\centering
\includegraphics[scale=0.7,angle=-90]{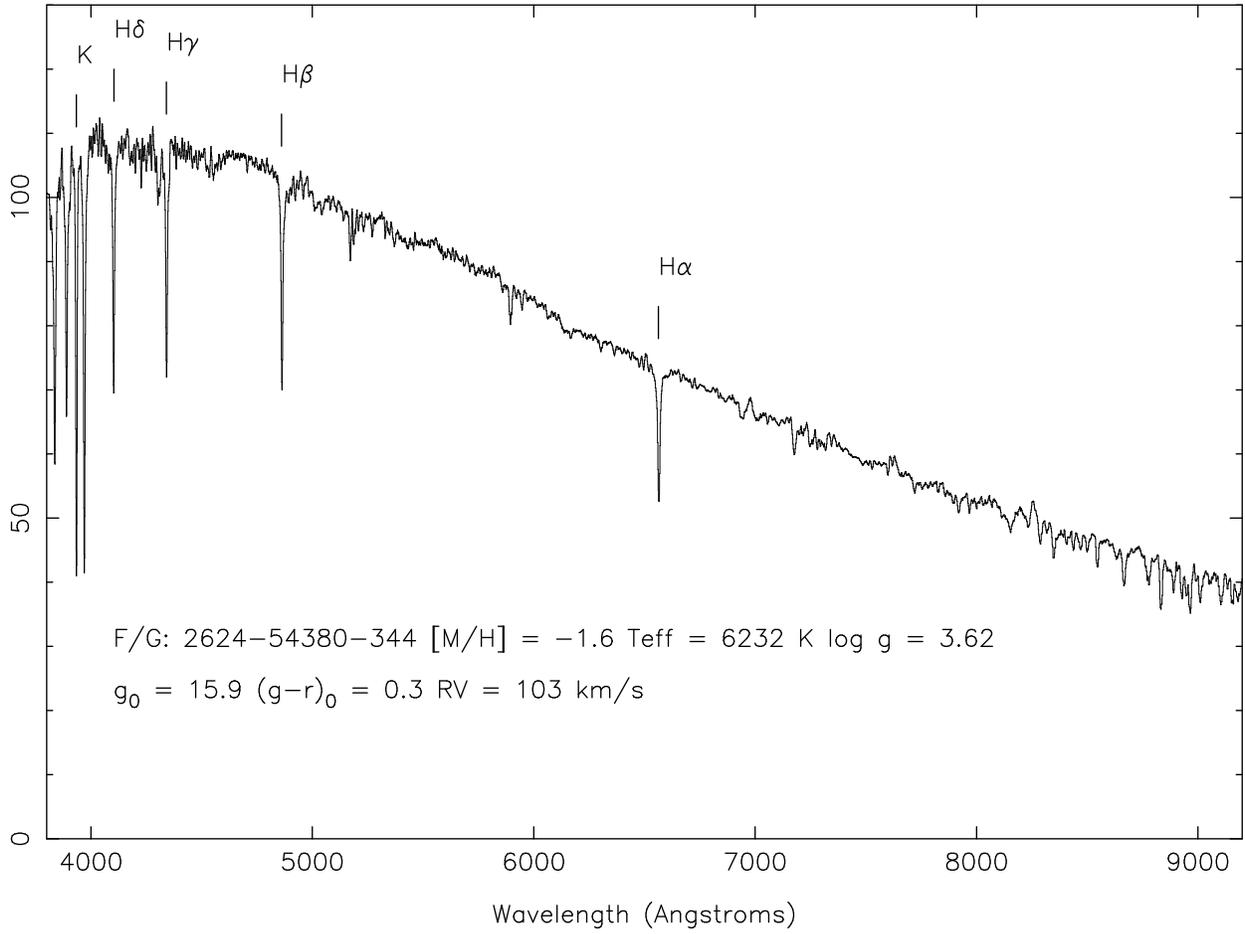}
\caption[FG] {
\footnotesize
a sample SEGUE F/G star. This category samples with only a simple 
$(g-r)_0$ color cut, and is therefore nearly unbiased.
\label{figfg}}
\end{figure}

%fig G
\begin{figure}
\centering
\includegraphics[scale=0.7,angle=-90]{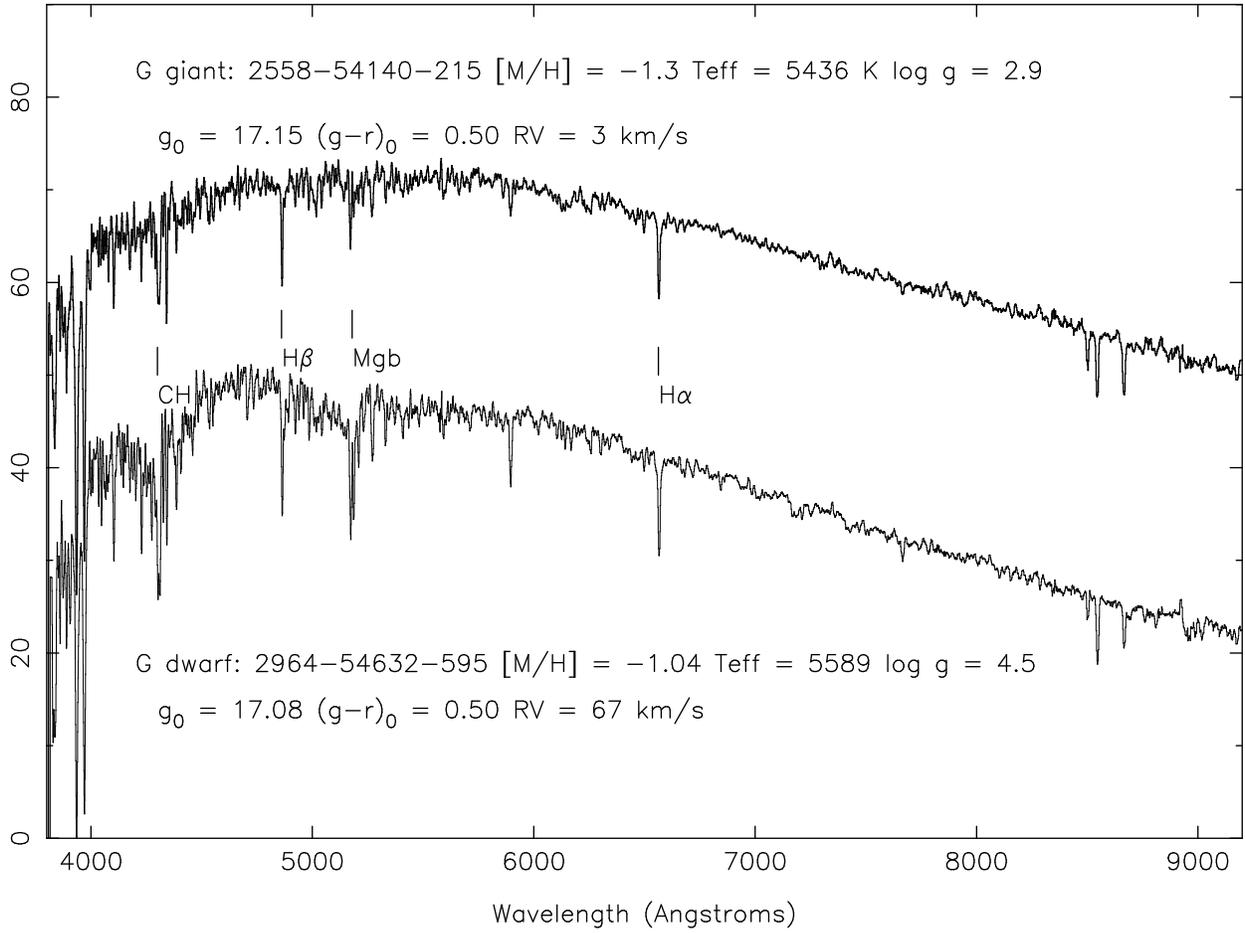}
\caption[G] {
\footnotesize
Upper: the weak Mg$b$/H feature at $\lambda 5140$ indicates that
this G star is actually a higher luminosity evolved subgiant, on its
way up the red giant branch.  Approximately 7\% of the SEGUE G star sample
shows low surface gravity.
Lower: a more typical G dwarf star. 
\label{figg}}
\end{figure}

%fig
\begin{figure}
\centering
\includegraphics[scale=0.65,angle=-90]{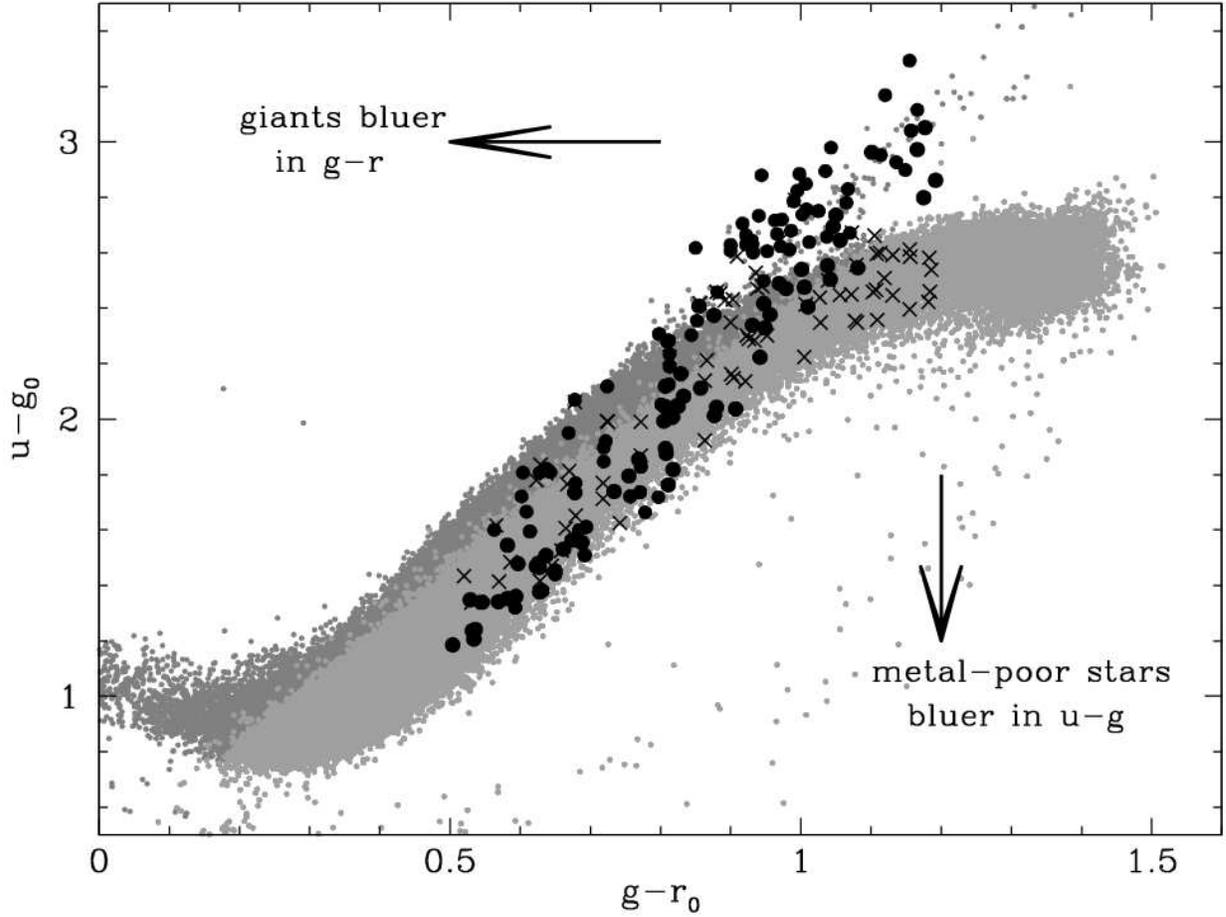}
\caption[Red Giant selection] {
\footnotesize
a demonstration of how the K giant locus crosses from the blue
($u-g$) side of the stellar locus at $g-r = 0.6$ over to the red
side of the locus at $g-r = 1.0$. M giants with TiO features appear
at $g-r > 1.3$ and $u-g > 3.0$.  The small dot shows the photometric
stellar locus in a SEGUE high latitude field (there are Sagittarius
M giants present at the end of the locus).  
 Stars with $l$-color greater than 0.07 are shown in light gray. 
Spectroscopically confirmed dwarfs are
shown with large crosses, giants with large filled circles.
}
%update the above figure caption
\end{figure}

%fig K + M III
\begin{figure}
\centering
\includegraphics[scale=0.7,angle=-90]{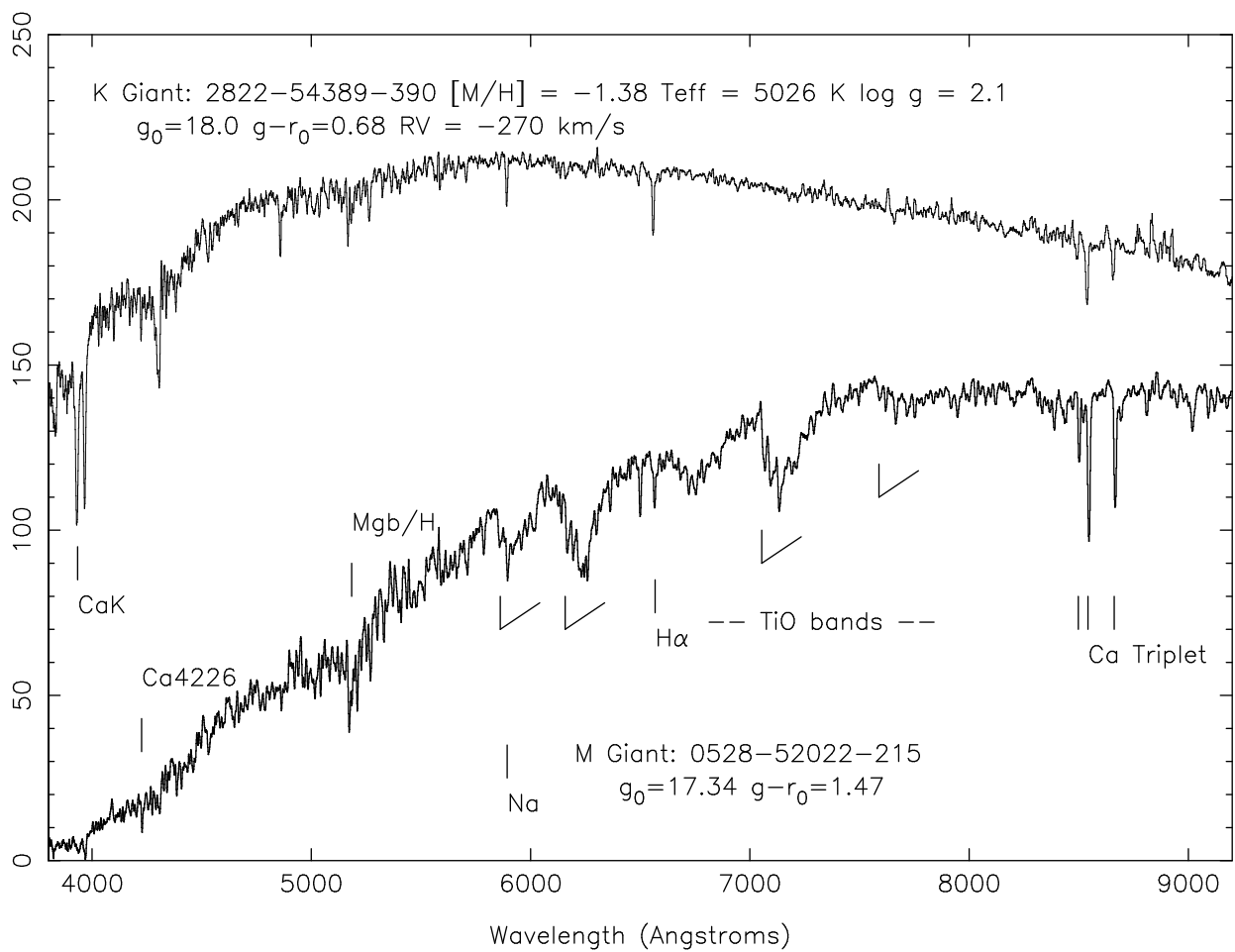}
\caption[KIII] {
\footnotesize
Top: a sample SEGUE K giant star, weak Mg$b$/H feature, the weak Ca 4226
and the weak Na I $\lambda 5890$.
Bottom: a sample SEGUE red K giant star.  Even though this spectrum shows
strong $\rm TiO$ bands, usually found in early M dwarfs, it is clearly a giant,
due to its relatively weak Mg$b$/H, nearly absent Na I$\lambda 5890$,
weak Ca $\lambda 4226$, and strong Ca IR triplet.  This star has
an absolute $M_g = -1$ and is at a distance of 45 kpc from the Sun in the
Sagittarius tidal stream's Northern leading tail.
\label{figkiii}}
\end{figure}

%fig K+M V
\begin{figure}
\centering
\includegraphics[scale=0.7,angle=-90]{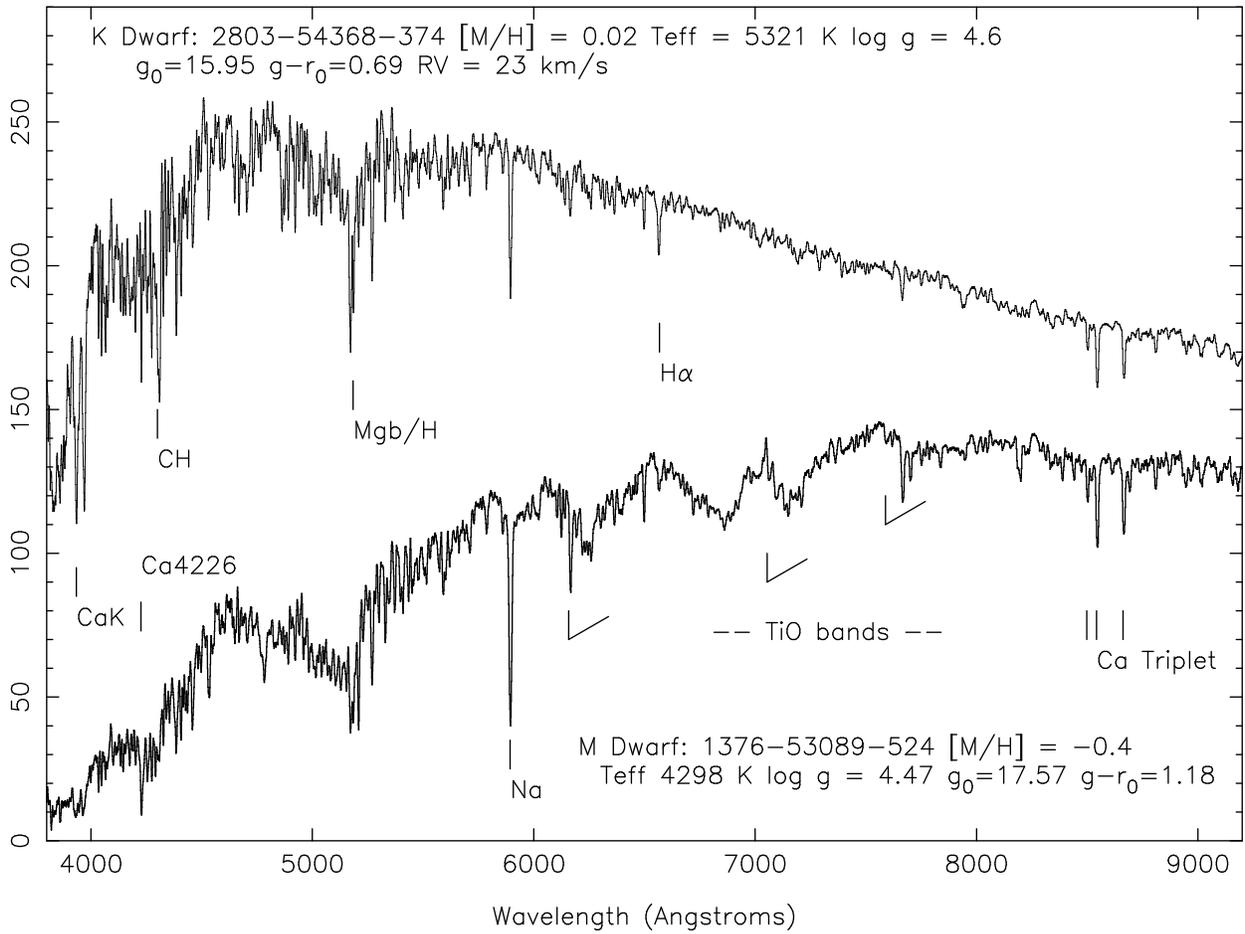}
\caption[KV] {
\footnotesize
Top: a sample SEGUE K dwarf, note the strong Mg$b$/H and Na I.
Lower: a sample SEGUE M dwarf, note the strong MgH, Na I 
and TiO molecular bands.
\label{figkv}}
\end{figure}

%fig sdM
\begin{figure}
\centering
\includegraphics[scale=0.7,angle=-90]{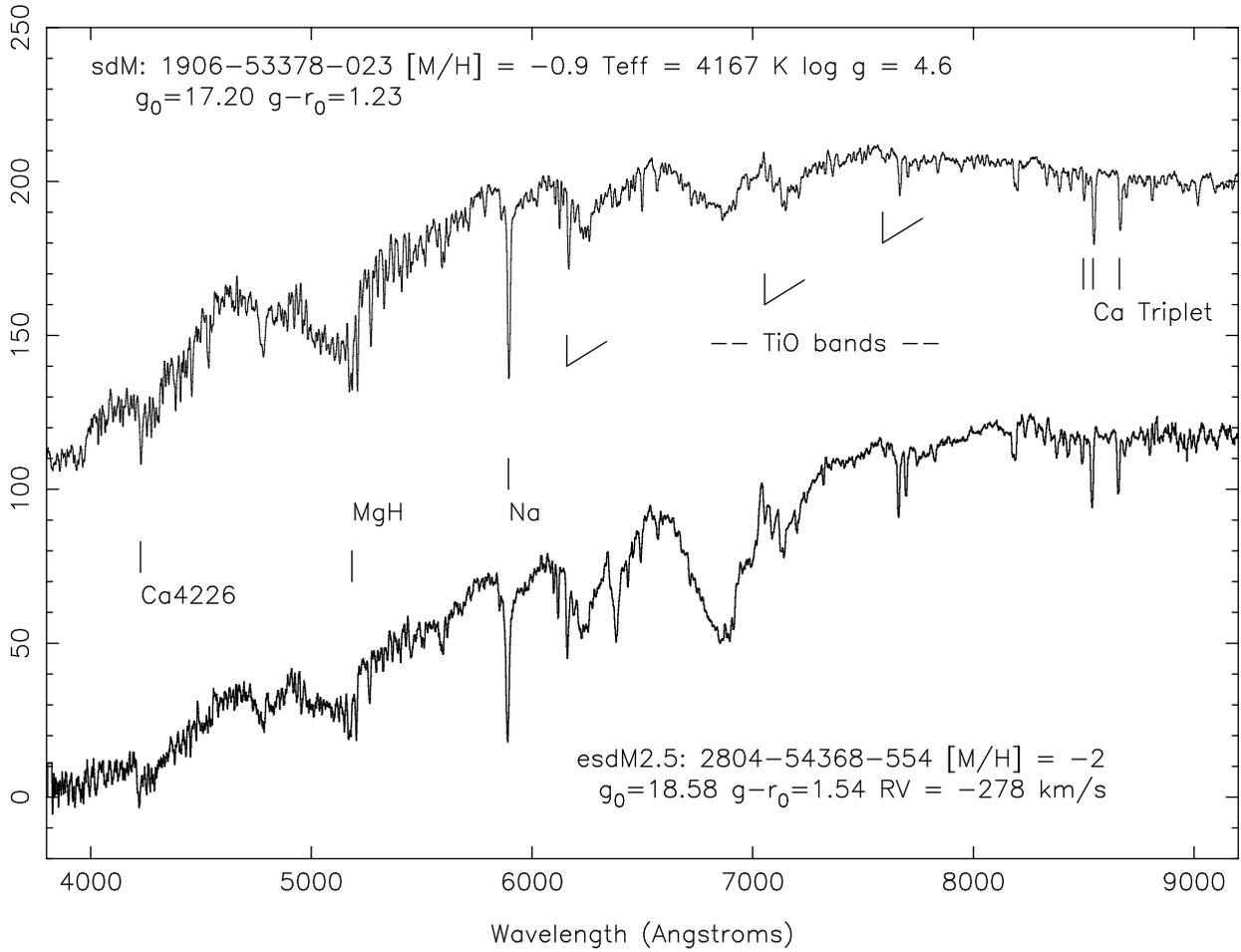}
\caption[sdM] {
\footnotesize
a sample SEGUE M subdwarf, the SEGUE estimate of metallicity is uncertain 
at these cool temperatures.
An AMNH proper motion and color selected SEGUE esdM2.5 extreme subdwarf.
Note that the TiO bands are nearly gone.
%what's that feature at 6900 A? --expert please XX
\label{figsdm}}
\end{figure}

%fig WDMS
\begin{figure}
\centering
\includegraphics[scale=0.7,angle=-90]{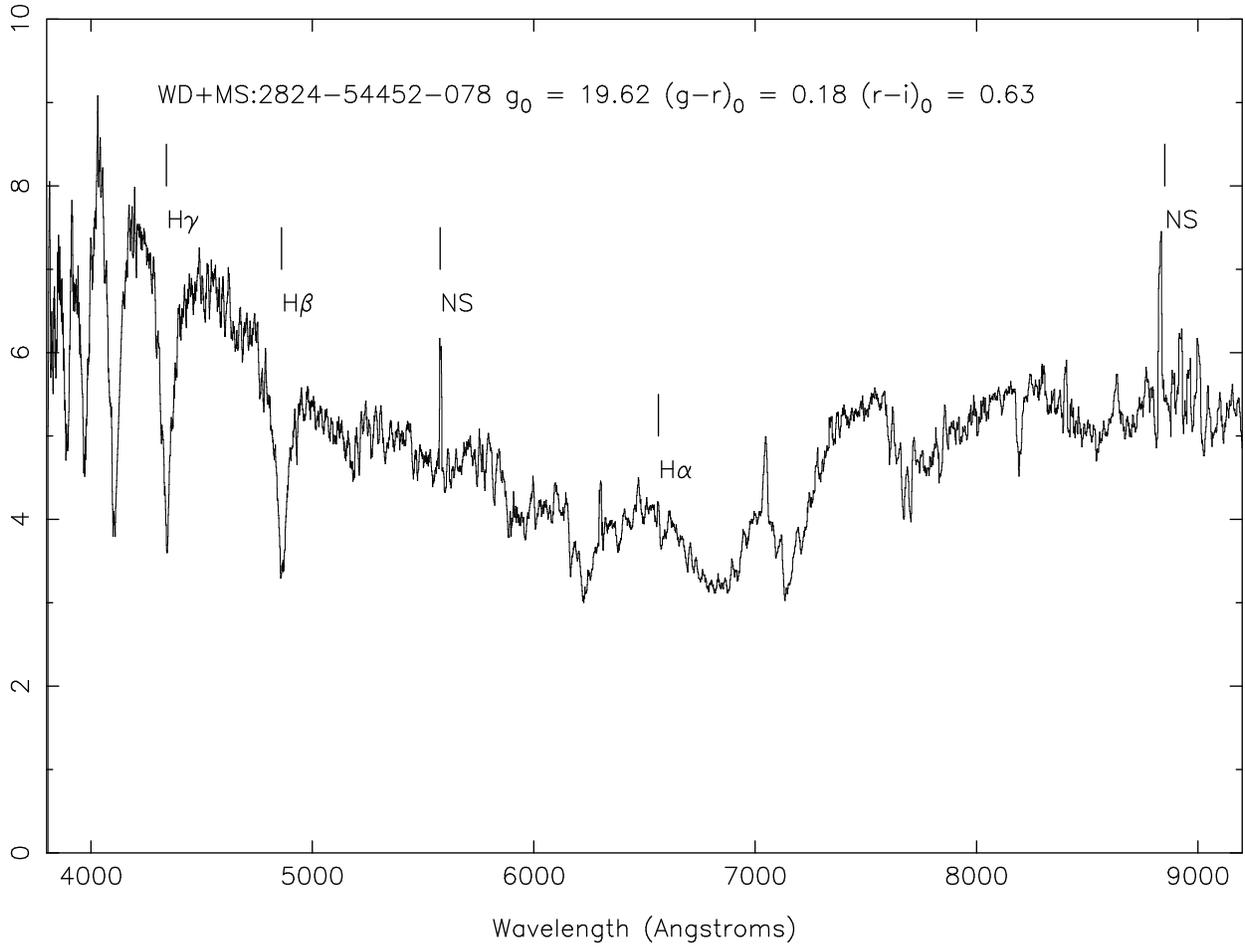}
\caption[WD/MS] {
\footnotesize
a sample SEGUE WD plus MS binary. 
Stellar parameters determined from modeling the SEGUE data are
a WD  of $\rm T_{\rm eff} \sim 19000~K$, a WD
mass of $\sim 0.6 ~\rm M_\odot$ and a spectral type
M4 for the companion.
\label{figmswd}}
\end{figure}

%fig L
\begin{figure}
\centering
\includegraphics[scale=0.7,angle=-90]{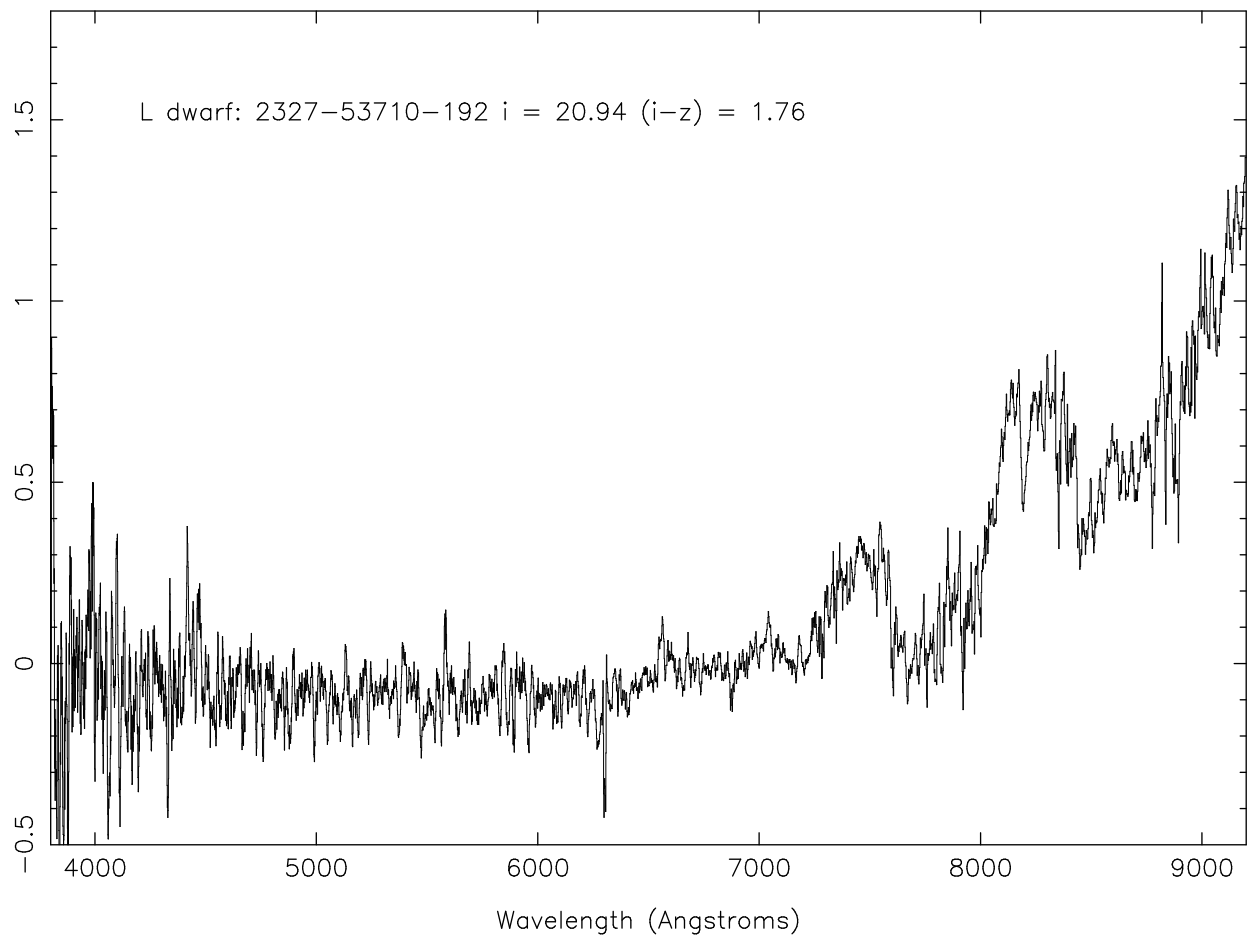}
\caption[L] {
\footnotesize
a sample SEGUE brown dwarf of approximate spectral type L0 (or M9). 
The $u$, $g$ and $r$ band flux is nearly completely missing from these
objects.
\label{figl}}
\end{figure}

%fig LL
\begin{figure}
\centering
\includegraphics[scale=0.7,angle=-90]{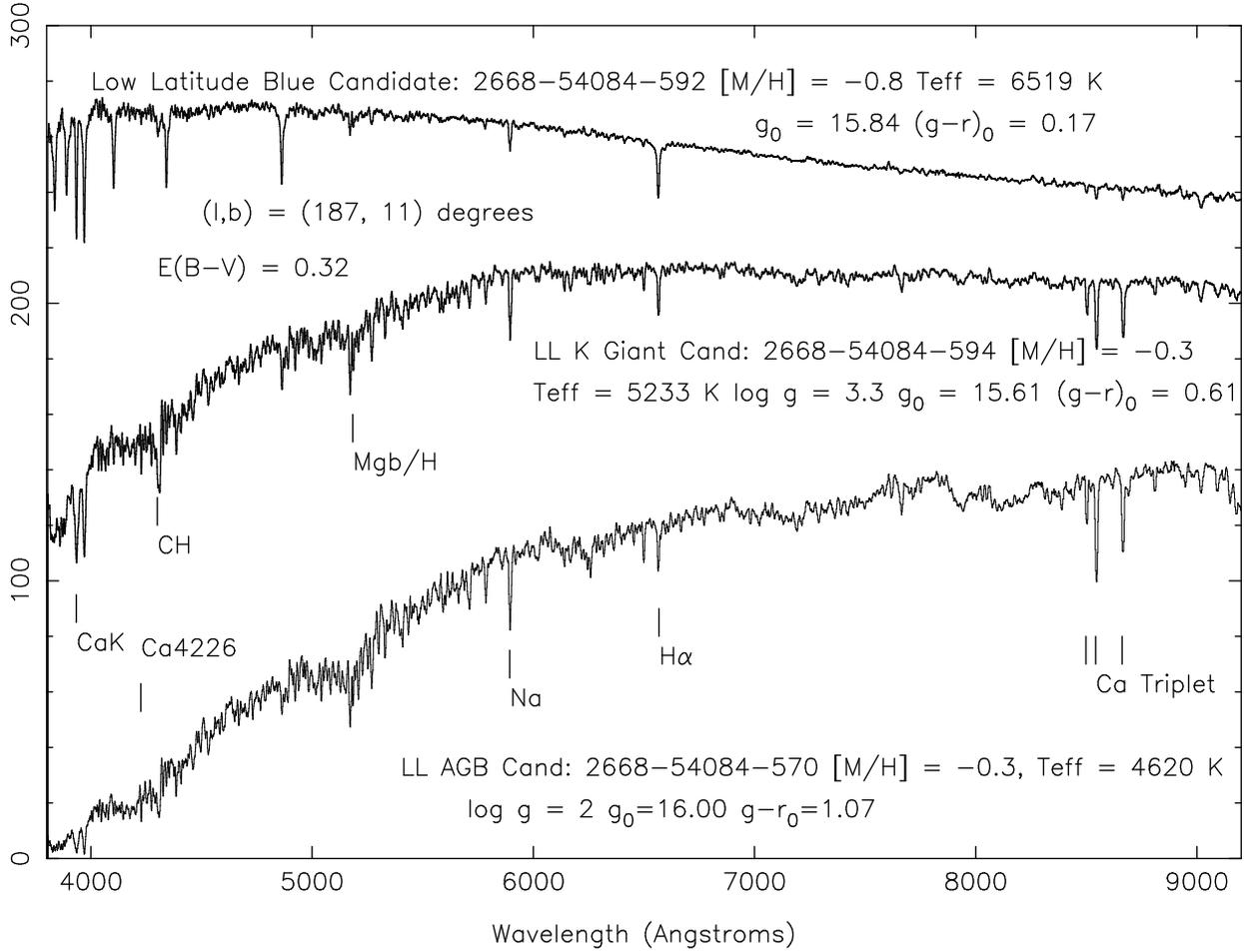}
\caption[Low-latitude selected objects] {
\footnotesize
a sample of three SEGUE low-latitude target-selection objects from
the same SEGUE plate.  Upper: a blue-tip candidate object, clearly of spectral
type F.  Middle: a K giant candidate.  Lower: a candidate AGB object.
\label{figll}}
\end{figure}

%fig Sample
\begin{figure}
\centering
\includegraphics[scale=0.65]{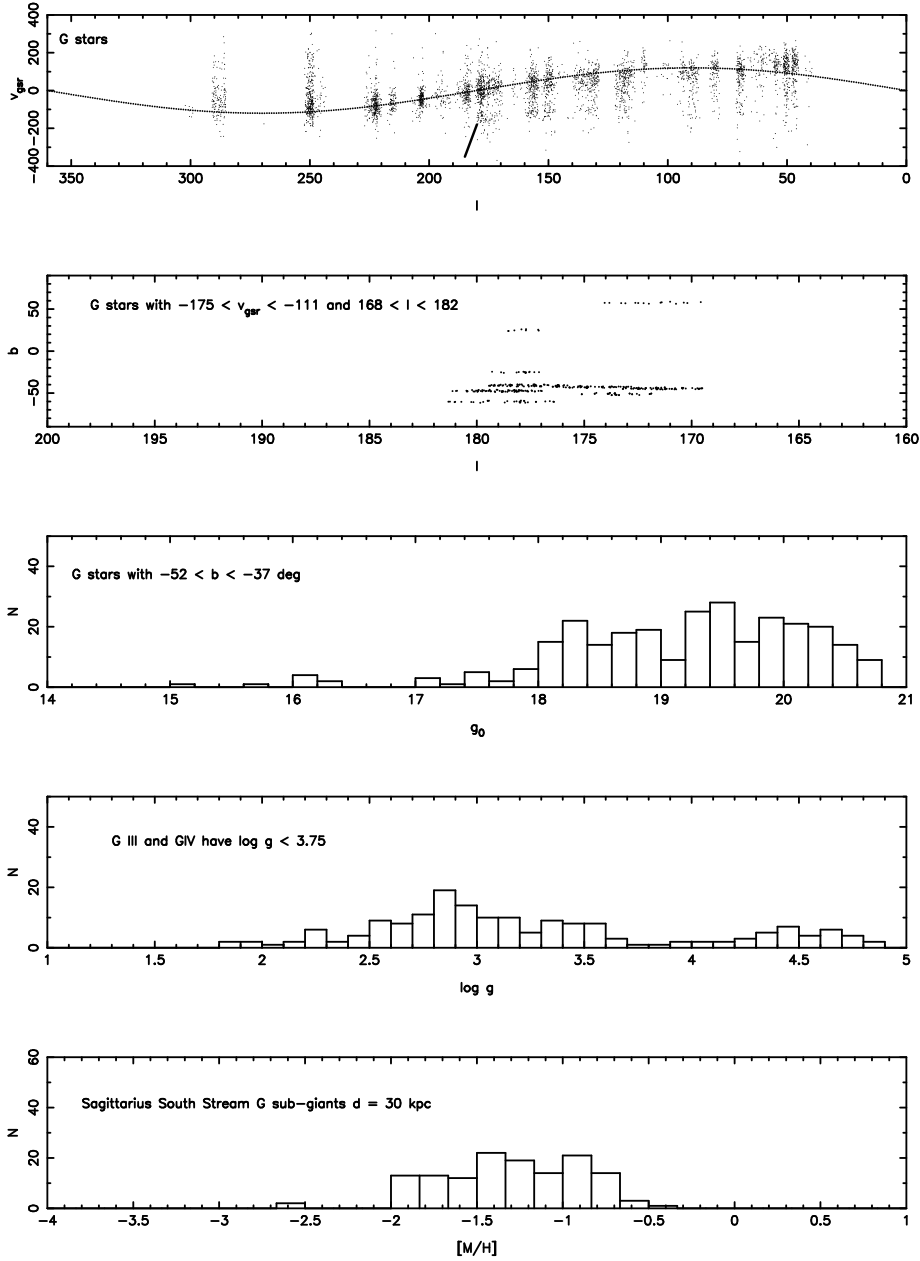}
\caption[Sample Query] {
\footnotesize
the five panels here show a step-by-step analysis of the SEGUE G star 
spectroscopic sample to isolate an interesting subpopulation.
Top: $(l,v_{gsr})$ plot of 61,343 measured SEGUE G star parameters from
the CAS database. A sinusoidal line, representing a rotating thick disk 
population is indicated with a black curve. 
Subsets of stars which stand out from the black curve are candidate 
halo dwarf galaxy or stream structures. A feature of interest 
for further study is highlighted with a black mark pointing to
$(l,v_{gsr}) = (170^\circ,-160\rm \>km~s^{-1}$). There are 
several other interesting
features which will not be explored further here.
Second from top: the set of stars selected in ($l,v_{gsr}$) is plotted
in $(l,b)$ to localize the interesting structure in position on the
sky.
Middle: the magnitude distribution of these stars is quite broad,
centered near $g\sim 19$, but covering in excess of a 2 mag range.
It is possible that they are all at the same approximate distance only
if we are sampling a steep subgiant branch population.
Second from bottom: a histogram of the SSPP measured surface gravity
of this sample indicates that nearly all these objects ($\rm log~g < 3.75$)
are in fact subgiants rather than dwarfs.  
Bottom:  the metallicity distribution of these objects indicates a 
metallicity of $\rm [M/H] = -1.4\pm 0.5$, with individual object errors
on metallicity of about 0.3.  The implied distance to these objects,
if they are one population, is about 30 kpc.  The spatial location
of these objects is consistent with previous studies of the Sagittarius
tidal stream Southern trailing tail.
}
\end{figure}

\end{document}